\documentclass[fleqn,usenatbib]{mnras}

\usepackage{newtxtext,newtxmath}

\usepackage[T1]{fontenc}
\usepackage{ae,aecompl}

\usepackage{graphicx}	
\usepackage{siunitx}
\usepackage{sidecap}

\title[J08205+0008 revisited]{A quantitative in-depth analysis of the prototype sdB+BD system SDSS\,J08205+0008 revisited in the \textit{Gaia} era}

\author[Schaffenroth et al.]{V. Schaffenroth$^{1}$ \thanks{E-mail: schaffenroth@astro.physik.uni-potsdam.de},
S. L. Casewell$^{2}$,
D. Schneider$^{3}$,
D. Kilkenny$^{4}$,
S. Geier$^{1}$,\and
U. Heber$^{3}$,
A. Irrgang$^{3}$,
N. Przybilla$^{5}$
T. R. Marsh$^{6}$,
S. P. Littlefair$^{7}$,
V. S. Dhillon$^{7,8}$
\\
$^{1}$Institute for Physics and Astronomy, University of Potsdam, Karl-Liebknecht-Str. 24/25, 14476 Potsdam, Germany\\
$^{2}$Department of Physics and Astronomy, University of Leicester, University Road, Leicester LE1 7RH, UK\\
$^{3}$Dr. Karl Remeis-Observatory \& ECAP, Astronomical Institute, Friedrich-Alexander-University Erlangen-Nuremberg,\\ Sternwartstr. 7, 96049 Bamberg, Germany\\
$^{4}$Department of Physics \& Astronomy, University of the Western Cape, Private Bag X17, Bellville 7535, South Africa\\
$^{5}$Institut f\"ur Astro- und Teilchenphysik, Universit\"at Innsbruck, Technikerstrasse 25, 6020, Innsbruck, Austria\\
$^{6}$Department of Physics, University of Warwick, Gibet Hill Road, Coventry CV4 7AL, UK\\
$^{7}$Department of Physics and Astronomy, University of Sheffield, Sheffield S3 7RH, UK\\
$^{8}$Instituto de Astrof\'{i}sica de Canarias (IAC), E-38200 La Laguna, Tenerife, Spain
}

\date{Accepted 2020 November 20. Received 2020 October 07; in original form 2020 August 04}

\pubyear{2020}

\begin{document}
\label{firstpage}
\pagerange{\pageref{firstpage}--\pageref{lastpage}}
\maketitle

\begin{abstract}
 Subdwarf B stars are core-helium burning stars located on the extreme horizontal branch. Extensive mass loss on the red giant branch is necessary to form them. 
 It has been proposed that substellar companions could lead to the required mass-loss when they are engulfed in the envelope of the red giant star. J08205+0008 was the first example of a hot subdwarf star with a close, substellar companion candidate to be found. 
Here we perform an in-depth re-analysis of this important system with much higher quality data allowing additional analysis methods. From the higher resolution spectra obtained with ESO-VLT/XSHOOTER we derive the chemical abundances of the hot subdwarf as well as its rotational velocity. 
Using the {\it Gaia} parallax and a fit to the spectral energy distribution in the secondary eclipse, tight constraints to the radius of the hot subdwarf 
are derived. From a long-term photometric campaign 
we detected a significant period decrease of $-3.2(8)\cdot 10^{-12} \,\rm dd^{-1}$. This can be explained by the non-synchronised hot subdwarf star being spun up by tidal interactions forcing it to become synchronised. From the rate of period decrease we could derive the synchronisation timescale to be 4 Myr, much smaller than the lifetime on EHB. 
By combining all different methods we could constrain the hot subdwarf to a mass of $0.39-0.50\,\rm M_\odot$ and a radius of $R_{\rm sdB}=0.194\pm0.008\,\rm R_\odot$, and the companion to $0.061-0.071\rm\,M_\odot$ with a radius of $R_{\rm comp}=0.092 \pm 0.005\,\rm R_\odot$, below the hydrogen burning limit. We therefore confirm that the companion is most likely a massive brown dwarf.
\end{abstract}

\begin{keywords}
Stars: fundamental parameters -- Stars: atmospheres -- Stars: abundances -- Stars: subdwarfs -- Stars: horizontal-branch -- Stars: low-mass
\end{keywords}



\section{Introduction}
Subluminous B stars (subdwarf B stars or sdBs) are stars with thin hydrogen envelopes, currently undergoing helium-core burning, which are found on the extreme horizontal branch (EHB). Their masses were determined to be around $0.47$\,M$_{\rm \odot}$ \citep{heber09, heber16}. About half of the known single-lined sdB stars are found to be members of short-period binaries \citep[P $\lesssim$ 30 d, most even with P $\lesssim$ 10 d,][]{maxted01,napiwotzki04a,kupfer15}. A large mass loss on the red giant branch (RGB) is required to form these stars, which can be caused by mass transfer to the companion, either via stable Roche lobe overflow or the formation and eventual ejection of a common envelope \citep{han02,han03}. 
For the existence of apparently single sdB stars binary evolution might play an important role as well, as such stars could be remnants of helium white dwarf mergers \citep{webbink84,ibentutukov84} or from engulfing a substellar object, which might get destroyed in the process \citep{soker98,nelemans98}.

Eclipsing sdB+dM binaries (HW\,Vir systems) having short orbital periods ($0.05-1\,{\rm d}$) and low companion masses between $0.06$\,M$_{\rm \odot}$ and $0.2$\,M$_{\rm \odot}$ \citep[see][for a summary of all known HW Vir systems]{schaffenroth18,schaffenroth19} have been known for decades \citep{menzies} and illustrate that objects close to the nuclear burning limit of $\sim 0.070-0.076$\,M$_{\rm \odot}$ for an object of solar metallicity and up to $0.09\,\rm M_{\rm \odot}$ for metal-poor objects \citep[see][for a review]{2014AJ....147...94D} can eject a common envelope and lead to the formation of an sdB. The light travel-time technique was used to detect substellar companion candidates to sdB stars \citep[e.g.][and references therein]{beuermann12, kilkenny12}. However, in these systems the substellar companions
have wide orbits and therefore cannot have influenced the evolution of the host star.

The short-period eclipsing HW~Vir type binary SDSS J082053.53+000843.4, hereafter J08205+0008, was discovered as part of the MUCHFUSS project \citep{geier11a, geier11b}. \citet{geier11c} derived an orbital solution based on time resolved medium resolution spectra from SDSS \citep{abazajian09} and ESO-NTT/EFOSC2. The best fit orbital period was $P_{\rm orb}=P=0.096\pm0.001\,{\rm d}$ and the radial velocity (RV) semi-amplitude $K=47.4\pm1.9\,{\rm km\,s^{-1}}$ of the sdB. An analysis of a light curve taken with Merope on the Mercator telescope allowed them to constrain the inclination of the system to $85.8^{\rm \circ}\pm0.16$. 

The analysis resulted in two different possible solutions for the fundamental parameters of the sdB and the companion. As the sdB sits on the EHB the most likely solution is a core-He burning object with a mass close to the canonical mass for the He flash of $0.47 \,\rm M_\odot$. Population synthesis models \citep{han02,han03} predict a mass range of $M_{\rm sdB}=0.37-0.48\,\rm M_\odot$, which is confirmed by asteroseismological measurements \citep{fontaine12}. A more massive ($2-3\,\rm M_\odot$) progenitor star would ignite the He core under non-degenerate conditions and lower masses down to $0.3\,\rm M_\odot$ are possible. Due to the shorter lifetime of the progenitors such lower mass hot subdwarfs would also be younger. Higher masses for the sdB were ruled out as contemporary theory did not predict that. By a combined analysis of the spectrum and the light curve the companion was derived to have a mass of $0.068\pm0.003\,\rm M_\odot$. However, the derived companion radius for this solution was significantly larger than predicted by theory.

The second solution that was consistent with the atmospheric parameters was a post-RGB star with an even lower mass of only $0.25\,\rm M_\odot$. Such an object can be formed whenever the evolution of the star on the RGB is interrupted due to the ejection of a common envelope before the stellar core mass reaches the mass, which is required for helium ignition. Those post-RGB stars, also called pre-helium white dwarfs, cross the EHB and evolve directly to white dwarfs. 
In this case the companion was determined to have a mass of $0.045\pm0.003\,\rm M_\odot$ and the radius was perfectly consistent with theoretical predictions.

The discovery of J08205+0008  was followed by the discovery of two more eclipsing systems with brown dwarf (BD) companions, J162256+473051 \citep{schaffenroth14} and V2008-1753 \citep{schaffenroth15}, both with periods of less than 2 hours. Two non-eclipsing systems were also discovered by \citet{schaffenroth14a}, and a subsequent analysis of a larger population of 26 candidate binary systems by \citet{schaffenroth18} suggests that the  fraction of sdB stars with close substellar companions is as high as 3 per cent, much higher than the $0.5\pm0.3$ per cent that is estimated for brown dwarf companions to white dwarfs (e.g. \citealt{steele11}). Seven  of the nine known white dwarf-brown dwarf systems have primary masses within the mass range for a He-core burning hot subdwarf and might therefore have evolved through this phase before. 

In this paper, we present new phase-resolved spectra of J08205+0008 obtained with ESO-VLT/UVES and XSHOOTER and high cadence light curves with ESO-NTT/ULTRACAM. Combining these datasets, we have refined the radial velocity solution and light curve fit. We performed an in-depth analysis of the sdB atmosphere and a fit of the spectral energy distribution using the ULTRACAM secondary eclipse measurements to better constrain the radius and mass of the sdB primary and the companion. We also present our photometric campaign using the SAAO/1m-telescope and BUSCA mounted at the Calar Alto/2.2m telescope which has been underway for more than 10 years now, and which has allowed us to derive variations of the orbital period.

\section{Spectroscopic and photometric data}

\subsection{UVES spectroscopy}
We obtained time-resolved, high resolution ($R\simeq40\,000$) spectroscopy of J08205+0008  with ESO-VLT/UVES \citep{dekker04} on the night of 2011-04-05 as part of program 087.D-0185(A). In total 33 single spectra with exposure times of $300\,{\rm s}$ were taken consecutively to cover the whole orbit of the binary.  We used the 1" slit in seeing of $\sim$ 1" and airmass ranging from 1.1. to 1.5. The spectra were taken using cross dispersers CD\#2 and CD\#3 on the blue and red chips respectively to cover a wavelength range from \SI{3300}{\angstrom} to \SI{6600}{\angstrom} with two small gaps ($\simeq$ \SI{100}{\angstrom}) at \SI{4600}{\angstrom} and \SI{5600}{\angstrom}. 

The data reduction was done with the UVES reduction pipeline in the \textsc{midas} package \citep{midas}. In order to ensure an accurate normalisation of the spectra, two spectra of the DQ type white dwarf WD\,0806$-$661 were also taken \citep{subasavage09}. Since the optical spectrum of this carbon-rich white dwarf is featureless, we divided our data by the co-added and smoothed spectrum of this star.

The individual spectra of J08205+0008 were then radial velocity corrected using the derived radial velocity of the individual spectra as described in Sect. \ref{rvs} and co-added for the atmospheric analysis. 
In this way, we increased the signal-to-noise ratio to S/N\,$\sim$\,90, which was essential for the subsequent quantitative analysis.

\subsection{XSHOOTER spectroscopy}
We obtained time resolved spectra of J08205+0008 with ESO-VLT/XSHOOTER \citep{vernet11} as part of programme 098.C-0754(A). The data were observed on the night of 2017-02-17 with 300~s exposure times in nod mode and in seeing of $0.5-0.8$". We obtained 24 spectra covering the whole orbital phase (see Fig. \ref{change of atmospheric parameters vs. orbital phase}) in each of the UVB ($R\sim$\,5400), VIS ($R\sim$\,8900) and NIR ($R\sim$\,5600) arms with the $0.9-1.0$" slits. The spectra were reduced using the ESO \textsc{reflex} package \citep{reflex} and the specific XSHOOTER routines in nod mode for the NIR arm, and in stare mode for the UVB and VIS arms.

To correct the astronomical observations for atmospheric absorption features in the VIS and NIR arms, we did not require any observations of telluric standard stars, as we used the \texttt{molecfit} software, which is based on fitting synthetic transmission spectra calculated by a radiative transfer code to the astronomical data \citep{2015A&A...576A..77S, 2015A&A...576A..78K}. The parameter set-up (fitted molecules, relative molecular column densities, degree of polynomial for the continuum fit, etc.) for the telluric absorption correction evaluation of the NIR-arm spectra were used according to Table 3 of \citet{2015A&A...576A..78K}. Unfortunately, the NIR arm spectra could not be used after the telluric corrections since the signal-to-noise (S/N) ratio and the fluxes are too low. Figure \ref{VIS X-Shooter arm before and after telluric correction with molecfit} shows an example comparison between the original and the telluric absorption corrected XSHOOTER VIS arm spectra. The quality of the telluric correction is sufficient to allow us to make use of the hydrogen Paschen series for the quantitative spectral analysis.

Accurate radial velocity measurements for the single XSHOOTER spectra were performed within the analysis program SPAS \citep{2009PhDT.......273H}, whereby selected sharp metal lines listed in Table \ref{list of lines detected} were used. We used a combination of Lorentzian, Gaussian and straight line (in order to model the slope of the continuum) function to fit the line profiles of the selected absorption lines. 
After having corrected all single spectra by the averaged radial velocities, a co-added spectrum was created in order to achieve S/N\,$\sim 460/260$ in the UVB and VIS channels, respectively.

The co-added spectrum then was normalized also within SPAS. Numerous anchor points were set where the stellar continuum to be normalized was assumed. In this way, the continuum was approximated by a spline function. To obtain the normalized spectrum, the original spectrum was divided by the spline. 

\subsection{ULTRACAM photometry}\label{ULTRACAM}
Light curves in the SDSS $u'g'r'$ filters were obtained simultaneously using the ULTRACAM instrument \citep{dhillon} on the 3.5m-ESO-NTT at La Silla. The photometry was taken on the night of 2017-03-19 with airmass $1.15-1.28$ as part of programme 098.D-679 (PI; Schaffenroth). The data were taken in full frame mode with 1$\times$1 binning and the slow readout speed with exposure times of 5.75~s resulting in 1755 frames obtained over the full orbit of the system. The dead-time between each exposure was only 25 msec. We reduced the data using the HiperCam pipeline (\url{http://deneb.astro.warwick.ac.uk/phsaap/hipercam/docs/html}). The flux of the sources was determined using aperture photometry with an aperture scaled variably according to the full width at half-maximum. The flux relative to a comparison star within the field of view (08:20:51.941 +00:08:21.64) was determined to account for any variations in observing conditions.  This reference star has SDSS magnitudes of $u'$=15.014$\pm$0.004, $g'$=13.868$\pm$0.003, $r'$=13.552$\pm$0.003 which were used to provide an absolute calibration for the light curve.

\subsection{SAAO photometry}\label{saao}


All the photometry was obtained on the 1m (Elizabeth) telescope at the Sutherland
site of the South African Astronomical Observatory (SAAO). Nearly all observations
were made with the STE3 CCD, except for the last two (Table H1), which were made
with the STE4 camera. The two cameras are very similar with the only difference being the pixel size as the STE3 is $512\times512$  pixels in size and the STE4 is  $1024\times1024$. We used a  $2\times2$ pre-binned mode for each CCD resulting in a read-out time of around 5 and 20s, respectively, so that
with typical exposure times around 10-12s, the time resolution of STE4 is only about
half as good as STE3. Data reduction and eclipse analysis were carried out as outlined
in \citep{kilkenny11}; in the case of J08205+008, there are several useful comparison 
stars, even in the STE3 field, and - given that efforts were made to observe eclipses           near the meridian - usually there were no obvious "drifts" caused by differential
extinction effects. In the few cases where such trends were seen, these were removed
with a linear fit to the data from just before ingress and just after egress. The stability of the
procedures (and the SAAO time system over a long time base) is demonstrated by the
constant-period system AA Dor \citep[Fig.1 of ][]{kilkenny14}  and by the intercomparisons      in Fig. 8 of \citet{baran18}, for example.

\subsection{BUSCA photometry}\label{calaralto}
Photometric follow-up data were also taken with the Bonn University
Simultaneous CAmera \cite[BUSCA; see][]{busca}, which is mounted to the
2.2 m-telescope located at the Calar Alto Observatory in Spain.
This instrument observes in four bands simultaneously giving a very accurate eclipse measurement and good estimate of the errors. The four different bands we used in our observation are given solely by the intrinsic transmission curve given by the
beam splitters (UB, BB, RB, IB, \url{http://www.caha.es/CAHA/Instruments/BUSCA/bands.txt}) and the efficiency of the CCDs, as no filters where used to ensure that all the visible light is used most efficiently. 

The data were taken during one run on  25 Feb 2011 and 1 Mar 2011. We used an exposure time of 30~s. Small windows were defined around the target and four comparison stars to decrease the read-out time from 2 min to 15 s. As comparison stars we used stars with similar magnitudes ($\Delta m< 2$mag) in all SDSS bands from $u$ to $z$, which have been pre-selected using the SDSS DR 9 skyserver (\url{http://skyserver.sdss.org/dr9/en/}). The data were reduced using IRAF\footnote{http://iraf.noao.edu/}; a standard CCD reduction was performed using the IRAF tools for bias- and flatfield-correction. Then the light curves of the target and the comparison stars were extracted using the aperture photometry
package of DAOPHOT. The final light was constructed by dividing the light curve of the target by the light curves of the comparison stars.

\section{Analysis}

\subsection{The hybrid LTE/NLTE approach and spectroscopic analysis}\label{The spectroscopic analysis technique}
Both the co-added UVES and XSHOOTER (UVB and VIS arm) spectra were analyzed using the same hybrid local thermodynamic equilibrium (LTE)/non-LTE (NLTE) model atmospheric approach. This approach has been successfully used to analyze B-type stars (see, for instance, \citealt{2006BaltA..15..107P, 2006A&A...445.1099P, 2011JPhCS.328a2015P}; \citealt{2007A&A...467..295N, Nieva_2008}) and is based on the three generic codes \textsc{atlas12} \citep{1996ASPC..108..160K}, \textsc{detail}, and \textsc{surface} (\citealt{1981PhDT.......113G}; \citealt{Butler_1985}, extended and updated). 

Based on the mean metallicity for hot subdwarf B stars according to \citet{naslim13}, metal-rich and line-blanketed, plane-parallel and chemically homogeneous model atmospheres in hydrostatic and radiative equilibrium were computed in LTE within \textsc{atlas12}. 
Occupation number densities in NLTE for hydrogen, helium, and for selected metals (see Table \ref{summary of model atoms used for the hybrid LTE/NLTE approach}) were computed with \textsc{detail} by solving the coupled radiative transfer and statistical equilibrium equations. The emergent flux spectrum was synthesized afterwards within \textsc{surface}, making use of realistic line-broadening data.
Recent improvements to all three codes \citep[see][for details]{2018A&A...615L...5I} with regard to NLTE effects on the atmospheric structure as well as the implementation of the occupation probability formalism \citep{1994A&A...282..151H} for H\,{\sc i} and He\,{\sc ii} and new Stark broadening tables for H \citep{2009ApJ...696.1755T} and He\,{\sc i} \citep{1997ApJS..108..559B} are considered as well. For applications of these models to sdB stars see \citet{schneider18}. 


We included spectral lines of H and \ion{He}{i}, and in addition, various metals in order to precisely measure the projected rotational velocity ($v\sin i$), radial velocity ($v_{\rm rad}$), and chemical abundances of J08205+0008. The calculation of the individual model spectra is presented in detail in \citet{2014A&A...565A..63I}.
In Table \ref{Hybrid LTE/NLTE model grid used for the quantitative spectral analysis of SDSS J08205+0008}, the covered effective temperatures, surface gravities, helium and metal abundances for the hybrid LTE/NLTE model grid used are listed.

The quantitative spectral analysis followed the methodology outlined in detail in \citet{2014A&A...565A..63I}, that is, the entire useful spectrum and all 15 free parameters ($T_{\rm eff}$, $\log g$, $v_{\rm rad}$, $v\sin i$, $\log{n(\text{He})}:=\log{\left[\frac{\text{N(He)}}{\text{N(all elements)}}\right]}$, plus abundances of all metals listed in Table \ref{summary of model atoms used for the hybrid LTE/NLTE approach}) were simultaneously fitted using standard $\chi^2$ minimization techniques. Macroturbulence $\zeta$ and microturbulence $\xi$ were fixed to zero because there is no indication for additional line-broadening due to these effects in sdB stars \citep[see, for instance,][]{geier:2012,schneider18}.

\begin{table}
\caption{Metal abundances of J08205+0008 derived from XSHOOTER and UVES.$^\dagger$}\label{abundance table}
\centering
\begin{tabular}{lll}
\hline\hline
Parameter & XSHOOTER & UVES\\
\hline
$\log{n(\text{C})}$ & $-4.38\pm0.05$ & $-4.39^{+0.04}_{-0.03}$\\
$\log{n(\text{N})}$ & $-4.00^{+0.03}_{-0.02}$ & $-3.98\pm0.03$\\
$\log{n(\text{O})}$ & $-4.01^{+0.05}_{-0.06}$ & $-3.86^{+0.07}_{-0.06}$\\
$\log{n(\text{Ne})}$ & $\leq -6.00$ & $\leq -6.00$\\
$\log{n(\text{Mg})}$ & $-4.98^{+0.05}_{-0.04}$ & $-5.03\pm0.05$\\
$\log{n(\text{Al})}$ & $-6.20\pm0.03$ & $\leq -6.00$\\
$\log{n(\text{Si})}$ & $-5.13\pm0.04$ & $-5.17^{+0.07}_{-0.08}$\\
$\log{n(\text{S})}$ & $-5.31^{+0.11}_{-0.10}$ & $-5.12^{+0.06}_{-0.08}$\\
$\log{n(\text{Ar})}$ & $-5.54^{+0.15}_{-0.27}$ & $-5.32^{+0.19}_{-0.23}$\\
$\log{n(\text{Fe})}$ & $-4.39\pm0.04$ & $-4.41^{+0.04}_{-0.05}$\\
\hline
\multicolumn{3}{l}{$\dagger$: Including 1$\sigma$ statistical and systematic errors.}\\
\multicolumn{3}{l}{$\log{n(\text{X})}:=\log{\left[\frac{\text{N(X)}}{\text{N(all elements)}}\right]}$}
\end{tabular}
\end{table}\noindent

\subsection{Effective temperature, surface gravity, helium content and metal abundances} \label{Effective temperature, surface gravity, and helium content}
\begin{figure*}
 \begin{center}
 \includegraphics[trim = 0cm 0cm 0cm 0cm, clip, scale=0.49]{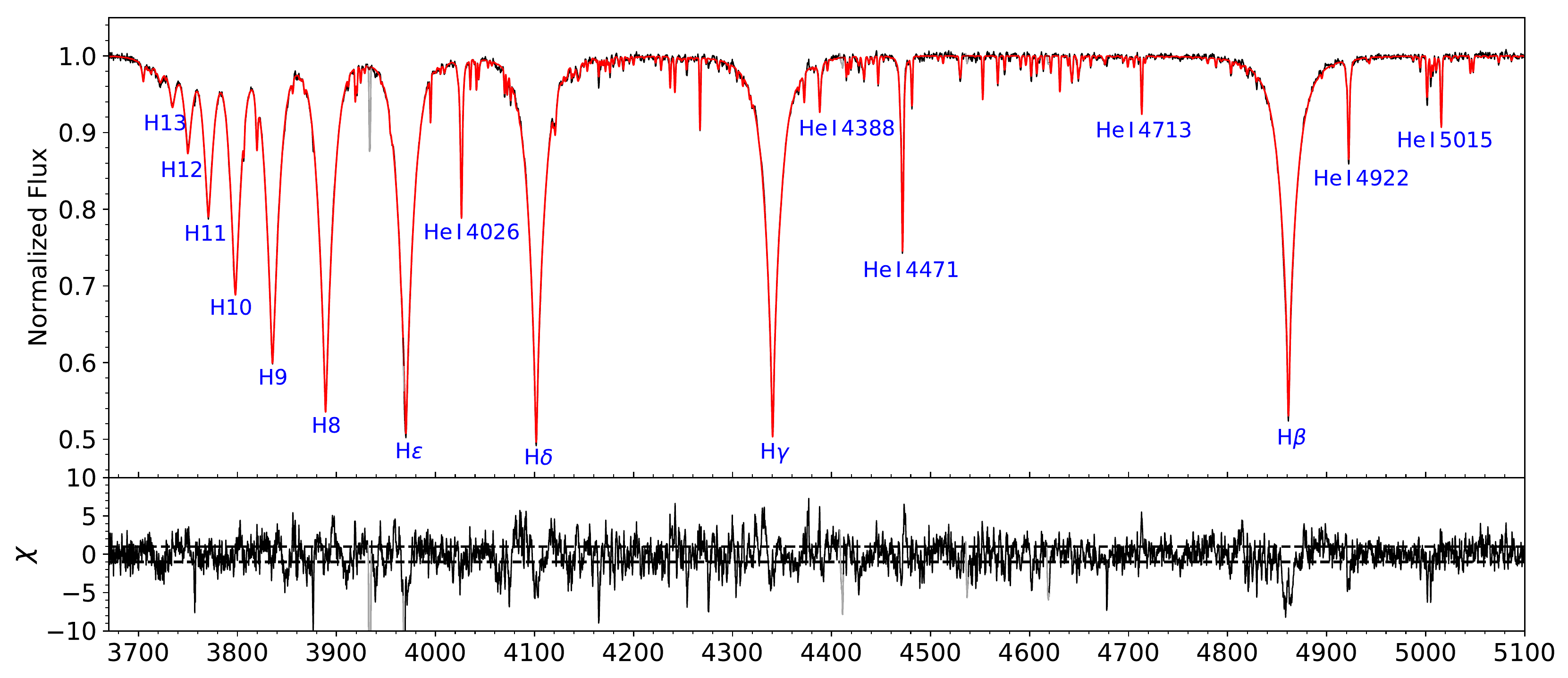}
  \includegraphics[trim = 0cm 0cm 0cm 0cm, clip, scale=0.49]{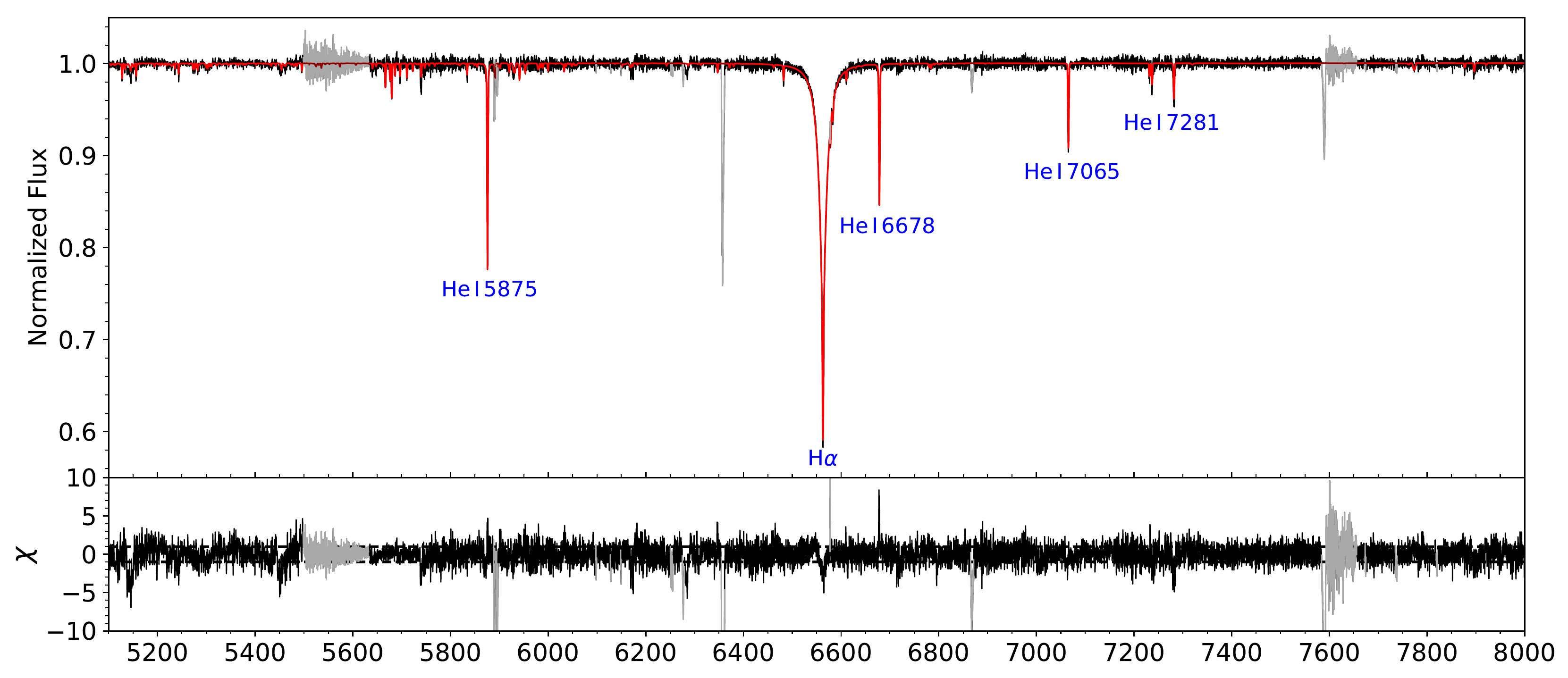}
  \includegraphics[trim = 0cm 0cm 0cm 0cm, clip, scale=0.49]{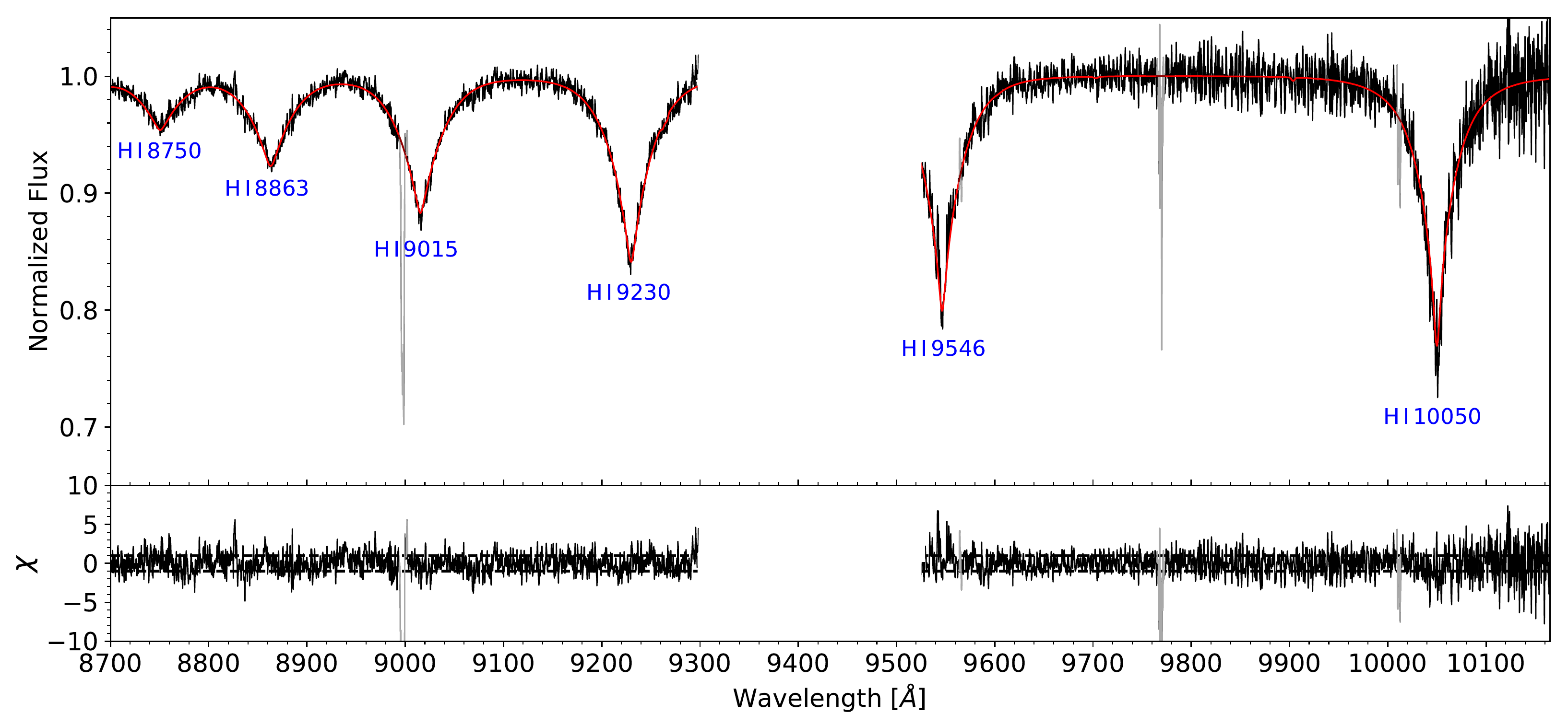}
    \caption{Comparison between observation (solid black line) and global best fit (solid red line) for selected spectral ranges in the co-added XSHOOTER spectrum of J08205+0008. Prominent hydrogen and $\ion{He}{i}$ lines are marked by blue labels and the residuals for each spectral range are shown in the bottom panels, whereby the dashed horizontal lines mark mark deviations in terms of $\pm1\sigma$, i.e., values of $\chi=\pm1$ (0.2\% in UVB and 0.4\% in VIS, respectively).} Additional absorption lines are caused by metals (see Fig. \ref{SDSS0820_metal_line_profiles}). Spectral regions, which have been excluded from the fit, are marked in grey (observation) and dark red (model), respectively. Since the range between $\ion{H}{i}$ \SI{9230}{\angstrom} and $\ion{H}{i}$ \SI{9546}{\angstrom} strongly suffers from telluric lines (even after the telluric correction with \texttt{molecfit}), it is excluded from the figure.
     \label{XSHOOTER hydrogen and helium lines 1}
     \end{center}
 \end{figure*}

The excellent match of the global best fit model spectrum to the observed one is demonstrated in Fig. \ref{XSHOOTER hydrogen and helium lines 1} for selected spectral ranges in the co-added XSHOOTER spectrum of J08205+0008 (UVB + VIS arm).

The wide spectral range covered by the XSHOOTER spectra allowed, besides the typical hydrogen Balmer series and prominent $\ion{He}{i}$ lines in the optical, Paschen lines to be included in the fit, which provides additional information that previously could not be used in sdB spectral analysis, but provides important consistency checks.\\

In the framework of our spectral analysis, we also tested for variations of the atmospheric parameters over the orbital phase as seen in other reflection effect systems \citep[e.g.][]{heber04,schaffenroth13}. As expected, due to the relatively weak reflection effect of less than 5\%, the variations were within the total uncertainties given in the following and can therefore be neglected (see also Fig. \ref{change of atmospheric parameters vs. orbital phase} for details).

 
  \begin{figure}
 \begin{center}
 \includegraphics[width=\linewidth]{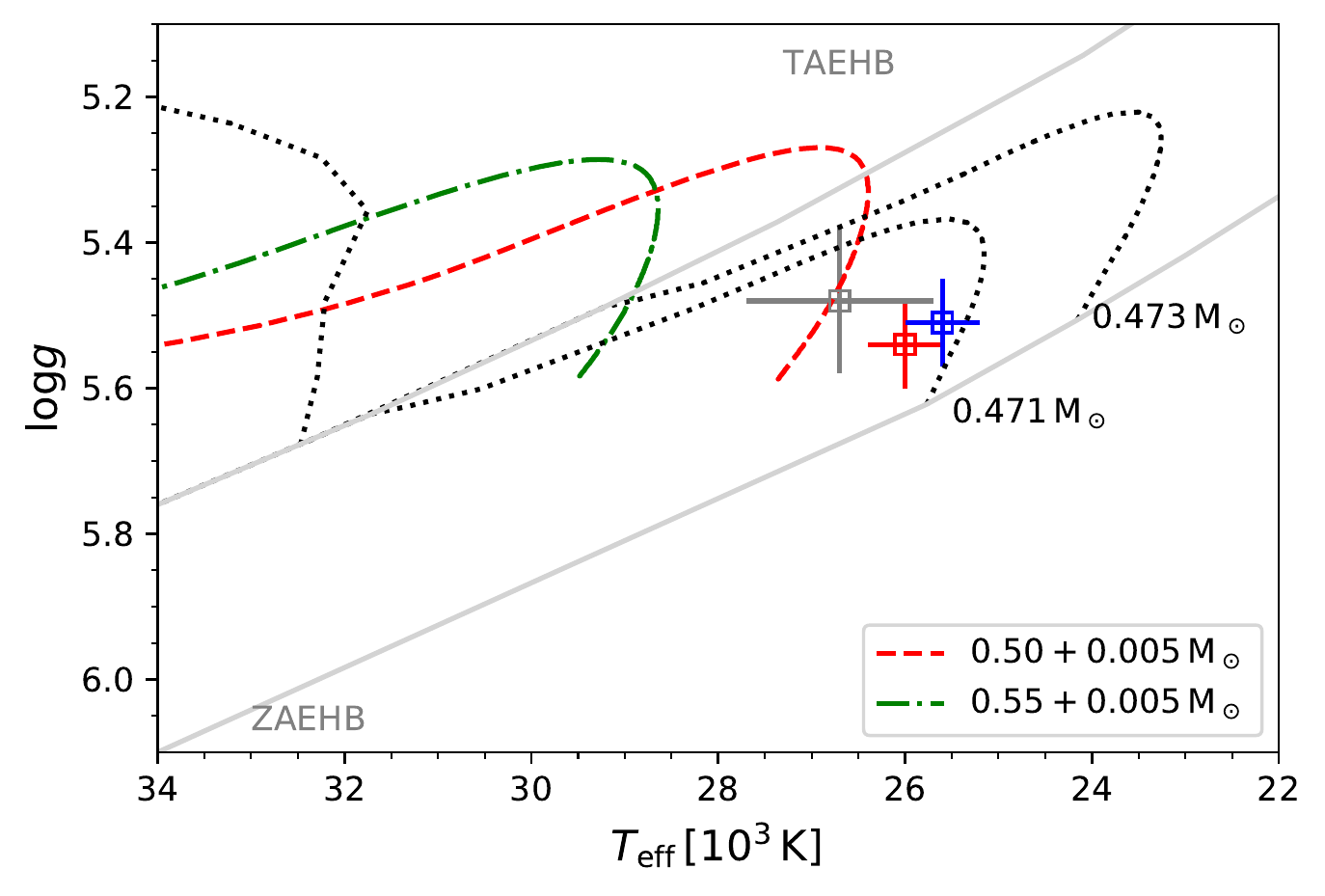}
    \caption{
    $T_{\text{eff}}-\log{(g)}$ diagram for J08205+0008. While the blue square represents the UVES solution, the red square results from XSHOOTER. The grey square marks the LTE solution of \citet{geier11c}. The zero-age (ZAEHB) and terminal-age horizontal  branch (TAEHB) for a canonical mass sdB are shown in grey as well as evolutionary tracks for a canonical mass sdB with different envelope masses from \citet{1993ApJ...419..596D} with black dotted lines. Additionally we show evolutionary tracks with solar metallicity for different sdB masses with hydrogen layers of $0.005\rm\,M_\odot$,  according to \citet{han02} to show the mass dependence of the EHB. The error bars include 1$\sigma$ statistical and systematic uncertainties as presented in the text (see Sect. \ref{Effective temperature, surface gravity, and helium content} for details).}
     \label{Kiel diagram}
     \end{center}
 \end{figure}
 
   \begin{figure}
 \begin{center}
 \includegraphics[trim = 0cm 0cm 0cm 0cm, clip, scale=0.45]{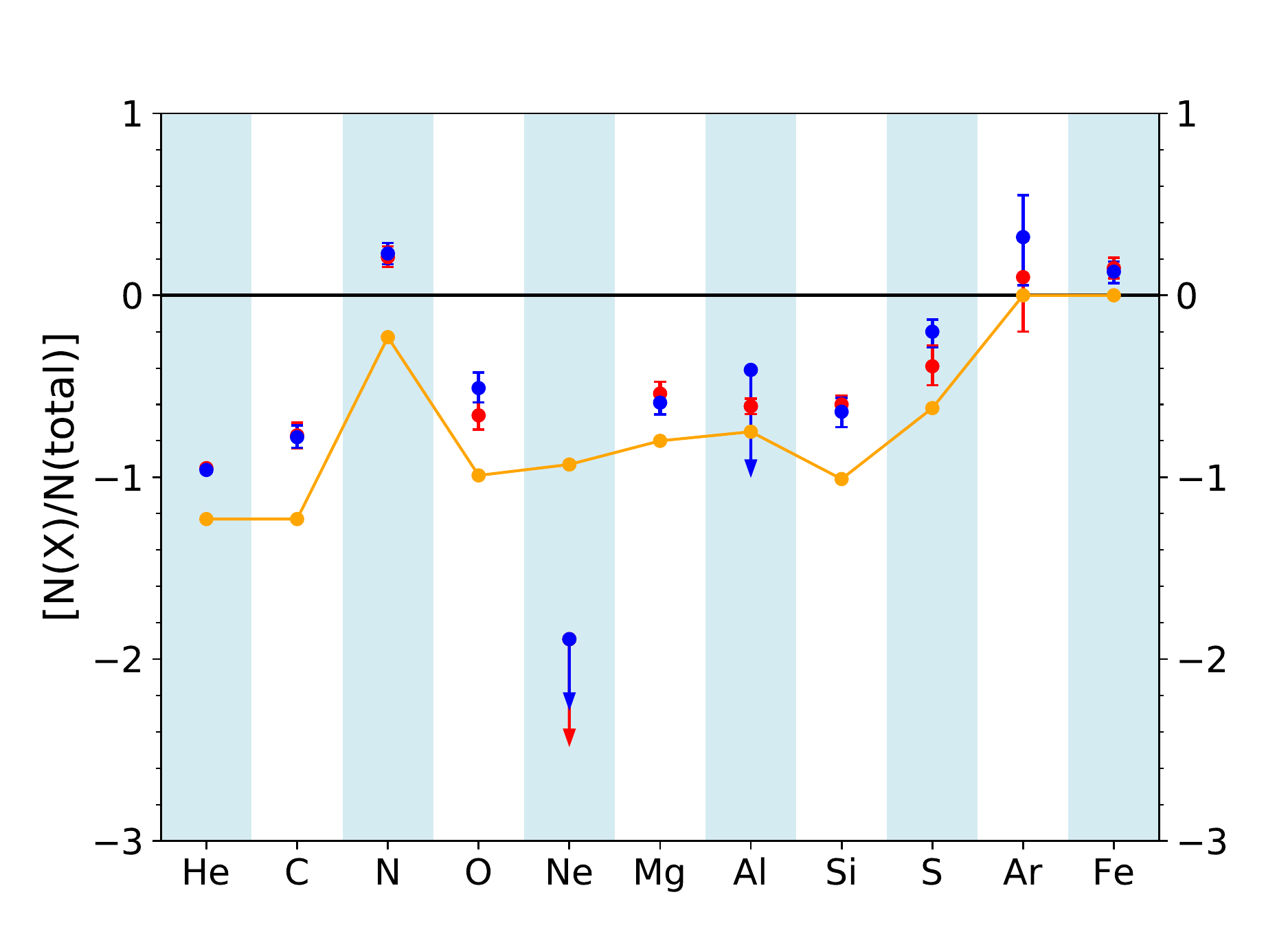}
    \caption{The chemical abundance pattern of J08205+0008 (red: XSHOOTER, blue: UVES) relative to solar abundances of \citet{asplund09}, represented by the black horizontal line. The orange solid line represents the mean abundances for hot subdwarf B stars according to \citet{naslim13} used as the metallicity for our quantitative spectral analysis. Upper limits are marked with downward arrows and $\left[\frac{\text{N}(\text{X})}{\text{N}(\text{total})}\right]:=\log_{10}{
    \left\{\frac{\text{N}(\text{X})}{\text{N}(\text{total})}\right\}}-\log_{10}{\left\{\frac{\text{N}(\text{X(solar)})}{\text{N}(\text{total})}\right\}}$.}
     \label{abundance}
     \end{center}
 \end{figure}

 \begin{figure*}
 \begin{center}
 \includegraphics[trim = 0cm 0cm 0cm 0cm, clip, scale=0.5]{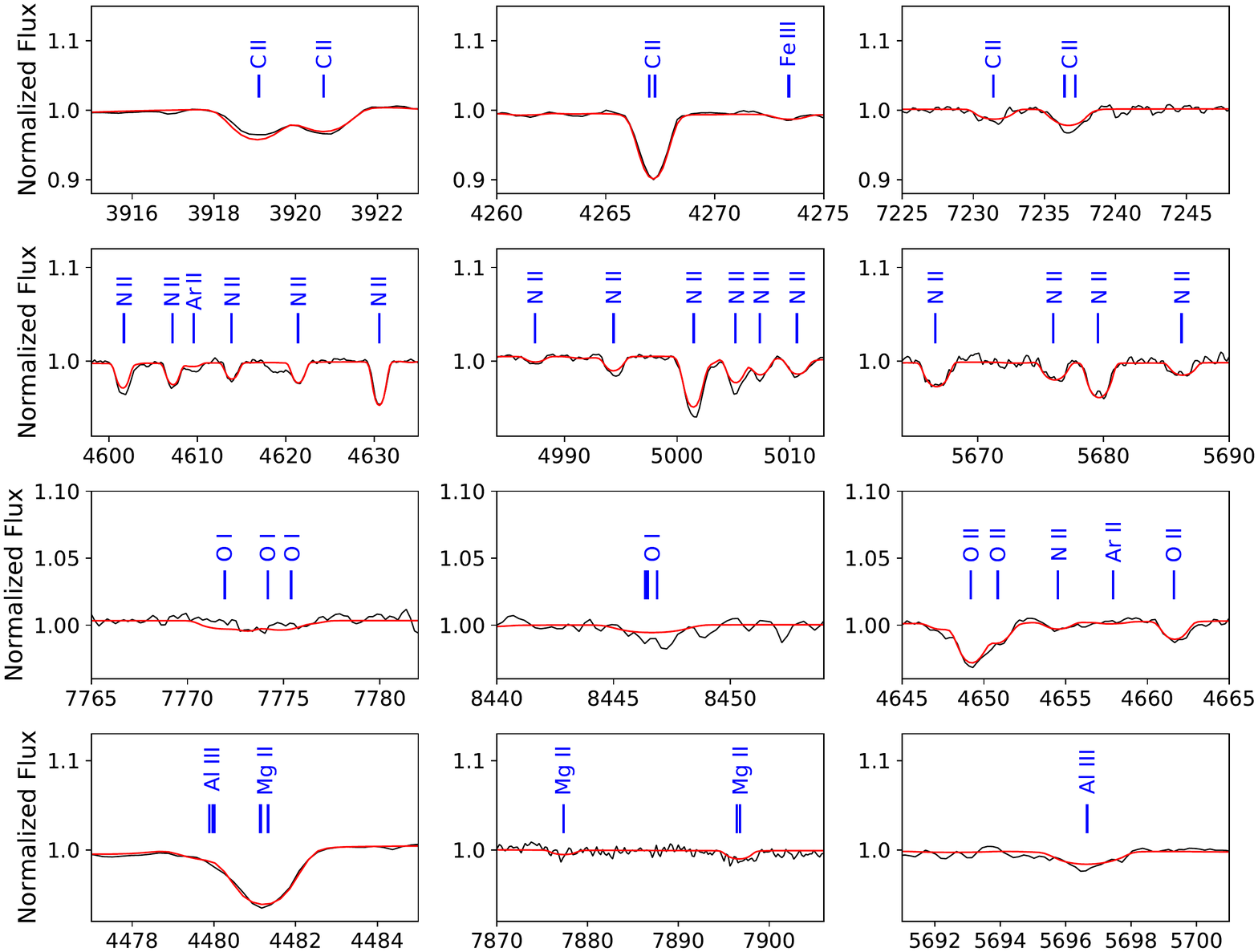}
  \includegraphics[trim = 0cm 0cm 0cm 0cm, clip, scale=0.5]{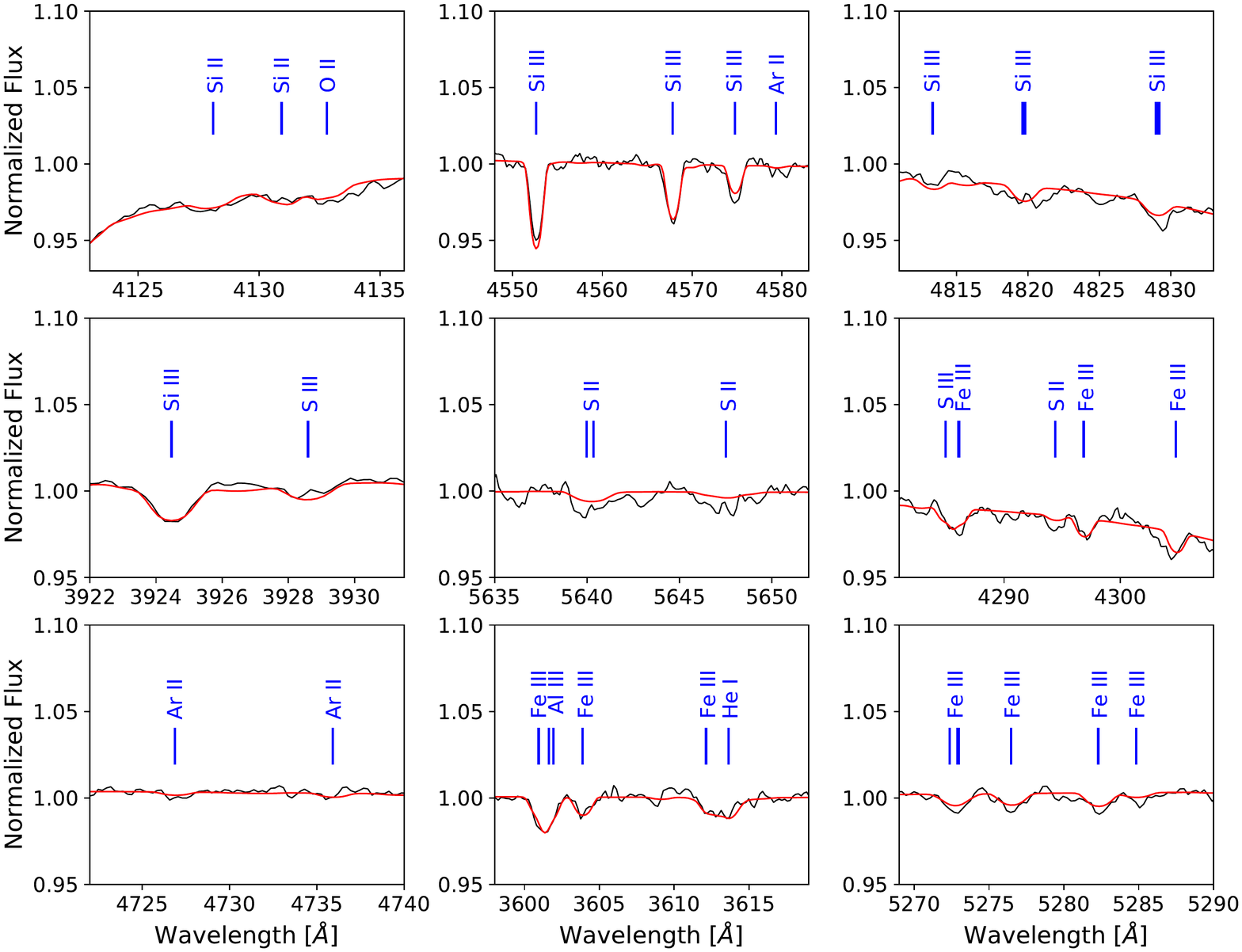}
    \caption{Selected metal lines in the co-added XSHOOTER spectrum of J08205+0008. The observed spectrum (solid black line) and the best fit (solid red line) are shown. Solid blue vertical lines mark the central wavelength positions and the ionization stages of the individual metal lines according to Table \ref{list of lines detected}.}
     \label{SDSS0820_metal_line_profiles}
     \end{center}
 \end{figure*}
 
The resulting effective temperatures, surface gravities, and helium abundances derived from XSHOOTER and UVES are listed in Table \ref{tab:par}. The results include 1$\sigma$ statistical errors and systematic uncertainties according to the detailed study of \citet{2005A&A...430..223L}, which has been conducted in the framework of the ESO Supernova Ia Progenitor Survey. For stars with two exposures or more, \citet{2005A&A...430..223L} determined a systematic uncertainty of $\pm$374\,K for $T_{\text{eff}}$, $\pm$\,0.049\,dex for $\log{(g)}$, and $\pm$\,0.044\,dex for $\log{n(\text{He})}$ (see Table 2 in \citealt{2005A&A...430..223L} for details). 

Figure \ref{Kiel diagram} shows the $T_{\text{eff}}-\log{(g)}$ diagram, where we compare the UVES and XSHOOTER results to predictions of evolutionary models for the horizontal branch for a canonical mass sdB with different envelope masses from \citet{1993ApJ...419..596D}, as well as evolutionary tracks
assuming solar metallicity and masses of $0.50$\,M$_{\rm \odot}$ and $0.55$\,M$_{\rm \odot}$ \citep{han02}. With $T_{\text{eff}}=26\,000\pm400$\,\si{\kelvin} and $\log{(g)}=5.54\pm0.05$ (XSHOOTER, statistical and systematic errors) and $T_{\text{eff}}=25\,600\pm400$\,\si{\kelvin} and $\log{(g)}=5.51\pm0.05$ (UVES, statistical and systematic errors), J08205+0008 lies within the EHB, as expected. Our final result ($T_{\text{eff}}=25\,800\pm290$\,\si{\kelvin}, $\log{(g)}=5.52\pm0.04$), the weighted average of the XSHOOTER and UVES parameters, is also in good agreement with the LTE results of \citet{geier11c}, which are $T_{\text{eff}}=26\,700\pm1000$\,\si{\kelvin} and $\log{(g)}=5.48\pm0.10$, respectively.

The determined helium content of J08205+0008 is $\log{n(\text{He})}=-2.06\pm0.05$ (XSHOOTER, statistical and systematic errors) and $\log{n(\text{He})}=-2.07\pm0.05$ (UVES, statistical and systematic errors), hence clearly subsolar (see \citealt{asplund09} for details). 
The final helium abundance ($\log{n(\text{He})}=-2.07\pm0.04$), the weighted average of XSHOOTER and UVES, therefore is comparable with \citet{geier11c}, who measured $\log{n(\text{He})}=-2.00\pm0.07$, and with the mean helium abundance for sdB stars from \citet{naslim13}, which is $\log{n(\text{He})}=-2.34$ (see also Fig. \ref{abundance}).


Moreover, it was possible to identify metals of various different ionization stages within the spectra (see Table \ref{list of lines detected} and Fig. \ref{SDSS0820_metal_line_profiles}) and to measure their abundances. Elements found in more than one ionization stage are oxygen ($\ion{O}{i/ii}$), silicon ($\ion{Si}{ii/iii}$), and sulfur ($\ion{S}{ii/iii}$), whereas carbon ($\ion{C}{ii}$), nitrogen ($\ion{N}{ii}$), magnesium ($\ion{Mg}{ii}$), aluminum ($\ion{Al}{iii}$), argon ($\ion{Ar}{ii}$), and iron ($\ion{Fe}{iii}$) are only detected in a single stage. We used the model grid in Table \ref{Hybrid LTE/NLTE model grid used for the quantitative spectral analysis of SDSS J08205+0008} to measure the individual metal abundances in both the co-added XSHOOTER and the UVES spectrum. We were able to fit the metal lines belonging to different ionization stages of the same elements similarly well (see Fig. \ref{SDSS0820_metal_line_profiles}). The corresponding ionization equilibria additionally constrained the effective temperature.\\
All metal abundances together with their total uncertainties are listed in Table 1. Systematic uncertainties were derived according to the methodology presented in detail in \citet{2014A&A...565A..63I} and cover the systematic uncertainties in effective temperature and surface gravity as described earlier.

The results of XSHOOTER and UVES are in good agreement, except for the abundances of oxygen, sulfur, and argon, where differences of $0.15$\,dex, $0.19$\,dex, and $0.22$\,dex, respectively, are measured. However, on average these metals also have the largest uncertainties, in particular argon, such that the abundances nearly overlap if the corresponding uncertainties are taken into account. According to Fig. \ref{abundance}, J08205+0008 is underabundant in carbon and oxygen, but overabundant in nitrogen compared to solar (\citealt{asplund09}), showing the prominent CNO signature as a remnant of the star's hydrogen core-burning through the CNO cycle. Aluminum and the alpha elements (neon, magnesium, silicon, and sulfur) are underabundant compared to solar. With the exception of neon, which is not present, the chemical abundance pattern of J08205+0008 generally follows the metallicity trend of hot subdwarf B stars (\citealt{naslim13}), even leading to a slight enrichment in argon and iron compared to solar. The latter may be explained by radiative levitation, which occurs in the context of atomic transport, that is, diffusion processes in the stellar atmosphere of hot subdwarf stars (\citealt{Greenstein_1967}; see \citealt{Michaud_Atomic_Diffusion_in_Stars_2015} for a detailed review).

Due to the high resolution of the UVES (and XSHOOTER) spectra, we were also able to measure the projected rotational velocity of J08205+0008 from the broadening of the spectral lines, in particular from the sharp metal lines, to $v\sin{i}=66.0\pm0.1\,{\rm km\,s^{-1}}$ (UVES, 1$\sigma$ statistical errors only) and $v\sin{i}=65.8\pm0.1\,{\rm km\,s^{-1}}$ (XSHOOTER, 1$\sigma$ statistical errors only).


\subsection{Search for chemical signatures of the companion} 
Although HW\,Vir type systems are known to be single-lined, traces of the irradiated and heated hemisphere of the cool companion have been found in some cases. \citet{wood99} discovered the H$\alpha$ absorption component of the companion in the prototype system HW\,Vir \citep[see also][]{edelmann08}. 

Metal lines in emission were found in the spectra of the hot sdOB star AA\,Dor by \citet{vuckovic16} moving in antiphase to the spectrum of the hot sdOB star indicating an origin near the surface of the companion. After the removal of the contribution of the hot subdwarf primary, which is dominating the spectrum, the residual
spectra showed more than 100 shallow emission lines originating from the heated side of the secondary, which show their maximum intensity close to the phases
around the secondary eclipse. They analysed the residual spectrum in order to model the irradiation of the low-mass companion by the hot subdwarf star.
The emission lines of the heated side of the secondary star allowed them to determine the radial velocity semi-amplitude of the centre-of-light. After the correction to the centre-of-mass of the secondary they could derive accurate masses of both components of the AA Dor system, which is consistent with a canonical sdB mass of $0.46\,\rm M_\odot$ and a companion of $0.079\pm0.002\rm\,M_\odot$ very close to the hydrogen burning limit. They also computed  a first generation atmosphere model
of the low mass secondary including irradiation effects.      

J08205+0008 is significantly fainter and cooler than AA Dor but with a much shorter period.
We searched the XSHOOTER spectra for signs of the low-mass companion of J08205+0008.
This was done by subtracting the spectrum in the secondary minimum where the companion is eclipsed from the spectra before and after the secondary eclipse where most of the heated atmosphere of the companion is visible. 
However, no emission or absorption lines from the companion were detected (see Fig. \ref{balmer_companion_uvb} and \ref{balmer_companion_vis}). Also, in the XSHOOTER NIR arm spectra, no emission lines could be found.

\subsection{Photometry: Angular diameter and interstellar reddening}\label{Photometric analysis}\label{SED fitting}

The angular diameter of a star is an important quantity, because it
allows the stellar radius to be determined, if the distance is known e.g. from
trigonometric parallax. The angular diameter can be determined by
comparing observed photometric magnitudes to those calculated from
model atmospheres for the stellar surface. 
Because of contamination by the reflection effect the apparent magnitudes of the hot subdwarf can be measured only during  
the  secondary  eclipse,  where  the  companion is completely eclipsed by the  larger  
subdwarf.  We performed a least squares fit to the flat bottom of the secondary  eclipse in the ULTRACAM light curves to  
determine the apparent magnitudes and 
derived $u'=14.926\pm$ 0.009mag, $g'=15.025\pm$ 0.004mag, and $r'=15.450\pm$ 0.011mag (1$\sigma$ statistical errors).

Because the star lies at low Galactic latitude (b=19$^\circ$) interstellar reddening is expected to be significant.
Therefore, both the angular diameter and the interstellar colour
excess have to be determined simultaneously. 
We used the reddening law of \citet{2019ApJ...886..108F} and matched a synthetic flux distribution calculated from the same grid of model 
atmospheres that where also used in the quantitative spectral analysis (see Sect. \ref{The spectroscopic analysis technique}) to the observed magnitudes as 
described in \citet{2018OAst...27...35H}. The $\chi^2$ based fitting routine uses two free parameters: the angular diameter $\theta$, 
which shifts the fluxes up and down according to $f(\lambda)=\theta^2 F(\lambda)/4$, where f($\lambda$) is the observed flux at the detector 
position and $F(\lambda)$ is the synthetic model flux at the stellar surface, and the color excess 
\footnote{\citet{2019ApJ...886..108F} use $E(44-55)$, the monochromatic equivalent of the usual $E(B-V)$ in the Johnson system, 
  using the wavelengths $\lambda$ =4400\,\AA  and 5000\,\AA, respectively. In fact, $E(44-55)$ is identical to $E(B-V)$ 
  for high effective temperatures as determined for J08205+0008.}.
The final atmospheric parameters and their respective uncertainties derived from the quantitative spectral analysis (see Sect. \ref{Effective temperature, surface gravity, and helium content}) 
result in an angular diameter of $\theta= 6.22\, (\pm 0.15) \cdot 10^{-12}\,\rm rad$  and an interstellar reddening of 
$E(B-V)=0.041 \pm 0.013$ mag.
The latter is consistent with values from reddening maps of \citet{1998ApJ...500..525S} and \citet{2011ApJ...737..103S}: 0.039 mag and 0.034 mag, 
respectively. 

In addition, ample photometric measurements of J08205+0008 are available in different filter systems, covering the spectral range all the way 
from the ultraviolet (GALEX) through the optical (e.g. SDSS) to the infrared (2MASS; UKIDDS;  WISE). However, those measurements are mostly averages of 
observations taken at multiple epochs or single epoch measurements at unknown orbital phase. 
Therefore, those measurements do not allow us to determine the angular diameter of the sdB because of the contamination by light from the heated hemisphere of the companion. However, an average spectral energy distribution of the system
can be derived. This allows us to redetermine the interstellar reddening and to search for an infrared excess caused by light from the 
cool companion.

The same fitting technique is used in the analysis of the SED as
described above for the analysis of the ULTRACAM magnitudes. Besides
the sdB grid, a grid of synthetic spectra of cool stars \citep[$2300\,{\rm K} \le T_{\rm eff} \le 15 000\,\rm K$,][]{2013A&A...553A...6H} is used.
In addition to the angular diameter and reddening parameter, the temperature of the cool companion as well as the surface ratio are 
free parameters in the fit.  The fit results in $E(B-V) = 0.040 \pm
0.010$ mag, which is fully consistent with the one derived from the ULTRACAM photometry as well as with the reddening map. 
The apparent angular diameter is larger than that from ULTRACAM photometry by 2.8\%, which is caused by the contamination by 
light from the companion's heated hemisphere. The effective temperature of the companion is unconstrained and the best match is achieved for the surface ratio of zero, which means there is no signature from the 
cool companion.
In a final step we allow the effective temperature of the sdB to vary and determine it along with the angular diameter and the interstellar reddening, which results in $T_{\rm eff}$= 26900$^{+1400}_{-1500}$\,K in agreement with the spectroscopic result.

\begin{figure}
    \centering
    \includegraphics[width=1.05\columnwidth]{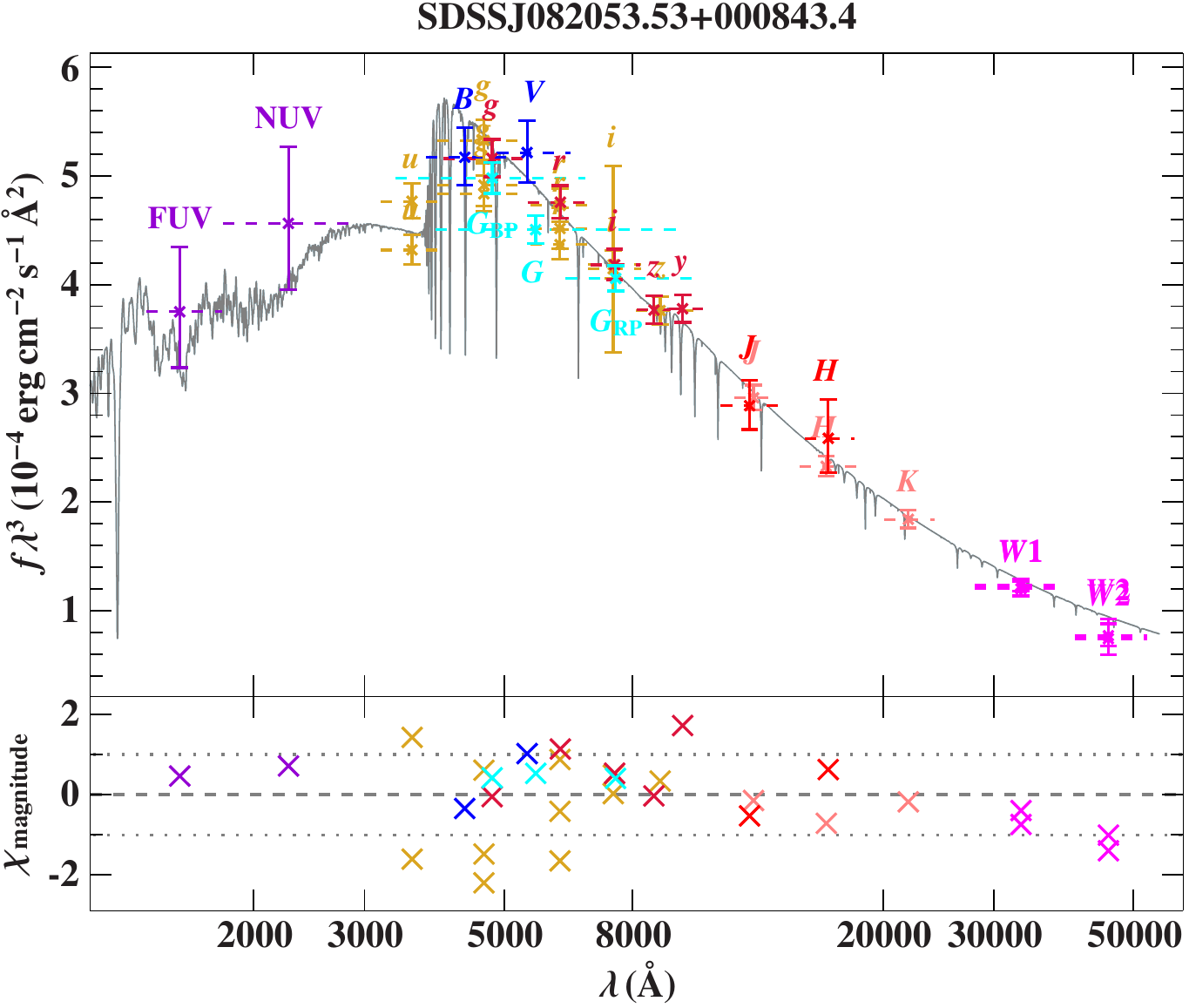}
    \caption{\label{fig:photometry_sed}Comparison of synthetic and
      observed photometry: \textit{Top panel:} Spectral
      energy distribution: Filter-averaged fluxes converted from
      observed magnitudes are shown in different colours. 
      The respective full width at tenth maximum are shown as dashed
      horizontal lines. The  best-fitting model, degraded to a
      spectral resolution of 6\,{\tiny\AA} is plotted in gray. 
      In order to reduce the steep SED slope the
      flux is multiplied by the wavelength cubed.
      \textit{Bottom panel:} Difference between synthetic and observed magnitudes
      divided by the corresponding uncertainties (residual $\chi$).
      The following color code is used for the different photometric
      systems: GALEX
      \citep[violet,][]{2017yCat.2335....0B}; SDSS
      \citep[golden,][]{2015ApJS..219...12A}; Pan-STARRS1
      \citep[dark red,][]{2017yCat.2349....0C}; Johnson
      \citep[blue,][]{2015AAS...22533616H};
      {\it Gaia}
      \citep[cyan,][with corrections and
      calibrations from \citet{2018A&A...619A.180M}]{2018A&A...616A...4E}; 2MASS
     \citep[red,][]{2003yCat.2246....0C}; UKIDSS 
      \citep[pink,][]{2007MNRAS.379.1599L}; 
      WISE \citep[magenta,][]{2014yCat.2328....0C,2019ApJS..240...30S}.}
    \end{figure}
    
\subsection{Stellar radius, mass and luminosity}\label{Stellar radius, mass and luminosity}
 
Since Gaia data release 2 \citep[DR2;][]{2018A&A...616A...1G}, trigonometric parallaxes are available for a
large sample of hot subdwarf stars, including J08205+0008
for which 10\% precision has been reached. We corrected for the Gaia DR2 parallax zero point offset of $-$0.029 mas  \citep{2018A&A...616A...2L}.

 Combining the parallax measurement
with the results from our quantitative spectral analysis ($\log{g}$ and $T_{\rm eff}$) and with the angular diameter $\theta$ derived
from ULTRACAM photometry, allows for the determination of the mass of the sdB primary in J08205+0008 via:
\begin{equation}
    M = \frac{g \theta^2}{4 G \varpi^2}\label{mass}
\end{equation}

The respective uncertainties of the stellar parameters are derived by Monte Carlo error propagation. The uncertainties are dominated by the error of the parallax measurement. Results are summarized in Table \ref{tab:par}. 
 Using the gravity and effective temperature derived by the spectroscopic analysis, the mass for the
sdB is $M= 0.48^{+0.12}_{-0.09}$ M$_\odot$ and its luminosity is $L=16^{+3.6}_{-2.8}$ L$_\odot$ in agreement with canonical models for EHB stars \citep[see Fig. 13][]{1993ApJ...419..596D}.
The radius of the sdB is calculated by the angular diameter and the parallax to
$R = 0.200^{+0.021}_{-0.018}$ R$_\odot$.

 \subsection{Radial velocity curve and orbital parameters}\label{rvs}

  \begin{figure}
 \begin{center}
 \includegraphics[width=\linewidth]{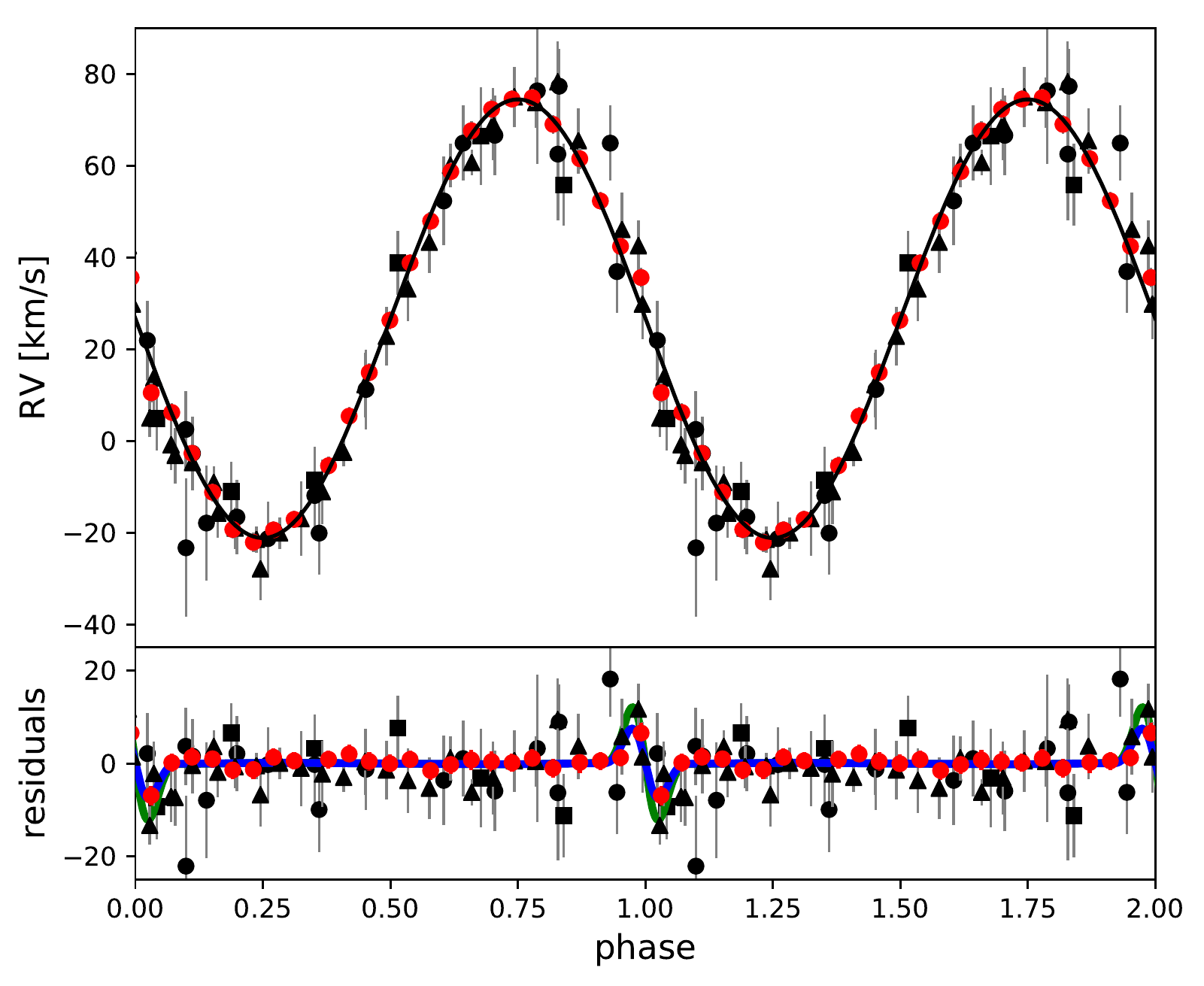}
    \caption{Radial velocity of J08205+0008 folded on the orbital period. 
    The residuals are shown together with a prediction of the Rossiter-McLaughlin effect using the parameters derived in this paper in blue and a model with a higher rotational velocity assuming bound rotation in green. The radial velocities were determined from spectra obtained with XSHOOTER (red circles), UVES (black triangles), EFOSC2 (black circles), and SDSS (black rectangles). The EFOSC2 and SDSS RVs have been corrected by a systematic shift (see text for details).}
     \label{rv}
     \end{center}
 \end{figure}

The radial velocities of the individual XSHOOTER spectra were measured by fitting all spectral features simultaneously to synthetic models as described in Sect.~\ref{The spectroscopic analysis technique}. 

Due to lower S/N of the individual UVES spectra, which were observed in poor conditions, only the most prominent features in the spectra are suitable for measuring the Doppler shifts. After excluding very poor quality spectra, radial velocities of the remaining 28 spectra were measured using the {\sc fitsb2} routine \citep{napiwotzki04b} by fitting a set of different mathematical functions to the hydrogen Balmer lines as well as He\,{\sc i} lines. The continuum is fitted by a polynomial, and the line wings and line core by a Lorentzian and a Gaussian function, respectively. The barycentrically corrected RVs together with formal $1\sigma$-errors are summarized in Table~\ref{RVs}. 

The orbital parameters $T_{\rm 0}$, period $P$, system velocity $\gamma$, and RV-semiamplitude $K$ as well as their uncertainties were derived with the same method described in \citet{geier11b}. To estimate 
the contribution of systematic effects to the total error budget additional to the statistic errors determined by the {\sc fitsb2} routine, we normalised the $\chi^{2}$ of the most probable solution by adding systematic errors to each data point $e_{\rm norm}$ until the reduced $\chi^{2}$ reached $\simeq1.0$.

Combining the UVES and XSHOOTER RVs we derived $T_{0}=57801.54954\pm0.00024\,{\rm d}$, $P=0.096241\pm 0.000003\,{\rm d}$, $K=47.9\pm0.4$\,\si{\kilo\metre\per\second} and the system velocity $\gamma=26.5\pm0.4$\,\si{\kilo\metre\per\second}. No significant systematic shift was detected between the two datasets and the systematic error added in quadrature was therefore very small $e_{\rm norm}=2.0$\,\si{\kilo\metre\per\second}. The gravitational redshift is significant at $1.6_{+0.05}^{-0.02}$\,\si{\kilo\metre\per\second} and might be important if the orbit of the companion could be measured by future high resolution measurements \citep[see, e.g.,][]{vos13}.

To improve the accuracy of the orbital parameters even more we then tried to combine them with the RV dataset from \citet{geier11c}, medium-resolution spectra taken with ESO-NTT/EFOSC2 and SDSS. A significant, but constant systematic shift of $+17.4$\,\si{\kilo\metre\per\second} was detected between the UVES+XSHOOTER and the SDSS+EFOSC2 datasets. Such zero-point shifts are common between low- or medium-resolution spectrographs. It is quite remarkable that both medium-resolution datasets behave in the same way. However, since the shift is of the same order as the statistical uncertainties of the EFOSC2 and SDSS individual RVs we refrain from interpreting it as real. 

Adopting a systematic correction of $+17.4$\,\si{\kilo\metre\per\second} to the SDSS+EFOSC2 dataset, we combined it with the UVES+XSHOOTER dataset and derived $T_{0}\,(\rm BJD_{TDB})=2457801.59769\pm0.00023\,{\rm d}$, $P=0.09624077\pm 0.00000001\,{\rm d}$, which is in perfect agreement with the photometric ephemeris, $K=47.8\pm0.4$\,\si{\kilo\metre\per\second} and $\gamma=26.6\pm0.4$\,\si{\kilo\metre\per\second}. This orbital solution is consistent with the solution from the XSHOOTER+UVES datasets alone. Due to the larger uncertainties of the SDSS+EFOSC2 RVs, the uncertainties of $\gamma$ and $K$ did not become smaller. The uncertainty of the orbital period on the other hand improved by two orders of magnitude due to the long timebase of 11 years between the individual epochs. Although this is still two orders of magnitude larger than the uncertainty derived from the light curve (see Sect.~\ref{timing}), the consistency with the light curve solution is remarkable. The RV curve for the combined solution phased to the orbital period is given in Fig.~\ref{rv}. Around phase 0 the Rossiter-McLaughin effect \citep{rossiter,mclaughlin} is visible. This effect is a RV deviation that occurs as parts of a rotating star are blocked out during the transit of the companion. The effect depends on the radius ratio and the rotational velocity of the primary. We can derive both parameters much more precisely with the spectroscopic and photometric analysis, but we plotted a model of this effect using our system parameters on the residuals of the radial velocity curve to show that is consistent. 

Except for the corrected system velocity, the revised orbital parameters of J08205+0008 are consistent with those determined by \citet{geier11c} ($P=0.096\pm 0.001\,{\rm d}$, $K=47.4\pm1.9$\,\si{\kilo\metre\per\second}), but much more precise.

 \subsection{Eclipse timing}\label{timing}
 \begin{figure}
    \centering
    \includegraphics[width=\linewidth]{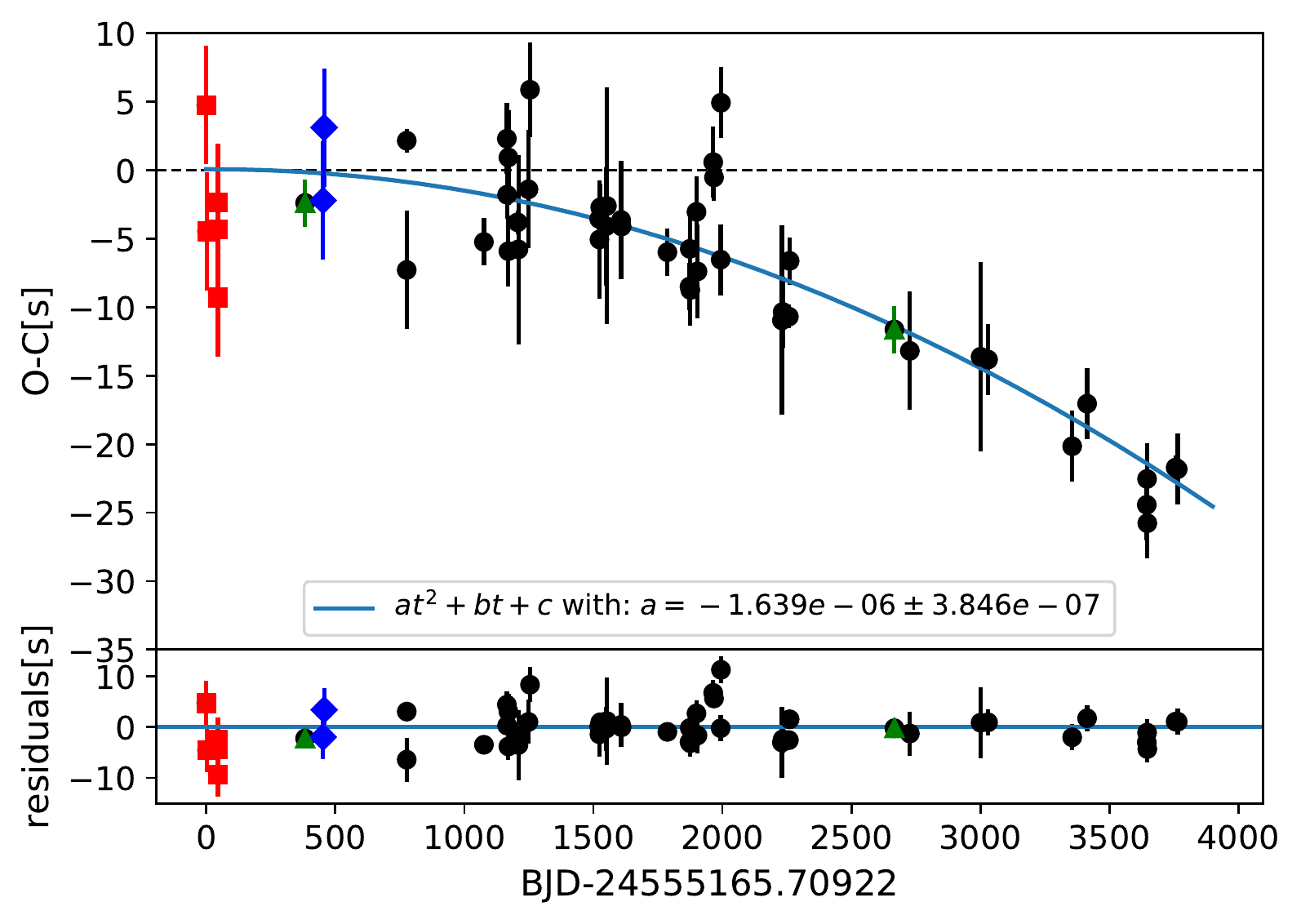}
    \caption{(O--C) diagram for J08205+0008 using eclipse times observed with Merope (red squares), BUSCA (blue diamonds), ULTRACAM (green triangles) and the SAAO-1m/1.9m telescope (black circles). The solid line represents a fit of a parabola to account for the period change of the orbital period. The derived quadratic term is given in the legend. The parameters of the fit are provided in the legend. In the lower panel the residuals between the observations and the best fit are shown.}
    \label{o-c}
\end{figure}
 
 Since the discovery that J08205+00008 is an eclipsing binary in November 2009, we have monitored the system regularly using  BUSCA mounted at the 2.2m-telescope in Calar Alto, Spain, ULTRACAM and the 1m in Sutherland Observatory (SAAO), South Africa. Such studies have been performed for several post-common envelope systems with sdB or white dwarf (WD) primaries and M dwarf companions \citep[see][for a summary]{lohr14}. In many of those systems period changes have been found.
 
The most convenient way to reveal period changes is to construct an observed minus calculated (O--C) diagram. Thereby we compare the observed mid-eclipse times (O) with the expected mid-eclipse times (C) assuming a fixed orbital period $P_0$ and using the mid-eclipse time for the first epoch $T_0$. Following \citet{kepler91}, if we expand the observed mid-eclipse of the Eth eclipse ($T_E$ with $E=t/P$) in a Taylor series
, we get the (O--C) equation:
\begin{equation}
    \mathrm{O-C}=\Delta T_0+\frac{\Delta P_0}{P_0}t+\frac{1}{2}\frac{\dot{P}}{P_0}t^2+...
\end{equation}
This means that with a quadratic fit to the O--C data we can derive the ephemeris $T_0$, $P$, and $\dot{P}$ in $BJD_{TDB}$.

Together with the discovery data observed with Merope at the Mercator telescope on La Palma \citep{geier11c} it was possible to determine timings of the primary eclipse over more than 10 years, as described in Sect. \ref{saao} and \ref{calaralto}.  
All measured mid-eclipse times can be found in Table \ref{ecl_time}.


 We used all eclipse timings to construct an O--C diagram, which is shown in Fig. \ref{o-c}. We used the ephemeris given in \citet{geier11c} as a starting value to find the eclipse numbers of each measured eclipse time and detrended the O-C diagram by varying the orbital period until no linear trend was visible to improve the determination of the orbital period. During the first $7-8$ years of observations, the ephemeris appeared to be linear. This was also found by \citet{pulley:18}. As their data show a large scatter, we do not use it in our analysis. However, in the last two years a strong quadratic effect was revealed. The most plausible explanation is a decrease in the orbital period of the system.
 This enabled us to derive an improved ephemeris for J08205+0008:
 \begin{eqnarray*}
 T_0&=&2455165.709211(1)\\
 P&=&0.09624073885(5)\rm\,d\\
 \dot{P}&=&-3.2(8)\cdot 10^{-12}\,\rm dd^{-1}
 \end{eqnarray*}
 

 \subsection{Light curve modeling}

 With the new very high quality ULTRACAM $u'g'r'$ light curves we repeated the light curve analysis of \citep{geier11c} obtaining a solution with much smaller errors.
 For the modeling of the light curve we used {\sc lcurve}, a code written to model detached and accreting binaries containing a white dwarf \citep[for details, see][]{copperwheat10}. It has been used to analyse several detached white dwarf-M dwarf binaries \citep[e.g.,][]{parsons_nnser}. Those systems show very similar light curves with very deep, narrow eclipses and a prominent reflection effect, if the primary is a hot white dwarf. Therefore, {\sc lcurve} is ideally suited for our purpose.
 
 The code calculates monochromatic light curves by subdividing each star into small elements with a geometry fixed by its radius as measured along the line from the center of one star towards the center of the companion. The flux of the visible elements is always summed up to get the flux at a certain phase. A number of different effects that are observed in compact and normal stars are considered, e.g. Roche distortions observed when a star is distorted from the tidal influence of a massive, close companion, as well as limb-darkening and gravitational darkening. Moreover, lensing and Doppler beaming, which are important for very compact objects with close companions, can be included. The Roemer delay, which is a light travel-time effect leading to a shift between primary and secondary eclipse times due to stars of different mass orbiting each other and changing their distance to us, and asynchronous orbits can be considered. The latter effects are not visible in our light curves and can hence be neglected in our case. 
 
  \begin{figure}
     \centering
     \includegraphics[width=1.05\linewidth]{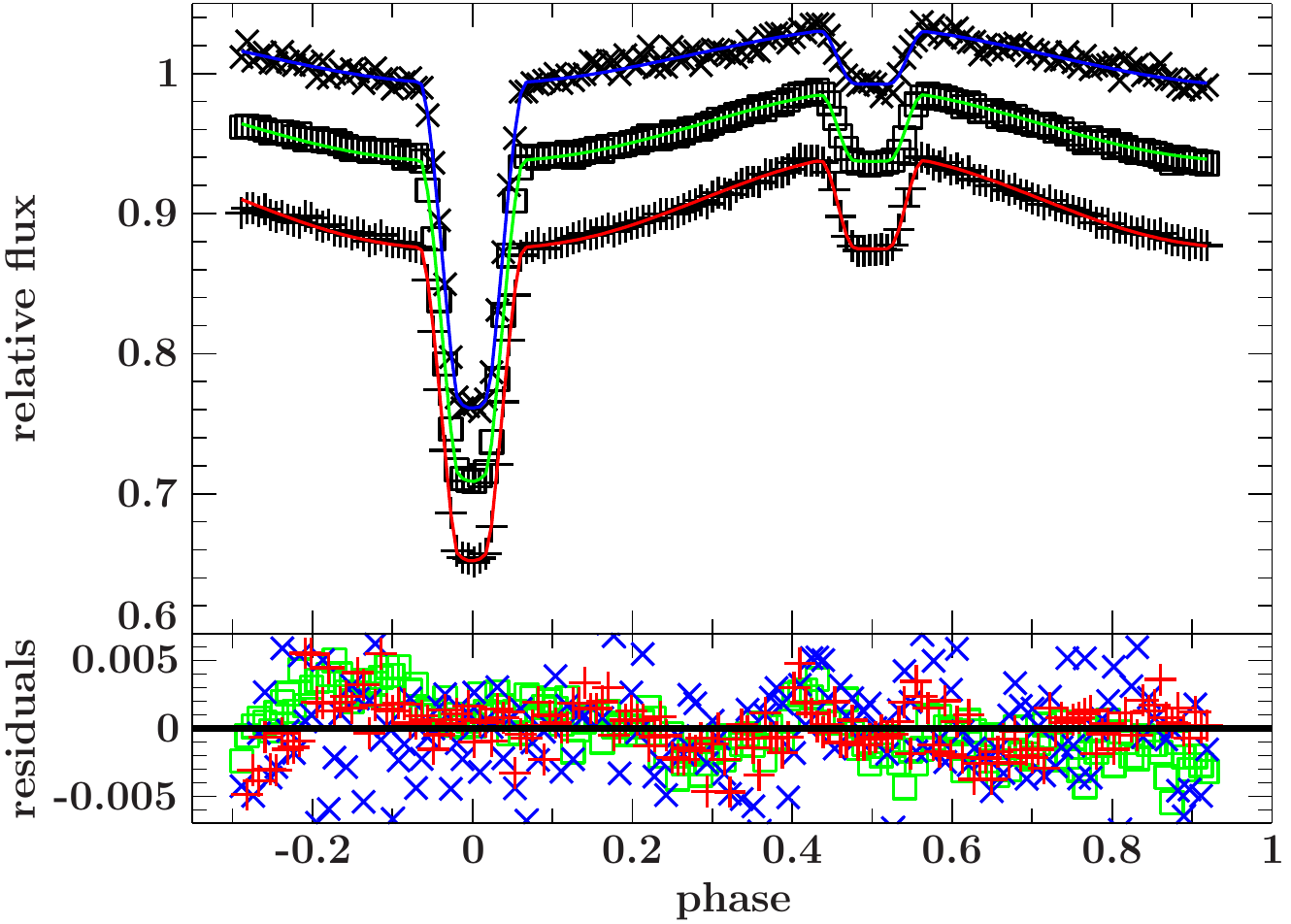}
     \caption{ULTRACAM $u'g'r'$ light curves of J08205+0008 together with the best fit of the most consistent solution. The light curves in the different filters have been shifted for better visualisation. The lower panel shows the residuals. The deviation of the light curves from the best fit is probably due to the fact that the comparison stars cannot completely correct for atmospheric effects due to the different colour and the crude reflection effect model used in the analysis is insufficient to correctly describe the shape of the reflection effect.}
     \label{lc_0820}
 \end{figure}
 
  \begin{figure*}
    \centering
    \includegraphics[width=1.0\linewidth]{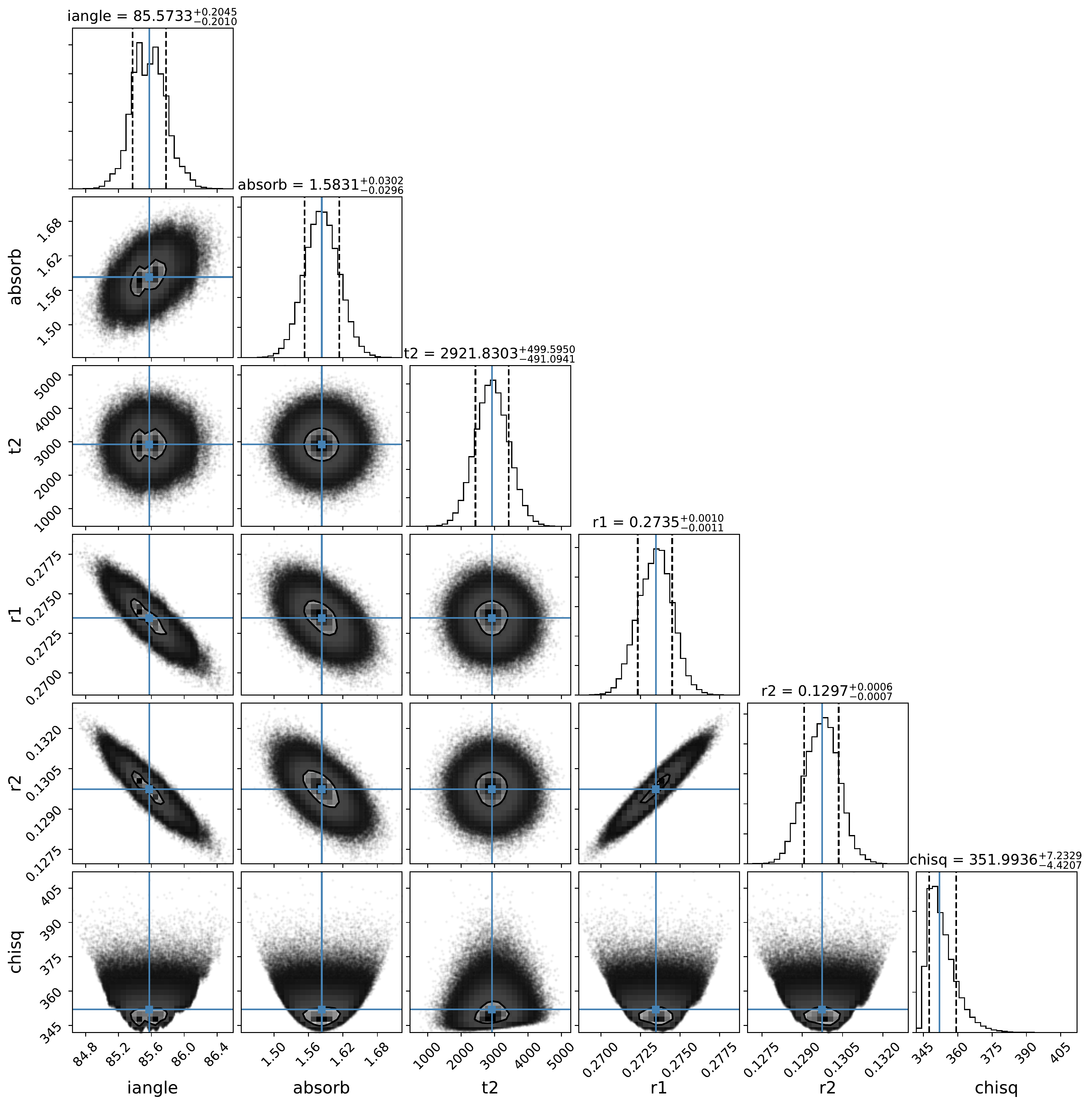}
    \caption{MCMC calculations showing the distributions of the parameter of the analysis of the ULTRACAM g'-band light curve.}
    \label{mcmc_r}
\end{figure*}

 As we have a prominent reflection effect it is very important to model this effect as accurately as possible. The reflection effect, better called the irradiation effect, results from the huge difference in temperature between the two stars, together with their small separation.
The (most likely) tidally locked companion is heated up on the side facing the hot subdwarf because of the strong irradiation by the hot primary. Therefore, the contribution of the companion to the total flux of the system varies with phase and increases as more of the heated side is visible to the observer. We use a quite simple model, which calculates the fluxes from the temperatures of both companions using a black body approximation. The irradiation is approximated by assigning a new temperature to the heated side of the companion 
 \begin{equation}
 \sigma T'^4_{\rm sec}=\sigma T^4_{\rm sec}+F_{\rm irr}=\sigma T^4_{\rm sec}\left[1+\alpha\left(\frac{T_{\rm prim}}{T^{\rm sec}}\right)^4\left(\frac{R_{\rm prim}}{a}\right)^2\right],
  \end{equation}
  with $\alpha$ being the albedo of the companion and $F_{\rm irr}$ the irradiating flux, accounting for the angle of incidence and distance from the hot subdwarf. The irradiated side is heated up to a temperature of $13\,000-15\,000$ K similar to HW Vir \citep{kiss00}, which is slightly hotter but has a longer period. Hence, the amplitude of the effect is increasing from blue to red as can be seen in Fig. \ref{lc_0820}, as the sdB is getting fainter compared to the companion in the red. If the irradiation effect is very strong, the description given above might not be sufficient, as the back of the irradiated star is completely unaffected in this description, but heat transport could heat it up, increasing the luminosity of unirradiated parts as well. This is not considered in our simple model. 

As the light curve model contains many parameters, not all  of them independent, we fixed as many parameters as possible (see Table \ref{param}). The temperature of the sdB was fixed to the temperature determined from the spectroscopic fit. We used the values determined by the coadded XSHOOTER spectra, as they have higher signal-to-noise. 
The gravitational limb darkening coefficients were fixed to the values expected for a radiative atmosphere for the primary \citep{von_zeipel} and a convective atmosphere for the secondary \citep{lucy} using a blackbody approximation to calculate the resulting intensities. For the limb darkening of the primary we adopted a quadratic limb darkening law using the tables by \citet{claret}. As the tables include only surface gravities up to $\log{g}=5$ we used the values closest to the parameters derived by the spectroscopic analysis. 

As it is a well-separated binary, the two stars are approximately spherical, which means the light curve is not sensitive to the mass ratio. Therefore, we computed solutions with different, fixed mass ratios. To localize the best set of parameters we used a \textsc{simplex} algorithm \citep{press92} varying the inclination, the radii, the temperature of the companion, the albedo of the companion (absorb), the limb darkening of the companion, and the time of the primary eclipse to derive additional mid-eclipse times. Moreover, we also allowed for corrections of a linear trend, which is often seen in the observations of hot stars, as the comparison stars are often redder and so the correction for the air mass is often insufficient. This is given by the parameter "slope". The model of the best fit is shown in Fig. \ref{lc_0820} together with the observations and the residuals.

To get an idea about the degeneracy of parameters used in the light curve solutions, as well as an estimation of the errors of the parameters we performed Markov-Chain Monte-
Carlo (MCMC) computations with \textsc{emcee} \citep{emcee} using the best solution we obtained with the \textsc{simplex} algorithm as a starting value varying the radii, the inclination, the temperature of the companion as well as the albedo of the companion. As a prior we constrained the temperature of the cool side of the companion to $3000\pm500$ K. Due to the large luminosity difference between the stars the temperature of the companion is not significantly constrained by the light curve. The computations were done for all three light curves separately.

For the visualisation we used the python package \textsc{corner} \citep[see Fig. \ref{mcmc_r}]{corner}. The results of the MCMC computations of the light curves of all three filters agree within the error (see Table \ref{param}). A clear correlation between both radii and the inclination is visible as well as a weak correlation of the albedo of the companion (absorb) and the inclination. This results from the fact that the companion is only visible in the combined flux due to the reflection effect and the eclipses and the amplitude of the reflection effect depends on the inclination, the radii, the separation, the albedo and the temperatures. Looking at the $\chi^2$ of the temperature of the companion we see that all temperatures give equally good solutions showing that the temperature can indeed not be derived from the light curve fit. The albedo we derived has, moreover, a value > 1, which has been found in other HW Vir systems as well and is due to the simplistic modeling of the reflection effect. The reason for the different distribution in the inclination is not clear to us. However, it is not seen in the other bands. It might be related to the insufficient correction of atmospheric effects by the comparison stars. 

\begin{table}\caption{Parameters of the light curve fit of the ULTRACAM u'g'r' band light curves}\label{param}
\begin{tabular}{llll}
	\hline\hline
	band & u'&g'&r'\\\hline
	\multicolumn{4}{c}{Fixed Parameters}\\\hline
	q&\multicolumn{3}{c}{0.147}\\
	$P$&\multicolumn{3}{c}{0.09624073885}\\
	$T_{\rm eff,sdB}$&\multicolumn{3}{c}{25800}\\
	$x_{1,1}$&0.1305&0.1004&0.0788\\
	$x_{1,2}$&0.2608&0.2734&0.2281\\
	$g_1$&\multicolumn{3}{c}{0.25}\\
	$g_2$&\multicolumn{3}{c}{0.08}\\
	\hline
	\multicolumn{4}{c}{Fitted parameters}\\\hline
	$i$&$85.3\pm0.6$&$85.6\pm0.2$&$85.4\pm0.3$\\
	$r_1/a$&$0.2772\pm0.0029$&$0.2734\pm0.0010$&$0.2748\pm0.0014$\\
	$r_2/a$&$0.1322\pm0.0018$&$0.1297\pm0.0006$&$0.1304\pm0.0008$\\
	$T_{\rm eff,comp}$&$3000\pm500$&$2900\pm500$&$3200\pm560$\\
	absorb&$1.54\pm0.08$&$1.58\pm0.03$&$2.08\pm0.05$\\
	$x_2$&0.70&0.78&0.84\\
	$T_0$ [MJD]&57832.0355&57832.0354&57832.0354\\
	slope&-0.000968&-0.002377&0.00013417\\
	$\frac{L_1}{L_1+L_2}$&0.992578&0.98735&0.97592\\
	\hline
\end{tabular}
\end{table}

\subsection{Absolute parameters of J08205+00008}\label{Absolute parameters of J08205+00008}
\begin{figure}
    \centering
    \includegraphics[width=\linewidth]{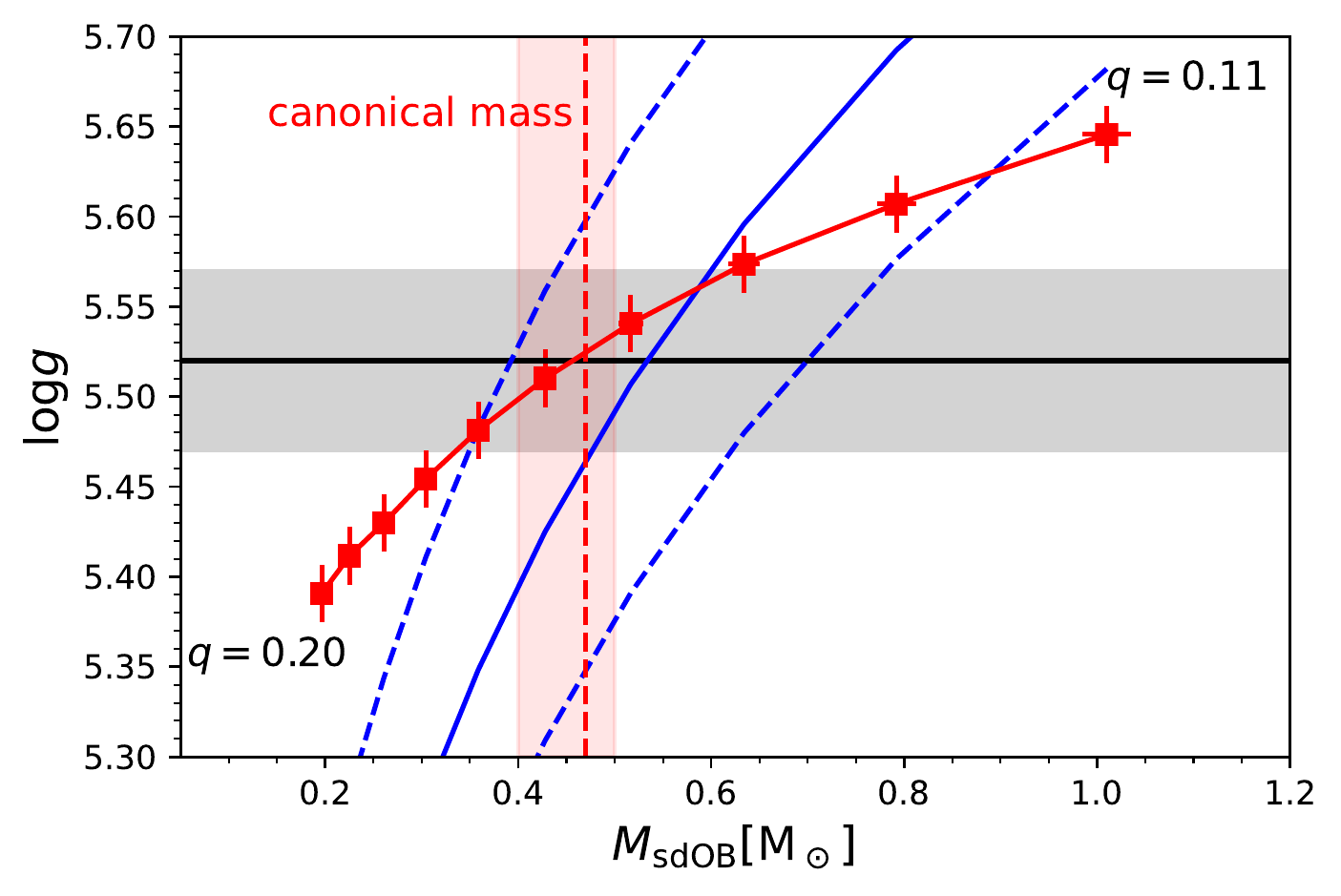}
    \caption{Mass of the sdB versus the photometric $\log g$ for J08205+0008 for different mass ratios from $0.11-0.20$ in steps of 0.01 (red solid line). The parameters were derived by combining the results from the analysis of the light curves and radial velocity curve. The grey area marks the spectroscopic $\log{g}$ that was derived from the spectroscopic analysis. 
    The blue dashed lines indicate the $\log g$ derived by the radius from the SED fitting and the {\it Gaia} distance for different sdB masses.
    The red area marks the mass range for the sdB for which we get a consistent solution by combining all different methods. The red vertical line represents the solution for a canonical mass sdB.}  
    \label{sdb}
\end{figure}
\begin{figure}
    \centering
    \includegraphics[width=\linewidth]{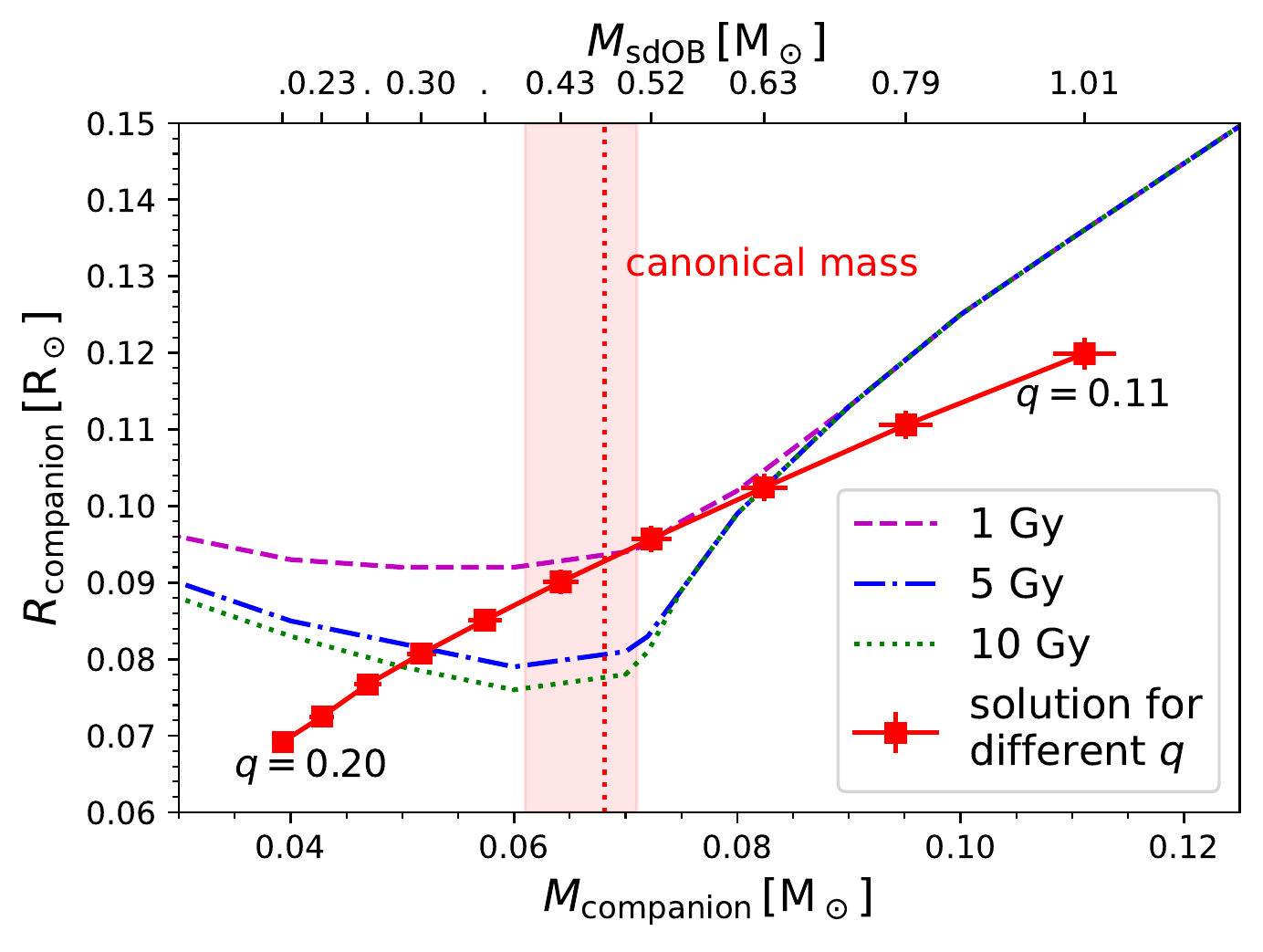}
    \caption{Comparison of theoretical mass-radius relations of low-mass stars \citep{baraffe,baraffe_2} to results from the light curve analysis of J08205+0008. We used tracks for different ages of 1 Gyr (dashed), 5 Gyr (dotted-dashed) and 10 Gyr (dotted). Each red square together with the errors represents a solution from the light curve analysis for a different mass ratio ($q = 0.11-0.20$ in steps of 0.01). The red vertical line represents the solution for a canonical mass sdB. The red area marks the mass range of the companion corresponding to the mass range we derived for the sdB.}
    \label{bd}
\end{figure}
As explained before, we calculated solutions for different mass ratios ($q=0.11-0.20$). We obtain equally good $\chi^2$ for all solutions, showing that the mass ratio cannot be constrained by the light curve fit as expected. Hence, the mass ratio needs to be constrained differently. 
However, the separation, which can be calculated from the mass ratio, period, semi-amplitude of the radial-velocity curve and the inclination, is different for each mass ratio. 
The masses of both companions can then be calculated from the mass function. From the relative radii derived from the light curve fit together with the separation, the absolute radii can be calculated. This results in different radii and masses for each mass ratio. 

As stated before, the previous analysis of \citet{geier11c} resulted in two possible solutions: A post-RGB star with a mass of 0.25 $\rm M_\odot$ and a core helium-burning star on the extreme horizontal branch with a mass of $\rm 0.47\,M_\odot$.
From the analysis of the photometry together with the Gaia magnitudes (see Sect. \ref{Stellar radius, mass and luminosity}) we get an additional good constraint on the radius of the sdB.  Moreover, the surface gravity was derived from the fit to the spectrum. This can be compared to the mass and radius of the sdB (and a photometric $\log g$: $g=GM/R^2$) derived in the combined analysis of radial velocity curve and light curve. This is shown in Fig. \ref{sdb}. We obtain a good agreement for of all three methods (spectroscopic, photometric, parallax-based) for an sdB mass between 0.39-0.60 $\rm M_\odot$. This means that we can exclude the post-RGB solution. The position of J0820 in the $T_{\rm eff}-\log{g}$ diagram, which is shown in Fig. \ref{Kiel diagram}, gives us another constraint on the sdB mass. By comparing the atmospheric parameters of J08205+0008 to theoretical evolutionary tracks calculated by \citet{han02} it is evident that the position is not consistent with sdB masses larger than $\sim 0.50\,\rm M_\odot$, which we, therefore, assume as the maximum possible mass for the sdB.

Accordingly, we conclude that the solution that is most consistent with all different analysis methods is an sdB mass close to the canonical mass ($0.39-0.50\,\rm M_\odot$). For this solution we have an excellent agreement of the parallax radius with the photometric radius only, if the parallax offset of $-0.029$ mas suggested by \citet{2018A&A...616A...2L} is used. Otherwise the parallax-based radius is too large. The companion has a mass of $0.061-0.71\,\rm M_\odot$, which is just below the limit for hydrogen-burning. Our final results can be found in Table \ref{tab:par}. The mass of the companion is below the hydrogen burning limit and the companion is hence most likely a massive brown dwarf.

We also investigated the mass and radius of the companion and compared it to theoretical calculations by \citet{baraffe} and \citet{baraffe_2} as shown in Fig.~\ref{bd}. It is usually assumed that the progenitor of the sdB was a star with about $1-2\,\rm M_\odot$ \citep{heber09,heber16}. Therefore, we expect that the system is already quite old (5-10 Gyrs). For the solutions in our allowed mass range the measured radius of the companion is about 20\% larger than expected from theoretical calculations. Such an effect, called inflation, has been observed in different binaries  and also planetary systems with very close Jupiter-like planets. A detailed discussion will be given later. This effect has already been observed in other hot subdwarf close binary systems \citep[e.g.][]{schaffenroth15}. 

\begin{figure}
    \centering
    \includegraphics[width=\linewidth]{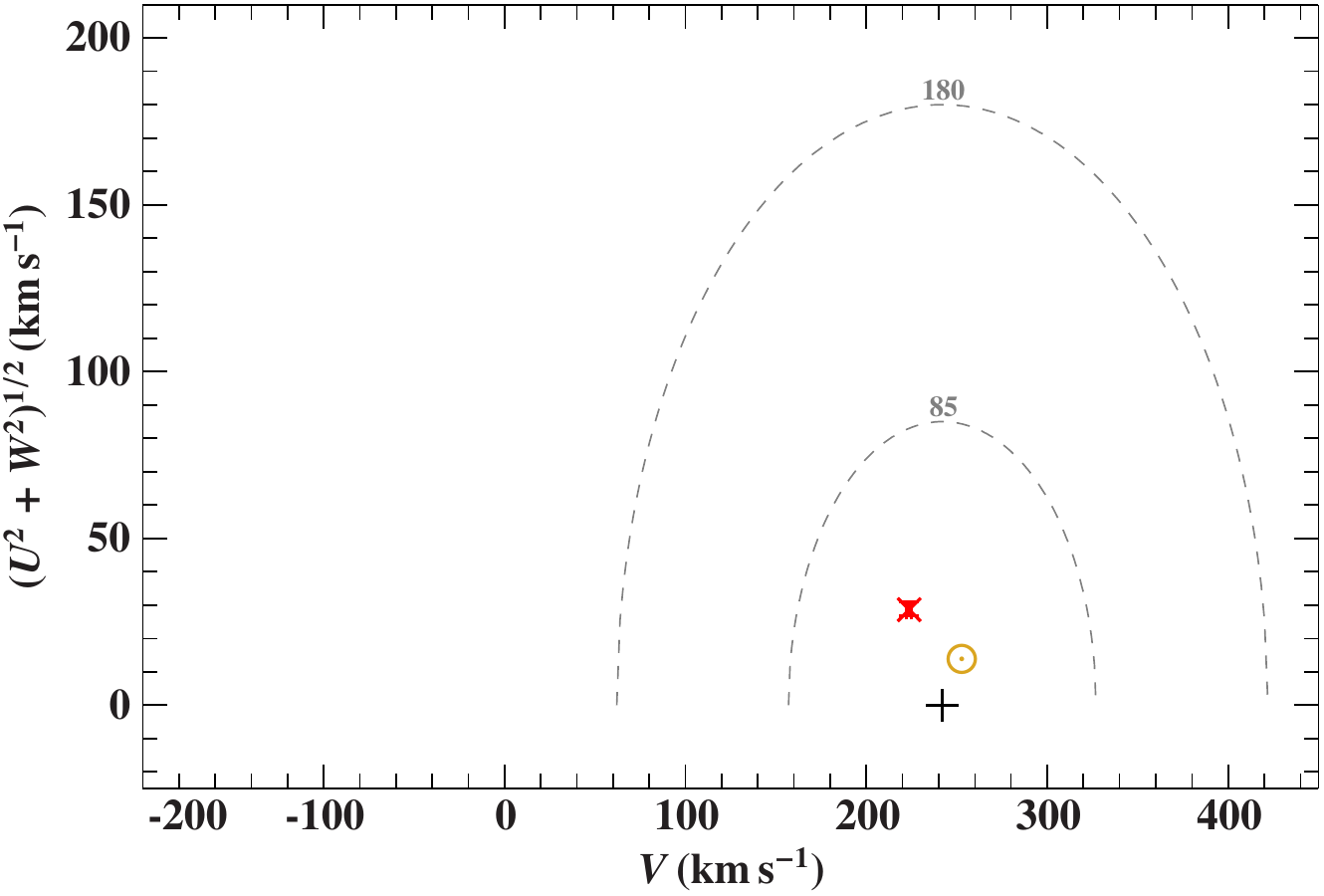}
    \caption{Toomre diagram of J08205+0008: the quantity $V$ is the velocity in direction of Galactic rotation, $U$ towards the Galactic center, and $W$ perpendicular to the Galactic plane. The two dashed ellipses mark boundaries for the thin (85\,km\,s$^{-1}$) and thick disk (180\,km\,s$^{-1}$) following Fuhrmann (2004). The red cross marks J08205+0008, the yellow circled dot the Sun, and the black plus the local standard of rest. The location of J08205+0008 in this diagram clearly hints at a thin disk membership. }
    \label{toomre}
\end{figure}

However, if the system would still be quite young with an age of about 1 Gyr, the companion would not be inflated. We performed a kinematic analysis to determine the Galactic population of J08205+0008. As seen in Fig. \ref{toomre} the sdB binary belongs to the thin disk where star formation is still ongoing and could therefore indeed be as young as 1 Gyr, if the progenitor was a $2\,\rm M_\odot$ star. About half of the sdO/Bs at larger distances from the Galactic plane (0.5 kpc) are found in the thin disk \citep{martin17}. However, it is unclear whether a brown dwarf companion can eject the evelope from such a massive 2 $\rm M_{\odot}$ star. Hydrodynamical simulations performed by \citet{kramer20} indicate that a BD companion of $\sim0.05-0.08\,\rm M_\odot$ might just be able to eject the CE of a lower mass ($1\,\rm M_\odot$) red giant.

\begin{table}
\caption{Parameters of J08205+0008. 
}\label{tab:par}
\vspace{0.5cm}
\begin{tabular}{lll}
\noalign{\smallskip}
\hline
\noalign{\smallskip}
\multicolumn{3}{l}{SPECTROSCOPIC PARAMETERS}\\
\noalign{\smallskip}
\hline
\noalign{\smallskip}
$\gamma$& [${\rm km\,s^{-1}}$] & {$26.5\pm0.4$}  \\ 
$K_1$ & [${\rm km\,s^{-1}}$] & {$47.8\pm0.4$} \\ 
$f(M)$& [$M_{\rm \odot}$] & {$0.0011\pm0.0001$}  \\
\noalign{\smallskip}
\hline
\noalign{\smallskip}
$T_{\rm eff,sdB}$ & [K] & {$25800\pm290^\ast$} \\  
$\log{g,sdB}$ & & {$5.52\pm0.04^\ast$}  \\ 
$\log{n(\text{He})}$ & & {$-2.07\pm0.04^\ast$}  \\ 
$v\sin{i}$ & [${\rm km\,s^{-1}}$] & $65.9\pm0.1^\dagger$\\ 

\noalign{\smallskip}
\hline
\noalign{\smallskip}
$a$ &[$\rm R_\odot$]&$0.71\pm0.02$\\
$M_{\rm 1}$&[$M_{\rm \odot}$]& $0.39 - 0.50$ \\
$M_{\rm 2}$ & [$M_{\rm \odot}$] & $0.061-0.071$  \\
\noalign{\smallskip}
\hline
\noalign{\smallskip}
\multicolumn{3}{l}{PHOTOMETRIC PARAMETERS}\\
\noalign{\smallskip}
\hline
\noalign{\smallskip}
$T_0$ & [BJD$_{TDB}$]&2455165.709211(1)\\
$P$&[d]&$0.09624073885(5)$\\
$\dot{P}$&dd$^{-1}$&$-3.2(8)\cdot 10^{-12}$\\
$i$&[$^\circ$]&$85.6\pm0.3$\\
$R_{\rm 1}$ & [$R_{\rm \odot}$] & $0.194\pm0.008$  \\
$R_{\rm 2}$ & [$R_{\rm \odot}$] & $0.092\pm 0.005$  \\
$\log g$ & &$5.52\pm0.03$\\
\noalign{\smallskip}
\hline
\noalign{\smallskip}
\multicolumn{3}{l}{SED FITTING}\\
\noalign{\smallskip}
\hline
\noalign{\smallskip}
$\varpi_{\text{Gaia}}$ & [mas] & $0.6899\pm0.0632^\dagger$  \\
$E(B-V)$ & [mag] & $0.040\pm0.010^\dagger$  \\
$\theta$ & [$10^{-12}$\,rad] & $6.22\pm0.15^\ast$  \\
$R_{\text{Gaia}}$ & [$R_{\odot}$] & 0.200$^{+0.021\ast}_{-0.018}$  \\
$M_{\text{Gaia}}$ & [$M_{\odot}$] & 0.48$^{+0.12\ast}_{-0.09}$  \\
$\log{(L_{\text{Gaia}}/L_{\odot})}$ &  & 16$^{+3.6\ast}_{-2.8}$  \\
\noalign{\smallskip}
\hline
\multicolumn{3}{l}{Gaia: Based on measured \textit{Gaia} parallax, but applying a zero}\\ \multicolumn{3}{l}{point offset of $-0.029$\,mas (see Sect. \ref{Stellar radius, mass and luminosity} for details).}\\
\multicolumn{3}{l}{$\dagger$: 1$\sigma$ statistical errors only.}\\
\multicolumn{3}{l}{$\ast$: Listed uncertainties result from statistical and systematic}\\ \multicolumn{3}{l}{errors (see Sects. \ref{Effective temperature, surface gravity, and helium content} and \ref{SED fitting} for details).}\\
\end{tabular}
\end{table}

\section{Discussion}

\subsection{Tidal synchronisation of sdB+dM binaries} \label{syncro}

In close binaries, the rotation of the components is often assumed to be synchronised to their orbital motion. In this case the projected rotational velocity can be used to put tighter constraints on the companion mass. \citet{geier10b} found that assuming tidal synchronisation of the subdwarf primaries in sdB binaries with orbital periods of less than $\simeq1.2\,{\rm d}$ leads to consistent results in most cases. In particular, all the HW\,Vir type systems analysed in the \citet{geier10b} study turned out to be synchronised. 

In contrast to this, the projected rotational velocity of J08205+0008 is much smaller than is required for tidal synchronisation. We can calculate the expected rotational velocity ($v_{\rm rot}$) using the inclination ($i$), rotational period ($P_{\rm rot}$) and the radius of the primary ($R_1$) from the light curve analysis if we assume the system is synchronised:
	\begin{equation}
	P_{\rm rot, 1}=\frac{2\pi R_1}{v_{\rm rot}}\equiv P_{\rm orb} \rightarrow v_{\rm synchro} \sin i=\frac{2\pi R_1\sin i}{P_{\rm orb}}.
	\end{equation}

Due to the short period of this binary, the sdB should spin with $v_{\rm syncro}\simeq102\,{\rm km\,s^{-1}}$ similar to the other known systems \citep[see][and references therein]{geier10b}. 


Other observational results in recent years also indicate that tidal synchronisation of the sdB primary in close sdB+dM binaries is not always established in contrast to the assumption made by \citet{geier10b}. New theoretical models for tidal synchronisation \citep{preece:18,preece:19} even predict that none of the hot subdwarfs in close binaries should rotate synchronously with the orbital period. 

From the observational point of view, the situation appears to be rather complicated.
\citet{geier10b} found the projected rotational velocities of the two short-period ($P=0.1-0.12\,{\rm d}$) HW\,Vir systems HS\,0705+6700 and the prototype HW\,Vir to be consistent with synchronisation. \citet{charpinet08} used the splitting of the pulsation modes to derive the rotation period of the pulsating sdB in the HW\,Vir-type binary PG\,1336$-$018 and found it to be consistent with synchronised rotation. This was later confirmed by the measurement of the rotational broadening \citep{geier10b}. 

However, the other two sdBs with brown dwarf companions J162256+473051 and V2008-1753 \citep{schaffenroth14,schaffenroth15} have even shorter periods of only 0.07 d and both show sub-synchronous rotation with 0.6 and 0.75 of the orbital period, respectively, just like J0820+0008. AA Dor on the other hand, which has a companion very close to the hydrogen burning limit and a longer period of 0.25 d, seems to be synchronised \citep[and references therein]{vuckovic16}, but it has already evolved beyond the EHB and is therefore older and has had more time to synchronise.

\citet{pablo11} and \citet{pablo12} studied three pulsating sdBs in reflection effect sdB+dM binaries with longer periods and again used the splitting of the pulsation modes to derive their rotation periods ($P\simeq0.39-0.44\,{\rm d}$). All three sdBs rotate much slower than synchronised. But also in this period range the situation is not clear, since a full asteroseismic analysis of the sdB+dM binary Feige\,48 ($P\simeq0.38\,{\rm d}$) is consistent with synchronised rotation. 

Since synchronisation timescales of any kind \citep{geier10b} scale dominantely with the orbital period of the close binary, these results seem puzzling. Especially since the other relevant parameters such as mass and structure of the primary or companion mass are all very similar in sdB+dM binaries. They all consist of core-helium burning stars with masses of $\sim0.5\,M_{\odot}$ and low-mass companions with masses of $\sim0.1\,M_{\odot}$. And yet 5 of the analysed systems appear to be synchronised, while 6 rotate slower than synchronised without any significant dependence on companion mass or orbital period. 
This fraction, which is of course biased by complicated selection effects, might be an observational indication that the synchronisation timescales of such binaries are of the same order as the evolutionary timescales.

It has to be pointed out that although evolutionary tracks of EHB stars exist, the accuracy of the derived observational parameters (usually $T_{\rm eff}$ and $\log{g}$) is not high enough to determine their evolutionary age on the EHB by comparison with those tracks as accurate as it can be done for other types of stars (see Fig.~\ref{Kiel diagram}). As shown in Fig. \ref{Kiel diagram}, the position of the EHB is also dependent on the core and envelope mass and so it is not possible to find a unique track to a certain position in the $T_{\rm eff}-\log{g}$ diagram and in most sdB systems the mass of the sdB is not constrained accurately enough.

\begin{figure}
 \begin{center}
 \includegraphics[width=\linewidth]{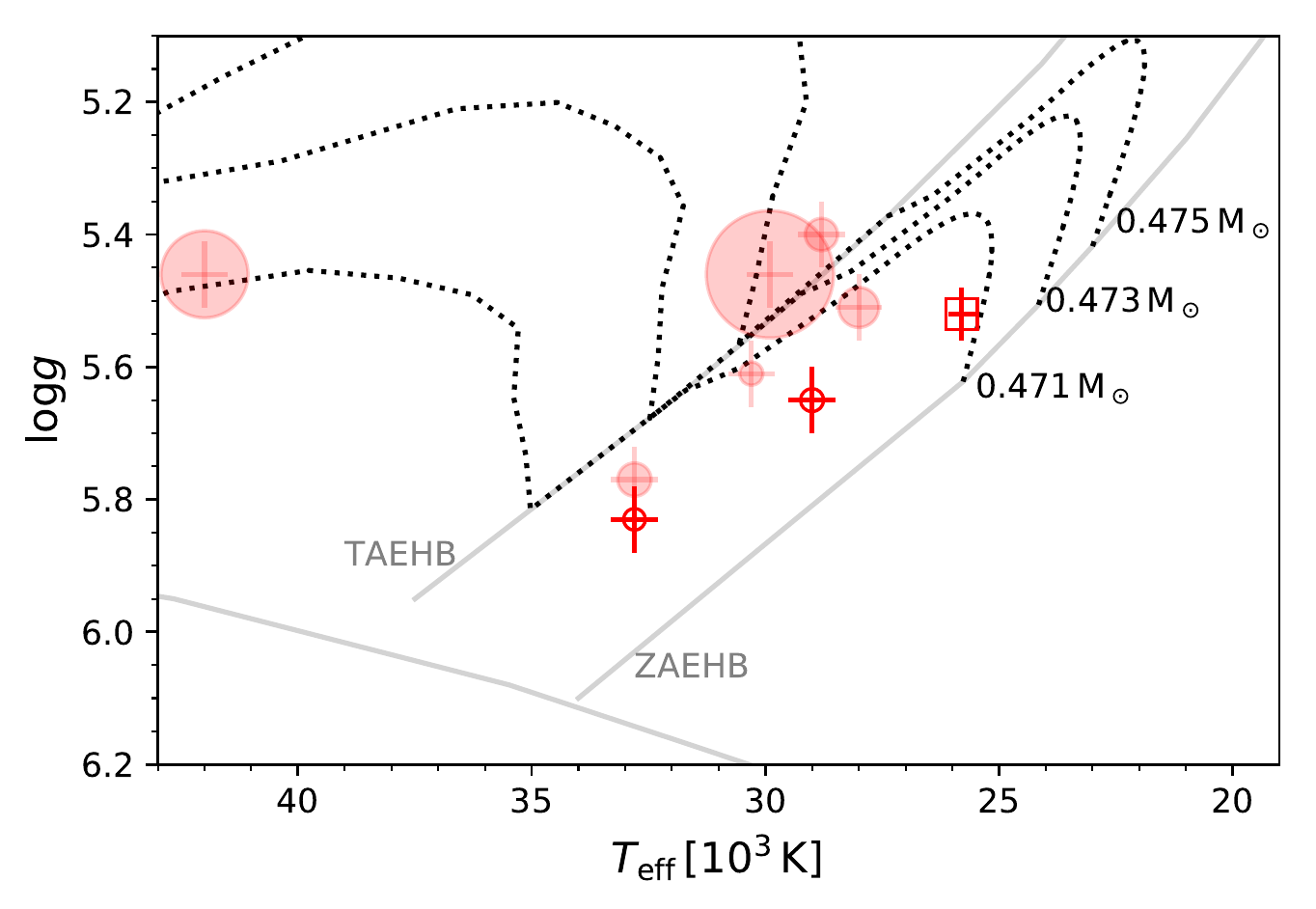}
    \caption{$T_{\text{eff}}-\log{(g)}$ diagram for the sdB+dM systems with known rotational periods mentioned in Sect. \ref{syncro}. The filled symbols represent synchronized systems, the open symbols, systems which are known to be non-synchronised. The square marks the position of J08205+0008. The sizes of the symbols scale with the orbital period, with longer periods having larger symbols.  Plotted error bars are the estimated parameter variations due to the reflection effect, as found e.g. in \citet{schaffenroth13}. The zero-age (ZAEHB) and terminal-age extreme horizontal branch (TAEHB) for a canonical mass sdB as well as evolutionary tracks for a canonical mass sdB with different envelope masses from \citet{1993ApJ...419..596D} are also shown.}
     \label{Kiel diagram synchro}
     \end{center}
 \end{figure}

\citet{2005A&A...430..223L} showed that sdB stars move at linear speed over the EHB and so the distance from the zero-age extreme horizontal branch (ZAEHB) represents how much time the star already spent on the EHB.
If we look at the position of the non-synchronised against the position of the synchronised systems in the $T_{\rm eff}-\log{g}$ diagram (Fig. \ref{Kiel diagram synchro}), it is obvious that all the systems, which are known to be synchronised, appear to be older. There also seems to be a trend that systems with a higher ratio of rotational to orbital velocity are further away from the ZAEHB. This means that the fraction of rotational to orbital period might even allow an age estimate of the sdB. 

The fact that the only post-EHB HW\,Vir system with a candidate substellar companion in our small sample (AA\,Dor) appears to be synchronised, while all the other HW\,Vir stars with very low-mass companions and shorter periods are not, fits quite well in this scenario. 
This could be a hint to the fact that for sdB+dM systems the synchronisation timescales are comparable to or even smaller than the lifetime on the EHB.
Hot subdwarfs spend $\sim100\,{\rm Myrs}$ on the EHB before they evolve to the post-EHB stage lasting $\sim10\,{\rm Myrs}$. So we would expect typical synchronisation timescales to be of the order of a few tens of millions of years, as we see both synchronised and unsynchronised systems.





 \subsection{A new explanation for the period decrease}
 There are different mechanisms of angular momentum loss in close binaries leading to a period decrease: gravitational waves, mass transfer (which can be excluded in a detached binary), or magnetic braking \citep[see][]{quian:08}.
 Here, we propose that tidal synchronisation can also be an additional mechanism to decrease the orbital period of a binary.
 
 From the rotational broadening of the stellar lines (see Sect. \ref{syncro}) we derived the rotational velocity of the subdwarf to be about half of what would be expected from the sdB being synchronised to the orbital period of the system. This means that the sdB is currently spun up by tidal forces until synchronisation is reached causing an increase in the rotational velocity. 
 As the mass of the companion is much smaller than the mass of the sdB, we assume synchronisation for the companion. 
 
 The total angular momentum of the binary system is given by the orbital angular momentum $J_{\rm orb}$ and the sum of the rotational angular momentum of the primary and secondary star $I_{\rm spin,1/2}$, with $\omega$ being the orbital angular velocity and $\Omega_i$ the rotational, angular velocity:
 \begin{align}
	J_{\rm tot}&=J_{\rm orb}+\sum_{i=1}^{2}I_{\rm spin,i} \\
	J_{\rm orb}&=(m_1a_1^2+m_2a_2^2)\,\omega = \frac{m_1m_2}{m_1+m_2}a^2\omega\\
	a^2&=\left(\frac{G(m1+m2)}{\omega^2}\right)^{2/3}\\
   I_{\rm spin,i}&=k_r^2M_iR_i^2\Omega_i
 \end{align}
 with $k_r^2$ the radius of gyration of the star. It refers to the distribution of the components of an object around its rotational axis. It is defined as $k_r^2=I/MR^2$, where $I$ is the moment of inertia of the star. \citet{geier10b} used a value of 0.04 derived from sdB models, which we adopt.

For now we neglect angular momentum loss due to gravitational waves and magnetic braking.
If we assume that the companion is already synchronised and its rotational velocity stays constant ($\frac{\rm d\Omega_{2}}{\mathrm{d}t}=0$) and that the masses and radii do not change, as we do not expect any mass transfer after the common envelope phase, we obtain 
	\begin{equation}
	\frac{dJ_{\rm tot}}{dt}=p_1\frac{\mathrm{d}\omega^{-1/3}}{\mathrm{d}t}+p_2\frac{\rm d\Omega_1}{\mathrm{d}t}=-p_1\frac{\dot{\omega}}{3\omega^{4/3}}+p_2\dot{\Omega}=0
	\end{equation}
	with 
	\begin{equation}
	p_1=\frac{m_1m_2G^{2/3}}{(m_1+m_2)^{1/3}}
	\end{equation}
	and
		\begin{equation}
	p_2=k_r^2m_1R_1^2
	\end{equation}
This shows that from an increase in the rotational velocity of the primary, which is expected from tidal synchronisation, we expect an increase of the orbital velocity, which we observe in the case of J08205+0008.
We can now calculate the current change of orbital velocity:
\begin{equation}
	\dot{\Omega}_1=\frac{p_1}{3p_2}\frac{\dot{\omega}}{\omega^{4/3}}=\frac{m_2G^{2/3}}{3k_r^2R_1^2(m_1+m_2)^{1/3}}\frac{\dot{\omega}}{\omega^{4/3}}
	\end{equation}

From this equation we can clearly see that rotational velocity change depends on the masses of both stars, the radius of the primary, the orbital velocity change and the current orbital velocity. An increasing rotational velocity causes an increasing orbital velocity and hence a period decrease.

\subsection{Synchronisation timescale}

If we assume that the observed period decrease is only due to the rotational velocity change, we can calculate the rate of the rotational velocity change and the timescale until synchronisation is reached.
 According to \citet{preece:18}, the change of rotational angular velocity is given by
\begin{equation}
    \frac{\mathrm{d}\Omega}{\mathrm{d}t}=\frac{\omega}{\tau_{\rm tide}}\left(1-\frac{\Omega}{\omega}\right)\frac{M_2}{M_1+M_2}\frac{a^2}{R^2k_r^2}\propto \left(1-\frac{\Omega}{\omega}\right)\label{eq}
\end{equation}
where $\tau_{\rm tide}$ is the tidal time-scale depending on the density, radius and mass of the star and the viscous time-scale of the convective region. The current position of J08205+0008 on the $T_{\rm eff}-\log g$ diagram and the mass we derived from our analysis suggest that the sdB is currently in the evolutionary phase of helium-burning. The lifetime of this phase is approximately 100 Myrs. So we do not expect the structure of the star to change significantly in the next few Myr. Because the moment of inertia of an sdB star is small compared to that of the binary orbit, the change in separation and angular velocity can be neglected.

Therefore, we can calculate the timescale until synchronisation is reached using the equation given in \citet{zahn:89}:
\begin{equation}
\frac{1}{T_{\rm sync}}=-\frac{1}{\Omega_1-\omega} \frac{\mathrm{d}\Omega_1}{\mathrm{d}t}
\end{equation}
Using our equation (\ref{eq}) and calculating and substituting the angular velocities by the periods we derive an expression for the synchronisation time scale:
\begin{equation}
    T_{\rm sync}=\left(1-\frac{2\pi R_1\sin i}{P_{\rm orb}v\sin i}\right)\frac{P_{\rm orb}^{2/3}v\sin i}{\dot{P}_{\rm orb}\sin i}\frac{3(2\pi)^{1/3}k_r^2R_1(m_1+m_2)^{1/3}}{m_2G^{2/3}}
\end{equation}
Using the orbital period, the masses, radii and inclination from our analysis, we calculate a synchronisation time $T_{\rm sync}$ of $2.1\pm0.1$ Myrs, well within the lifetime of a helium burning object on the extreme horizontal branch. The orbital period will change by about 200 s (3.5\%) in this 2 Myrs, which means a change in the separation of only 0.01 $\rm R_\odot$, which shows that our assumption of a negligible change in separation is valid. If we assume that the rotation after the common envelope phase was close to zero, the total timescale until the system reaches synchronisation is about 4 Myrs. This assumption is plausible as most red giant progenitors rotate slowly and the common envelope phase is very short-lived and so no change of the rotation is expected.

This means that this effect could significantly add to the observed period decrease. The fact that the synchronised systems appear to be older than the non-synchronised ones confirms that the synchronisation timescale is of the expected order of magnitude and it is possible that we might indeed measure the synchronisation timescale. 

As mentioned before \citet{preece:18} predict that the synchronisation timescales are much longer than the lifetime on the EHB and that none of the HW Vir systems should be synchronised.
\citet{preece:19} investigated also the special case of NY Vir, which was determined to be synchronised from spectroscopy and asteroseismolgy, and came to the conclusion that they cannot explain, why it is synchronised. They proposed that maybe the outer layers of the sdB were synchronised during the common envelope phase. However, observations show that synchronised sdB+dM systems are not rare, but that synchronisation occurs most likely during the phase of helium-burning,  
which shows that synchronisation theory is not yet able to predict accurate synchronization time scales.

\subsection{Orbital period variations in HW Vir systems}

As mentioned before, there are several mechanisms that can explain period changes in HW Vir systems. 
The period change due to gravitational waves is usually very small in HW Vir systems and would only be observable after observations for many decades \citep[e.g.][]{kilkenny14}. Using the equation given in \citet{kupfer20} with the system parameters derived in this paper, we predict an orbital period decay due to gravitational waves of $\dot{P}=4.5\cdot10^{-14}\,\rm ss^{-1}$.
The observed change in orbital period is hence about 100 times higher than expected by an orbital decay due to gravitational waves.

HW Vir and NY Vir have also been observed to show a period decrease of the same order of magnitude \citep{quian:08,kilkenny14} but have been found to rotate (nearly) synchronously. 
Both also show additionally to the period decrease a long-period sinusoidal signal \citep{Lee:09,Lee:14}. These additional variations in the O--C diagram have been interpreted as caused by  circumbinary planets in both cases, however the solutions were not confirmed with observations of longer baselines. Observations of more than one orbital period of the planet would be necessary to confirm it. The period decrease was explained to be caused by angular momentum loss due to magnetic stellar wind braking.

\begin{figure}
    \centering
    \includegraphics[width=\linewidth]{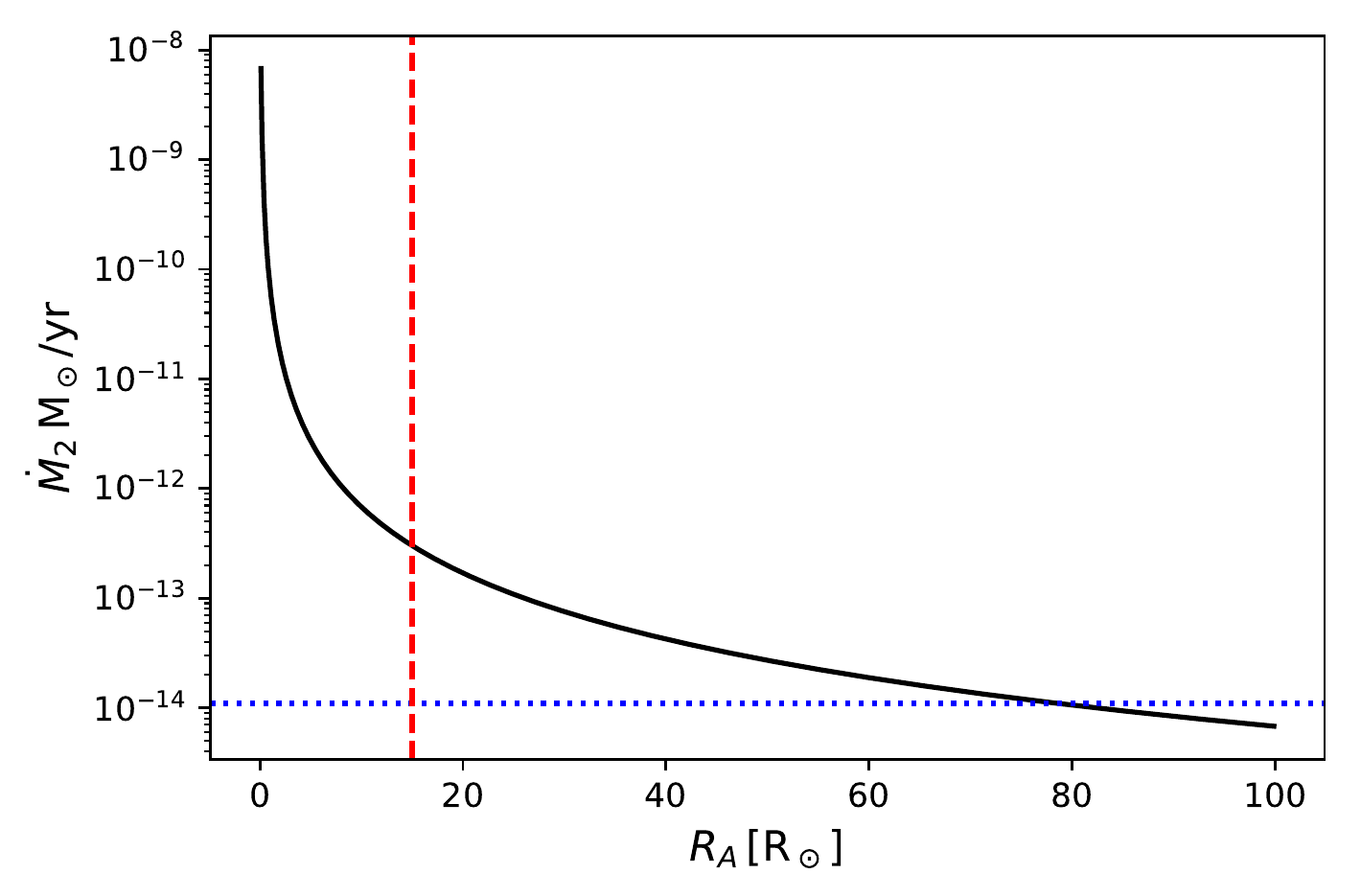}
    \caption{Correlation between the Alfv\'{e}n radius and the mass loss rate
for the companion of J08205+0008.  The red dashed line marks the Alfv\'{e}n radius for the Sun, the blue dotted line indicates the tidally enhanced mass-loss rate determined using the parameters of the sdB using the formula of \citet{tout:88}.}
    \label{mag_braking}
\end{figure}

Following the approach of \citet{qian:07} we calculated the relation between the mass-loss rate and the Alfv\'{e}n radius that would be required to account for the period decrease in J08205+0008 due to magnetic braking. This is shown in Fig. \ref{mag_braking}. Using the tidally enhanced mass-loss rate of \citet{tout:88} we derive that an Alfv\'{e}n radius of $75\,\rm R_\odot$ would be required to cause the period decrease we measure, much larger than the Alfv\'{e}n radius of the Sun. This shows that, as expected, the effect of magnetic braking in a late M dwarf or massive brown dwarf is very small at best and cannot explain the period decrease we derive.

\citet{bours:16} made a study of close white dwarf binaries and observed that the amplitude of eclipse arrival time variations in K dwarf and early M dwarf companions is much larger than in late M dwarf, brown dwarf or white dwarf companions, which do not show significant orbital period variations. They concluded that these findings are in agreement with the so-called Applegate mechanism, which proposes that variability in the binary orbits can be driven by magnetic cycles in the secondary stars. In all published HW Vir systems with a longer observational baseline of several years quite large period variations on the order of minutes have been detected \citep[see][for an overview]{zorotovic:13, pulley:18}, with the exception of AA Dor \citep{kilkenny14}, which still shows no sign of period variations after a baseline of about 40 years. Also the orbital period decrease in J08205+0008 is on the order of seconds and has only been found after 10 years of observation and no additional sinusoidal signals have been found as seen in many of the other systems. This confirms that the findings of \citet{bours:16} apply to close hot subdwarf binaries with cool companions. The fact that the synchronised HW Vir system AA Dor does not show any period variations also confirms our theory that the period variations in HW Vir systems with companions close to the hydrogen-burning limit might be caused by tidal synchronisation. In higher-mass M dwarf companions the larger period variations are likely caused by the Applegate mechanism and the period decrease can be caused dominantly by magnetic braking and additionally tidal synchronisation.

It seems that orbital period changes in HW Vir systems are still poorly understood and have also not been studied observationally to the full extent. More observations over long time spans of synchronised and non-synchronised short-period sdB binaries with companions of different masses will be necessary to understand synchronisation and orbital period changes of hot subdwarf binaries. Most likely it cannot be explained with just one effect and is likely an interplay of different effects.

\subsection{Inflation of brown dwarfs and low-mass M dwarfs in eclipsing WD or sdB binaries}
Close brown dwarf companions that eclipse main sequence stars are rare, with only 23 known to date \citep{carmichael20}. Consequently, brown dwarf companions to the evolved form of these systems are much rarer with only three (including J08205+0008) known to eclipse hot subdwarfs, and three known to eclipse white dwarfs. These evolved systems are old ($>$ 1 Gyr), and the brown dwarfs are massive, and hence not expected to be inflated \citep{thorngren18}.

\begin{figure*}
    \centering
    \includegraphics[width=1.05\linewidth]{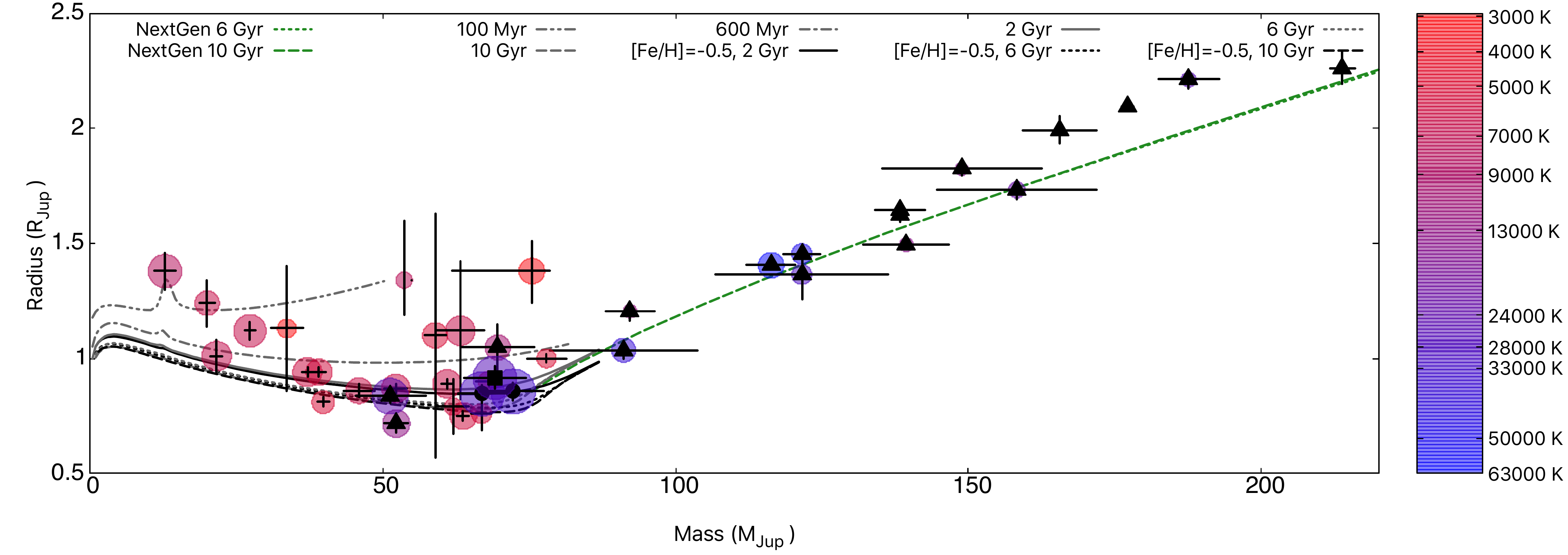}
    \caption{All known eclipsing binary white dwarfs with detached brown dwarf (triangles: \citealt{parsons17, littlefair14}) and late M dwarf companions (triangles) from \citet{parsons18}, hot subdwarfs with eclipsing brown dwarf companions (circles: \citealt{schaffenroth14a, schaffenroth15}) and all known eclipsing brown dwarf companions to main sequence stars (+: \citealt{carmichael20}). J08205+0008 is plotted as the filled square. The colour is proportional to the effective temperature of the primary in each system and the coloured circle size is proportional to the amount of total incident radiation the secondary receives. Also shown are the Sonora Bobcat brown dwarf evolutionary models of \citet{bobcat} for solar and sub-solar metallicity and the NextGen models \citep{baraffe97}.}
     \label{MR}
\end{figure*}

Surprisingly, of the three hot subdwarfs with brown dwarf companions, J08205+0008 is the one that receives the least irradiation - almost half that received by V2008-1753 and SDSSJ162256.66+473051.1, both of which have hotter primaries (32000~K, 29000~K) and shorter periods ($\sim$1.6~hr) than J08205+0008. This suggests that more irradiation, and more irradiation at shorter wavelengths does not equate to a higher level of inflation of a brown dwarf. Indeed this finding is consistent with that for brown dwarfs irradiated by white dwarfs, where the most irradiated object with a measured radius is SDSS J1205-0242B, in a 71.2 min orbit around a 23681~K white dwarf and yet the brown dwarf is not inflated \citep{parsons17}. The brown dwarf in this system only receives a hundredth of the irradiation that J08205+0008 does. However, WD1032+011, an old white dwarf ($T_{\rm eff} \sim$ 10000~K) with a high mass brown dwarf companion (0.0665 M$_{\odot}$) does appear to be inflated \citep{casewell:2020}. As can be seen from Figure \ref{MR}, the majority of the low mass brown dwarfs (M$<$35 M$_{\rm Jup}$) are inflated, irrelevant of how much irradiation they receive. For the few old (5-10 Gyr), higher mass inflated brown dwarfs, the mechanism leading to the observed inflation is not yet understood.

\subsection{Previous and future evolution of the system}
As stated before, stars with a cool, low-mass companion sitting on the EHB are thought to have formed by a common-envelope phase from a progenitor of up to two solar mass on the RGB. Due to the large mass ratio only unstable mass transfer is possible. If the mass transfer happened at the tip of the RGB, a core-helium burning object with about $0.5\,\rm M_\odot$ will be formed. If the mass transfer happened earlier then the core of the progenitor has not enough mass to start He-core burning and the pre-He WD will move to the WD cooling track crossing the EHB. Our analysis of J08205+0008 showed that a low-mass solution ($0.25\,M_\odot$, as discussed previously) can be excluded and that the primary star is indeed currently a core He-burning object. 

\citet{kupfer15} calculated the evolution of J08205+0008 and  considering only angular momentum loss due to gravitational waves and found that the companion will fill its Roche lobe in about 2.2 Gyrs and mass transfer is expected to start forming a cataclysmic variable.
We detected a significantly higher orbital period decrease in this system than expected from gravitational waves. Up to now, we could not detect any change in the rate of this period decrease. If we assume that the orbital period change is due to rotational period change 
until synchronisation is reached and afterwards the period decrease will be solely due to gravitational waves, 
we can calculate when the companion will fill its Roche lobe and accretion to the primary will start. To calculate the Roche radius the equation derived in \citet{eggleton83} was used:
\begin{equation}
    R_L=\frac{0.49q^{2/3}}{0.6q^{2/3}+\ln(1+q^{1/3})}a
\end{equation}
Using the values derived in our analysis we calculate that the Roche lobe of the companion will be filled at a system separation of 0.410 $\rm R_\odot$, 56\% of the current separation, which is reached at a period of 3525 s. 
From this we calculate a time scale of 1.8 Gyrs until the Roche lobe will be filled.

Systems with a mass ratio $q=M_2/M_1<2/3$, with $M_1$ being the mass of the accretor, are assumed to be able to undergo stable mass transfer. Our system has a mass ratio of $0.147\ll2/3$. 
The subdwarf will already have evolved to a white dwarf and a cataclysmic variable will be formed. It is expected that the period of an accreting binary with a hydrogen-rich donor star will decrease until a minimum period of $\simeq70$ min is reached at a companion mass around $0.06\,M_\odot$ and the period will increase again afterwards \citep{nelson18}. Such systems are called period bouncers. Our system comes into contact already close to the minimum period and should hence increase the period when the mass transfer starts.

The future of the system depends completely on the period evolution. A longer baseline of observations of this system is necessary to confirm that the period decrease is indeed stable and caused by the tidal synchronisation. 


\section{Conclusion and summary}

The analysis of J08205+0008 with higher quality data from ESO-VLT/XSHOOTER, ESO-VLT/UVES and ESO-NTT/ULTRACAM  allowed us to constrain the masses of the sdB and the companion much better by combining the analysis of the radial velocity curve and the light curve. We determine an sdB mass of $0.39-0.50\,\rm M_\odot$ consistent with the canonical mass and a companion mass of $0.061-0.071\,\rm M_\odot$ close to the hydrogen burning limit. Therefore, we confirm that the companion is likely be a massive brown dwarf. 

The atmospheric parameters and abundances show that J08205+0008 is a typical sdB and comparison with stellar evolution tracks suggest that the mass has to be less than $0.50\,M_\odot$ consistent with our solution and also the mass derived by a spectrophotometric method using Gaia parallaxes and the SED derived in the secondary eclipse, where the companion is not visible. 

If the sdB evolved from a $1\,\rm M_\odot$ star, the age of the system is expected to be around 10\, Gyrs. In this case the radius of the brown dwarf companion is about 20\% inflated compared to theoretical calculations. Such an inflation is observed in several sdB/WD+dM/BD systems but not understood yet. However, the inflation seems not to be caused by the strong irradiation.
The sdB binary belongs to the thin disk, as do about half of the sdB at this distance from the Galacic plane. This means that they also could be young, if they have evolved from a more massive progenitor. Then we get a consistent solution without requiring inflation of the companion. However, a brown dwarf companion might not be able to remove the envelope of a more massive progenitor.

We detected a significant period decrease in J0820+0008. This can be explained by the spin-up of the sdB due to tidal sychronisation. We calculated the synchronisation timescale to 4 Myrs well within the lifetime on the EHB. The investigation of the parameters of all known Vir systems with rotational periods (see Sect. \ref{syncro}) shows that the synchronised systems tend to be older, showing that the synchronisation timescale seems to be comparable but smaller than the lifetime on the EHB in contrast to current synchronisation theories.

By investigating the known orbital period variations in HW Vir systems, we can confirm the findings by \citet{bours:16} that period variations in systems with higher mass M dwarf companions seem to be larger. Hence, we conclude that the large period variations in those systems are likely caused by the Applegate mechanism and the observed period decreases dominantly by magnetic braking. In lower-mass companions close to the hydrogen-burning limit, on the other hand, tidal synchronisation spinning up the sdB could be responsible for the period decrease, allowing us to derive a synchronisation timescale.  


The results of our analysis are limited by the precision of the available trigonometric parallax. As the Gaia mission proceeds, the precision and accuracy of the trigonometric parallax will improve, which will narrow down the uncertainties of the stellar parameters.
A very important goal is to detect spectral signatures from the companion and to measure the radial velocity curve of the companion. We failed to do so, because the infrared spectra at hand are of insufficient quality. The future IR instrumentation on larger telescopes, such as the ESO-ELT, will be needed. A high precision measurement of the radial velocity curves of both components will then allow us to derive an additional constraint on mass and radius from the difference of the stars' gravitational redshifts \citep{vos13}.  
Such measurements will give an independent determination of the nature of the companion and will help to test evolutionary models for low mass star near the hydrogen burning limit via the mass-radius relation.

The combination of many different methods allowed us to constrain the masses of both components much better without having to assume a canonical mass for the sdB. This is only the fourth HW Vir system for which this is possible.





\section*{Acknowledgements}
D.S. is supported by the Deutsche Forschungsgemeinschaft (DFG) under grant HE 1356/70-1 and IR190/1-1.
V.S. is supported by the Deutsche Forschungsgemeinschaft, DFG through grant GE 2506/9-1. S.L.C. is supported by an STFC Ernest Rutherford Fellowship ST/R003726/1.
DK thanks the SAAO for generous allocations of telescope time and the National Research Foundation of South Africa and the University of the Western Cape for financial support. VSD, SPL and ULTRACAM are supported by the STFC.
 We thank J. E. Davis for the development of the \texttt{slxfig} module, which has been used to prepare figures in this work. \texttt{matplotlib} \citep{2007CSE.....9...90H} and \texttt{NumPy} \citep{2011CSE....13b..22V} were used in order to prepare figures in this work. This work has made use of data from the European Space Agency (ESA) mission
{\it Gaia} (\url{https://www.cosmos.esa.int/gaia}), processed by the {\it Gaia}
Data Processing and Analysis Consortium (DPAC, \url{https://www.cosmos.esa.int/web/gaia/dpac/consortium}). Funding for the DPAC has been provided by national institutions, in particular the institutions participating in the {\it Gaia} Multilateral Agreement. Based on observations at the Cerro Paranal Observatory of the European Southern Observatory (ESO) in Chile under the program IDs 087.D-0185(A), and 098.C-0754(A). Based on observations at the La Silla Observatory of the European Southern Observatory (ESO) in Chile under the program IDs 082.D-0649(A), 084.D-0348(A), and 098.D-679. 
This paper uses observations made at the South African Astronomical Observatory.
We made extensive use of NASAs Astrophysics Data System Abstract Service (ADS) and the SIMBAD and VizieR database, operated at CDS, Strasbourg, France.

\section*{Data availability statement}
Most data are incorporated into the article and its online supplementary material. All other data are available on request.




\bibliographystyle{mnras}
\bibliography{bibliog} 

\newpage
\appendix




\section{Telluric correction}\label{app:telluric}

\begin{figure*}
\begin{center}
\begin{minipage}[b]{0.5\linewidth}
\resizebox{\hsize}{!}{\includegraphics[trim = 0cm 0cm 0cm 0cm, clip, scale=0.5]{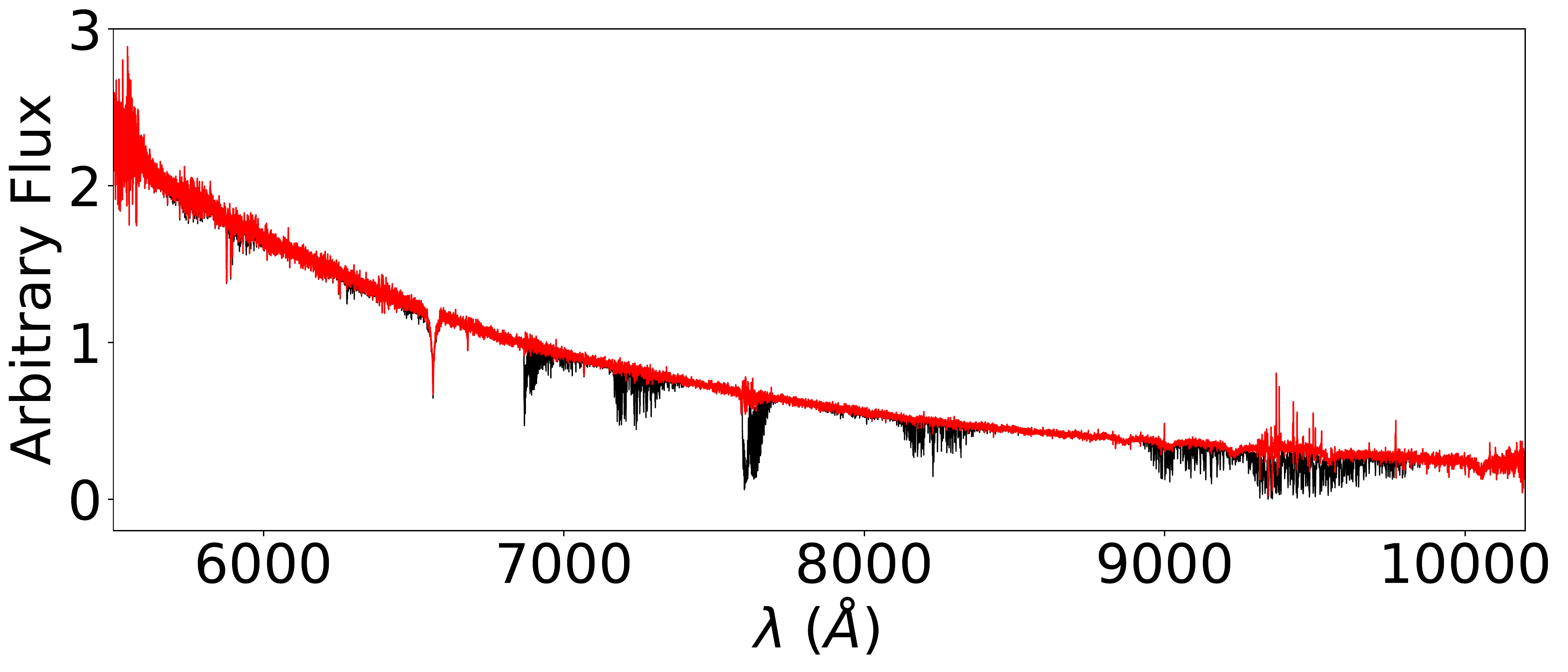}}
\centering
\end{minipage}\hfill
\begin{minipage}[b]{0.5\linewidth}
\resizebox{\hsize}{!}{\includegraphics[trim = 0cm 0cm 0cm 0cm, clip, scale=0.5]{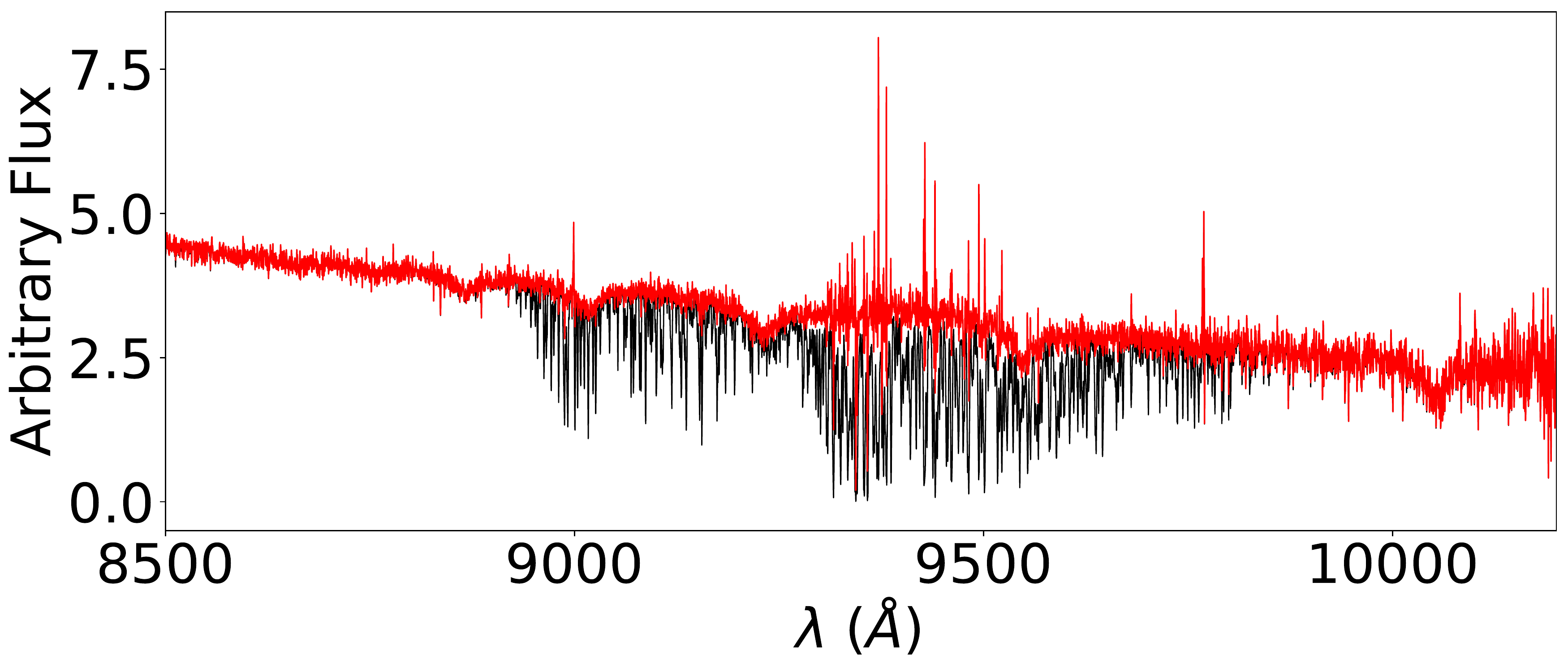}}
\centering
\end{minipage}\hfill
\caption{\textit{Left-hand panel}: Quality of telluric absorption correction for a full example spectrum of J08205+0008 taken with the VIS arm of the XSHOOTER spectrograph. The telluric absorption corrected spectrum (red) is shown in comparison with the original spectrum (black). Note that fluxes were scaled for illustrative purposes. \textit{Right-hand panel}: Same as left-hand panel, but for the spectral range of the hydrogen Paschen series.} 
\label{VIS X-Shooter arm before and after telluric correction with molecfit}
\end{center}
\end{figure*}\noindent

\section{Atmospheric model parameters}
\begin{table}
\caption{Model atoms for NLTE calculations used for the hybrid LTE/NLTE approach.}\label{summary of model atoms used for the hybrid LTE/NLTE approach}
\centering
\begin{tabular}{cc}
\hline\hline
Ion & Model atom\\
\hline
$\ion{H}{i}$ & \citet{Przybilla_2004}\\
$\ion{He}{i}$ & \citet{Przybilla_2005}\\
$\ion{C}{ii}$ & \citet{Nieva_2006, Nieva_2008}\\
$\ion{N}{ii}$ & \citet{Przybilla_2001a}$^\dagger$\\
$\ion{O}{i/ii}$ & \citet{Przybilla_2000}, \citet{Becker_1988}$^\dagger$\\
$\ion{Ne}{i/ii}$ & \citet{Morel_2008}$^\dagger$\\
$\ion{Mg}{ii}$ & \citet{Przybilla_2001b}\\
$\ion{Al}{iii}$ & Przybilla (in prep.)\\
$\ion{Si}{ii/iii/iv}$ & Przybilla \& Butler (in prep.)\\
$\ion{S}{ii/iii}$ & \citet{Vrancken_1996}$^\dagger$\\
$\ion{Ar}{ii}$ & Butler (in prep.)\\
$\ion{Fe}{ii/iii}$ & \citet{Becker_1998}, \citet{Morel_2006}$^\dagger$\\
\hline
\multicolumn{2}{l}{$\dagger$: Updated and corrected as described by \citet{Nieva_2012}.}
\end{tabular}
\end{table}\noindent

\begin{figure}
    \centering
    \includegraphics[trim = 0cm 0cm 0cm 0cm, clip, scale=0.273]{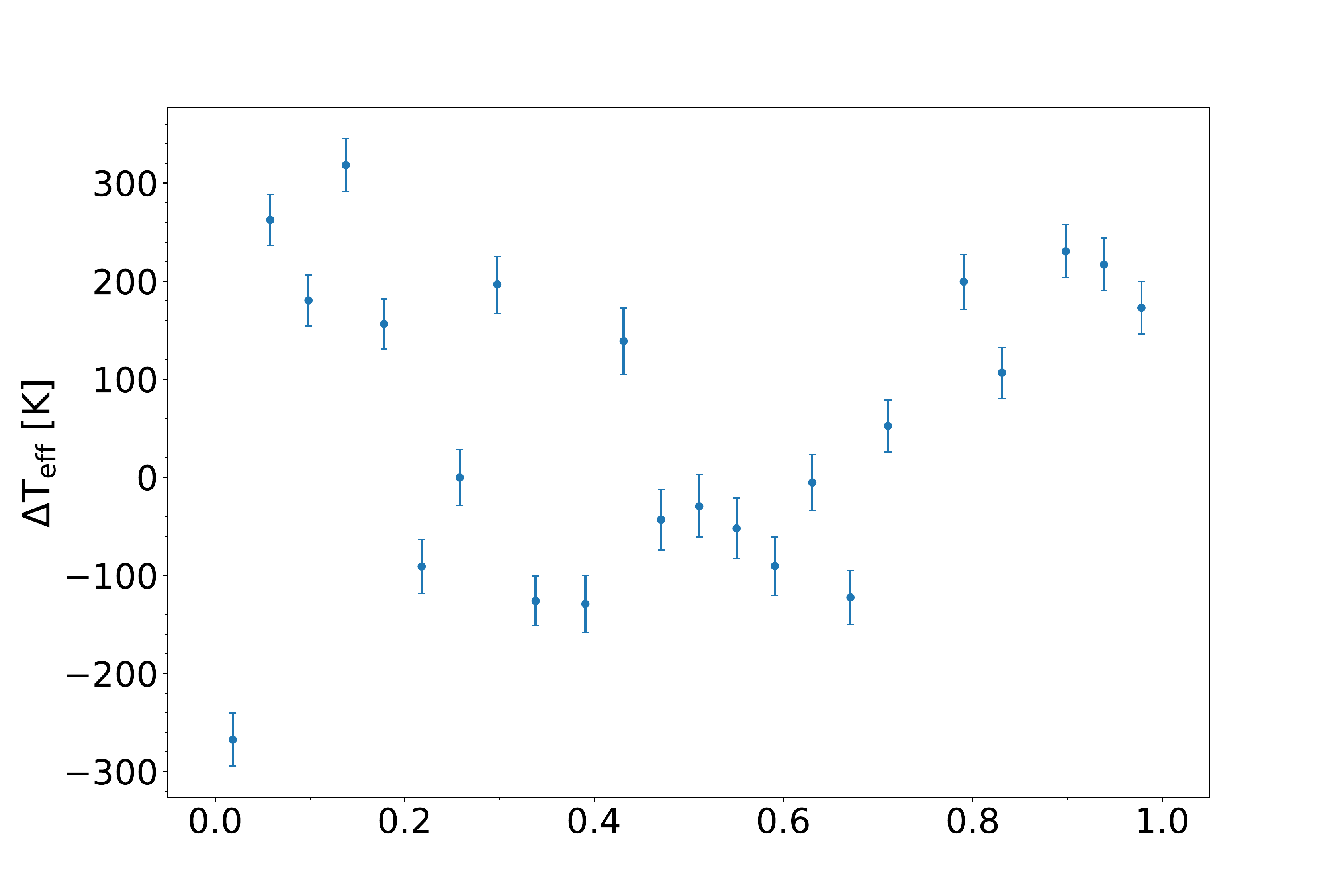}
    \includegraphics[trim = 0cm 0cm 0cm 0cm, clip, scale=0.273]{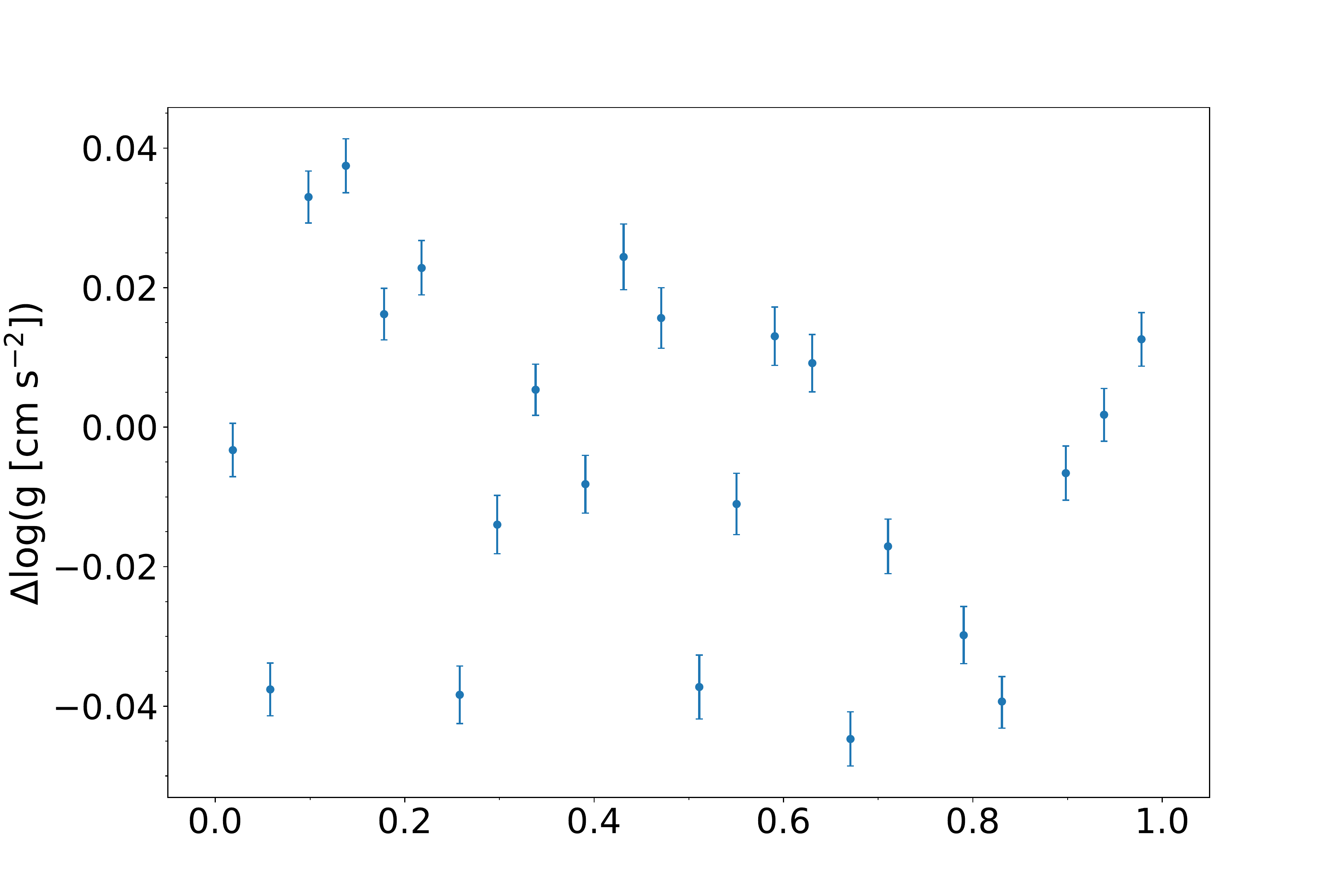}
    \includegraphics[trim = 0cm 0cm 0cm 0cm, clip, scale=0.273]{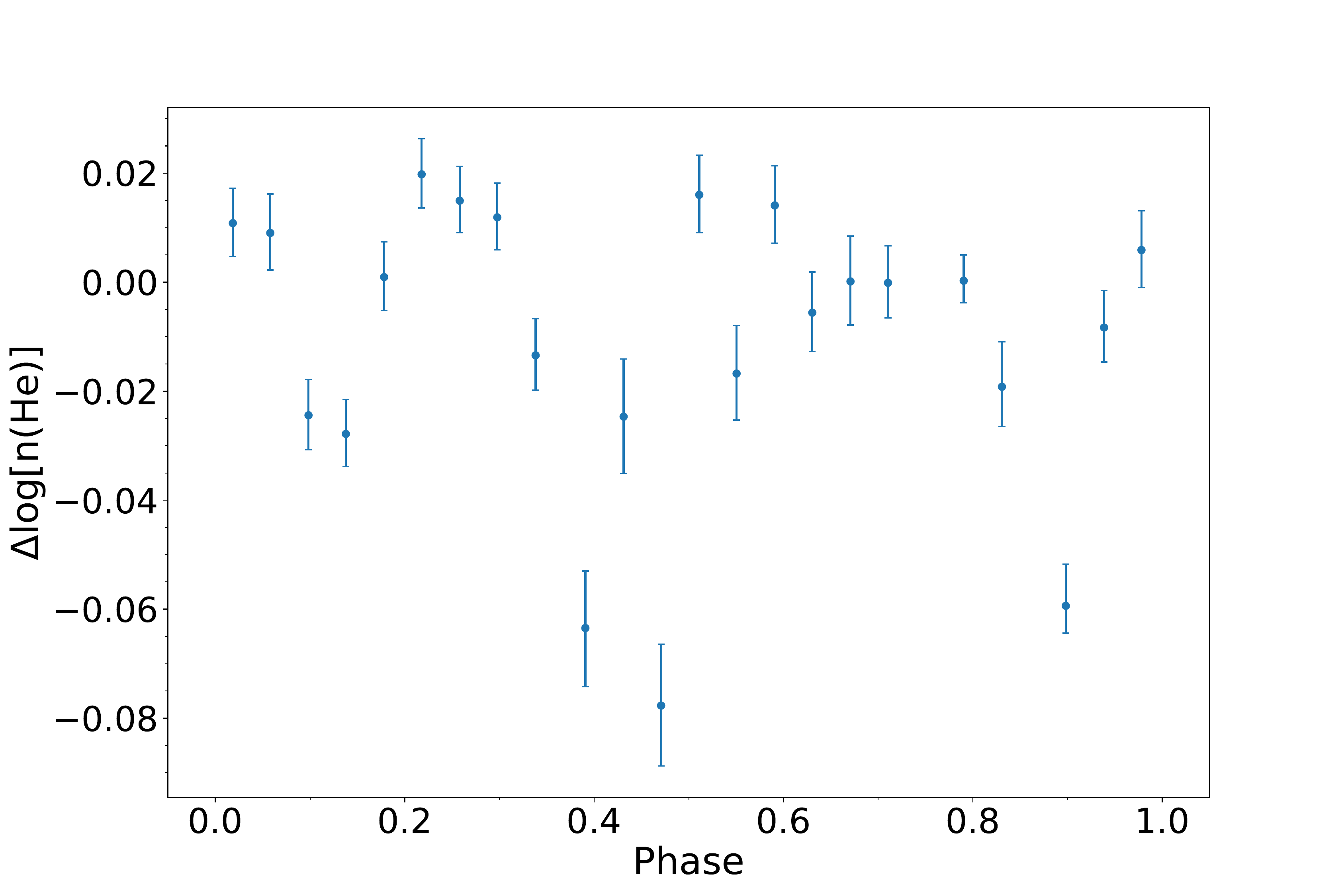}
    \caption{Change of the atmospheric parameters determined from the single XSHOOTER spectra plotted against the orbital phase. While the differences plotted on the y axes result from the subtraction of the best fit parameters derived from the co-added XSHOOTER spectrum from the determined parameters for the single spectra, the orbital phase was calculated based on the photometric solutions of $T_0$ and $P$ (see Table \ref{tab:par} for details). Due to the relatively weak reflection effect of less than 5\%, the variations measured for effective temperature (upper panel), surface gravity (middle panel), and helium abundance (lower panel) are of the order of the total uncertainties listed in Table \ref{tab:par} and therefore are not significant.}
    \label{change of atmospheric parameters vs. orbital phase}
\end{figure}

\begin{table}
\caption{Hybrid LTE/NLTE model grid used for the quantitative spectral analysis of J08205+0008.}\label{Hybrid LTE/NLTE model grid used for the quantitative spectral analysis of SDSS J08205+0008}
\centering
\begin{tabular}{ccc}
\hline\hline
Parameter & Grid size & Step size\\
\hline
$T_{\text{eff}}$ & \SI{25000}{\kelvin} to \SI{30000}{\kelvin} & \SI{1000}{\kelvin}\\
$\log{(g)}$ & 5.2 to 5.8 & 0.2\\
$\log{n(\text{He})}$ & -2.2 to -1.6 & 0.2\\
$\log{n(\text{C})}$ & -4.6 to -4.0 & 0.2\\
$\log{n(\text{N})}$ & -4.2 to -3.6 & 0.2\\
$\log{n(\text{O})}$ & -4.4 to -3.8 & 0.2\\
$\log{n(\text{Ne})}$ & -7.0 to -6.0 & 0.2\\
$\log{n(\text{Mg})}$ & -5.4 to -4.4 & 0.2\\
$\log{n(\text{Al})}$ & -7.0 to -6.0 & 0.2\\
$\log{n(\text{Si})}$ & -5.4 to -5.0 & 0.2\\
$\log{n(\text{S})}$ & -6.0 to -5.2 & 0.2\\
$\log{n(\text{Ar})}$ & -5.8 to -5.4 & 0.2\\
$\log{n(\text{Fe})}$ & -4.8 to -4.2 & 0.2\\
\hline
\end{tabular}
\end{table}\noindent

\section{Line fits}\label{app:fits}

 
 

\section{List of metal lines}\label{app:metallines}

\begin{table*}
\caption{List of selected metal lines in the co-added XSHOOTER and UVES spectra of J08205+0008.}\label{list of lines detected}
\centering
\begin{tabular}{cccccccc}
\hline\hline
El. + ion. stage & $\lambda$ [\si{\angstrom}] & El. + ion. stage & $\lambda$ [\si{\angstrom}] & El. + ion. stage & $\lambda$ [\si{\angstrom}] & El. + ion. stage & $\lambda$ [\si{\angstrom}]\\
\hline
$\ion{C}{ii}$ &	3918.97 & $\ion{N}{ii}$ & 5686.21 & $\ion{Al}{iii}$ & 4512.565 & $\ion{Fe}{iii}$ & 3600.943 \\ 	
$\ion{C}{ii}$ &	3920.68 & $\ion{N}{ii}$ & 5710.77 & $\ion{Al}{iii}$ & 5696.604 & $\ion{Fe}{iii}$ & 3603.890 \\ 
$\ion{C}{ii}$ &	4267.00	& $\ion{N}{ii}$ & 5927.81 & $\ion{Al}{iii}$ & 5722.730 & $\ion{Fe}{iii}$ & 3611.736 \\ 
$\ion{C}{ii}$ &	4267.26	& $\ion{N}{ii}$ & 5931.78 & $\ion{Si}{ii}$ & 3856.018 & $\ion{Fe}{iii}$ & 3999.325 \\
$\ion{C}{ii}$ &	5132.95 & $\ion{N}{ii}$ & 5940.24 & $\ion{Si}{ii}$ & 3862.595 & $\ion{Fe}{iii}$ & 4000.518 \\
$\ion{C}{ii}$ &	5133.28 & $\ion{N}{ii}$ & 5941.65 & $\ion{Si}{ii}$ & 4128.067 & $\ion{Fe}{iii}$ & 4005.573 \\
$\ion{C}{ii}$ & 5145.16 & $\ion{N}{ii}$ & 5952.39 & $\ion{Si}{ii}$ & 4130.893 & $\ion{Fe}{iii}$ & 4137.130 \\
$\ion{C}{ii}$ &	6151.265 & $\ion{N}{ii}$ & 5954.28 & $\ion{Si}{ii}$ & 6347.103 & $\ion{Fe}{iii}$ & 4139.350 \\
$\ion{C}{ii}$ &	6151.534 & $\ion{N}{ii}$ & 6150.75 & $\ion{Si}{ii}$ & 6371.359 & $\ion{Fe}{iii}$ & 4140.482 \\
$\ion{C}{ii}$ &	6461.95 & $\ion{N}{ii}$ & 6482.05 & $\ion{Si}{iii}$ & 3590.465 & $\ion{Fe}{iii}$ & 4164.916 \\
$\ion{C}{ii}$ &	6578.05	& $\ion{N}{ii}$ & 6610.56 & $\ion{Si}{iii}$ & 3806.526 & $\ion{Fe}{iii}$ & 4194.051 \\
$\ion{C}{ii}$ &	6582.88	& $\ion{O}{i}$ & 7771.94 & $\ion{Si}{iii}$ & 3806.7 & $\ion{Fe}{iii}$ & 4210.674 \\
$\ion{C}{ii}$ & 6779.94 & $\ion{O}{i}$ & 7774.17 & $\ion{Si}{iii}$ & 3806.779 & $\ion{Fe}{iii}$ & 4222.271 \\
$\ion{C}{ii}$ & 6780.59 & $\ion{O}{i}$ & 7775.39 & $\ion{Si}{iii}$ & 3924.468 & $\ion{Fe}{iii}$ & 4248.773 \\
$\ion{C}{ii}$ & 6783.91 & $\ion{O}{i}$ & 8446.25 & $\ion{Si}{iii}$ & 4552.622 & $\ion{Fe}{iii}$ & 4261.391 \\
$\ion{C}{ii}$ & 6791.47 & $\ion{O}{i}$ & 8446.36 & $\ion{Si}{iii}$ & 4567.84 & $\ion{Fe}{iii}$ & 4273.372 \\
$\ion{C}{ii}$ &	6800.69 & $\ion{O}{i}$ & 8446.76 & $\ion{Si}{iii}$ & 4574.757 & $\ion{Fe}{iii}$ & 4273.409 \\
$\ion{C}{ii}$ & 7231.33 & $\ion{O}{ii}$ & 3390.21 & $\ion{Si}{iii}$ & 4716.654 & $\ion{Fe}{iii}$ & 4286.091 \\
$\ion{C}{ii}$ & 7236.42	& $\ion{O}{ii}$ & 3712.74 & $\ion{Si}{iii}$ & 4813.333 & $\ion{Fe}{iii}$ & 4286.128 \\
$\ion{C}{ii}$ & 7237.17	& $\ion{O}{ii}$ & 3727.32 & $\ion{Si}{iii}$ & 4819.631 & $\ion{Fe}{iii}$ & 4286.164 \\
$\ion{N}{ii}$ & 3328.72 & $\ion{O}{ii}$ & 3911.96 & $\ion{Si}{iii}$ & 4819.712 & $\ion{Fe}{iii}$ & 4296.814 \\
$\ion{N}{ii}$ & 3329.70 & $\ion{O}{ii}$ & 3912.12 & $\ion{Si}{iii}$ & 4819.814 & $\ion{Fe}{iii}$ & 4296.851 \\ 
$\ion{N}{ii}$ & 3330.32 & $\ion{O}{ii}$ & 4069.62 & $\ion{Si}{iii}$ & 4828.95 & $\ion{Fe}{iii}$ & 4304.748 \\
$\ion{N}{ii}$ & 3331.31 & $\ion{O}{ii}$ & 4069.88 & $\ion{Si}{iii}$ & 4829.03 & $\ion{Fe}{iii}$ & 4304.767 \\
$\ion{N}{ii}$ & 3437.14 & $\ion{O}{ii}$ & 4072.16 & $\ion{Si}{iii}$ & 4829.111 & $\ion{Fe}{iii}$ & 4310.355 \\
$\ion{N}{ii}$ & 3995.00 & $\ion{O}{ii}$ & 4075.86 & $\ion{Si}{iii}$ & 4829.214 & $\ion{Fe}{iii}$ & 4419.596 \\
$\ion{N}{ii}$ & 4035.08 & $\ion{O}{ii}$ & 4132.80 & $\ion{Si}{iii}$ & 5696.49 & $\ion{Fe}{iii}$ & 4649.271 \\
$\ion{N}{ii}$ & 4041.31 & $\ion{O}{ii}$ & 4185.44 & $\ion{Si}{iii}$ & 5739.73 & $\ion{Fe}{iii}$ & 5063.421 \\
$\ion{N}{ii}$ & 4043.53 & $\ion{O}{ii}$ & 4189.58 & $\ion{S}{ii}$ & 3613.03 & $\ion{Fe}{iii}$ & 5073.903 \\
$\ion{N}{ii}$ & 4176.16 & $\ion{O}{ii}$ & 4189.79 & $\ion{S}{ii}$ & 5201.027 & $\ion{Fe}{iii}$ & 5086.701 \\
$\ion{N}{ii}$ & 4199.98 & $\ion{O}{ii}$ & 4366.89 & $\ion{S}{ii}$ & 5201.379 & $\ion{Fe}{iii}$ & 5194.160 \\
$\ion{N}{ii}$ & 4227.74 & $\ion{O}{ii}$ & 4395.93 & $\ion{S}{ii}$ & 5212.267 & $\ion{Fe}{iii}$ & 5272.369 \\
$\ion{N}{ii}$ & 4237.05 & $\ion{O}{ii}$ & 4414.46 & $\ion{S}{ii}$ & 5212.62 & $\ion{Fe}{iii}$ & 5272.900 \\
$\ion{N}{ii}$ & 4241.76 & $\ion{O}{ii}$ & 4414.90 & $\ion{S}{ii}$ & 5345.712 & $\ion{Fe}{iii}$ & 5272.975 \\
$\ion{N}{ii}$ & 4432.74 & $\ion{O}{ii}$ & 4452.38 & $\ion{S}{ii}$ & 5346.084 & $\ion{Fe}{iii}$ & 5276.476 \\
$\ion{N}{ii}$ & 4433.48 & $\ion{O}{ii}$ & 4590.97 & $\ion{S}{ii}$ & 5428.655 & $\ion{Fe}{iii}$ & 5282.297 \\
$\ion{N}{ii}$ & 4447.03 & $\ion{O}{ii}$ & 4595.96 & $\ion{S}{ii}$ & 5432.797 & $\ion{Fe}{iii}$ & 5284.827 \\
$\ion{N}{ii}$ & 4601.48 & $\ion{O}{ii}$ & 4596.18 & $\ion{S}{ii}$ & 5639.977 & $\ion{Fe}{iii}$ & 5288.887 \\
$\ion{N}{ii}$ & 4601.69 & $\ion{O}{ii}$ & 4638.86 & $\ion{S}{ii}$ & 5640.346 & $\ion{Fe}{iii}$ & 5289.304 \\
$\ion{N}{ii}$ & 4607.15 & $\ion{O}{ii}$ & 4649.13 & $\ion{S}{ii}$ & 5647.02 & $\ion{Fe}{iii}$ & 5290.071 \\
$\ion{N}{ii}$ & 4613.87 & $\ion{O}{ii}$ & 4650.84 & $\ion{S}{iii}$ & 3632.024 & $\ion{Fe}{iii}$ & 5293.780 \\
$\ion{N}{ii}$ & 4621.39 & $\ion{O}{ii}$ & 4661.63 & $\ion{S}{iii}$ & 3662.008 & $\ion{Fe}{iii}$ & 5295.027 \\
$\ion{N}{ii}$ & 4630.54 & $\ion{O}{ii}$ & 4676.23 & $\ion{S}{iii}$ & 3717.771 & $\ion{Fe}{iii}$ & 5298.114 \\
$\ion{N}{ii}$ & 4643.09 & $\ion{O}{ii}$ & 4698.44 & $\ion{S}{iii}$ & 3928.595 & $\ion{Fe}{iii}$ & 5299.926 \\
$\ion{N}{ii}$ & 4654.53 & $\ion{O}{ii}$ & 4699.01 & $\ion{S}{iii}$ & 4253.589 & $\ion{Fe}{iii}$ & 5302.602 \\
$\ion{N}{ii}$ & 4779.72 & $\ion{O}{ii}$ & 4699.22 & $\ion{S}{iii}$ & 4284.979 & $\ion{Fe}{iii}$ & 5306.757 \\
$\ion{N}{ii}$ & 4780.44 & $\ion{O}{ii}$ & 4941.07 & $\ion{S}{iii}$ & 4294.402 & $\ion{Fe}{iii}$ & 5310.337 \\
$\ion{N}{ii}$ & 4781.19 & $\ion{O}{ii}$ & 4943.01 & $\ion{Ar}{ii}$ & 3603.904 & $\ion{Fe}{iii}$ & 5340.535 \\
$\ion{N}{ii}$ & 4788.14 & $\ion{Mg}{ii}$ & 4481.126 & $\ion{Ar}{ii}$ & 4013.856 & $\ion{Fe}{iii}$ & 5363.764 \\
$\ion{N}{ii}$ & 4803.29 & $\ion{Mg}{ii}$ & 4481.15 & $\ion{Ar}{ii}$ & 4072.004 & $\ion{Fe}{iii}$ & 5375.566 \\
$\ion{N}{ii}$ & 4987.38 & $\ion{Mg}{ii}$ & 4481.325 & $\ion{Ar}{ii}$ & 4072.325 & $\ion{Fe}{iii}$ & 5535.475 \\
$\ion{N}{ii}$ & 4994.36 & $\ion{Mg}{ii}$ & 7877.054 & $\ion{Ar}{ii}$ & 4072.384 & $\ion{Fe}{iii}$ & 5573.424 \\
$\ion{N}{ii}$ & 5001.13 & $\ion{Mg}{ii}$ & 7896.04 & $\ion{Ar}{ii}$ & 4372.095 & $\ion{Fe}{iii}$ & 5813.302 \\
$\ion{N}{ii}$ & 5001.47 & $\ion{Mg}{ii}$ & 7896.366 & $\ion{Ar}{ii}$ & 4372.490 & $\ion{Fe}{iii}$ & 5833.938 \\
$\ion{N}{ii}$ & 5005.15 & $\ion{Al}{iii}$ & 3601.630 & $\ion{Ar}{ii}$ & 4545.052 & $\ion{Fe}{iii}$ & 5848.744 \\
$\ion{N}{ii}$ & 5007.33 & $\ion{Al}{iii}$ & 3601.927 & $\ion{Ar}{ii}$ & 4579.349 & $\ion{Fe}{iii}$ & 5920.394 \\
$\ion{N}{ii}$ & 5010.62 & $\ion{Al}{iii}$ & 3612.355 & $\ion{Ar}{ii}$ & 4609.567 & $\ion{Fe}{iii}$ & 6032.673 \\
$\ion{N}{ii}$ & 5045.10 & $\ion{Al}{iii}$ & 4149.913 & $\ion{Ar}{ii}$ & 4657.901 & $\ion{Fe}{iii}$ & 7320.230 \\
$\ion{N}{ii}$ & 5073.59 & $\ion{Al}{iii}$ & 4149.968 & $\ion{Ar}{ii}$ & 4726.868 & $\ion{Fe}{iii}$ & 7920.559 \\
$\ion{N}{ii}$ & 5495.65 & $\ion{Al}{iii}$ & 4150.173 & $\ion{Ar}{ii}$ & 4735.905 & $\ion{Fe}{iii}$ & 7920.872 \\
$\ion{N}{ii}$ & 5666.63 & $\ion{Al}{iii}$ & 4479.885 & $\ion{Ar}{ii}$ & 4806.020 & $\ion{Fe}{iii}$ & 7921.186 \\
$\ion{N}{ii}$ & 5676.02 & $\ion{Al}{iii}$ & 4479.971 & $\ion{Ar}{ii}$ & 4965.079 & $\ion{Fe}{iii}$ & 7921.500 \\
$\ion{N}{ii}$ & 5679.56 & $\ion{Al}{iii}$ & 4480.000 & $\ion{Ar}{ii}$ & 6643.697 & $\ion{Fe}{iii}$ & 7921.814 \\
\hline
\end{tabular}
\end{table*}\noindent

\section{Comparison of the spectra with and without companion visible }

\begin{figure}
    \centering
    \includegraphics[width=\linewidth]{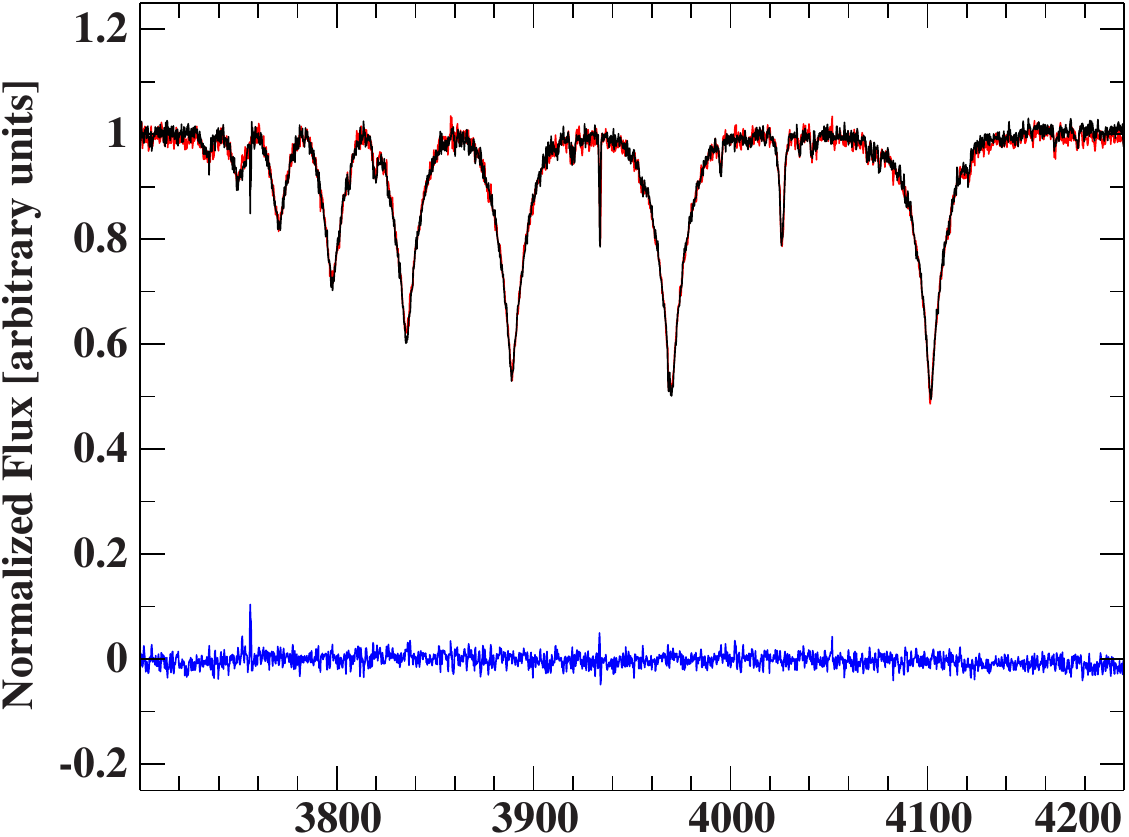}\vspace{2.25mm}
    \includegraphics[width=\linewidth]{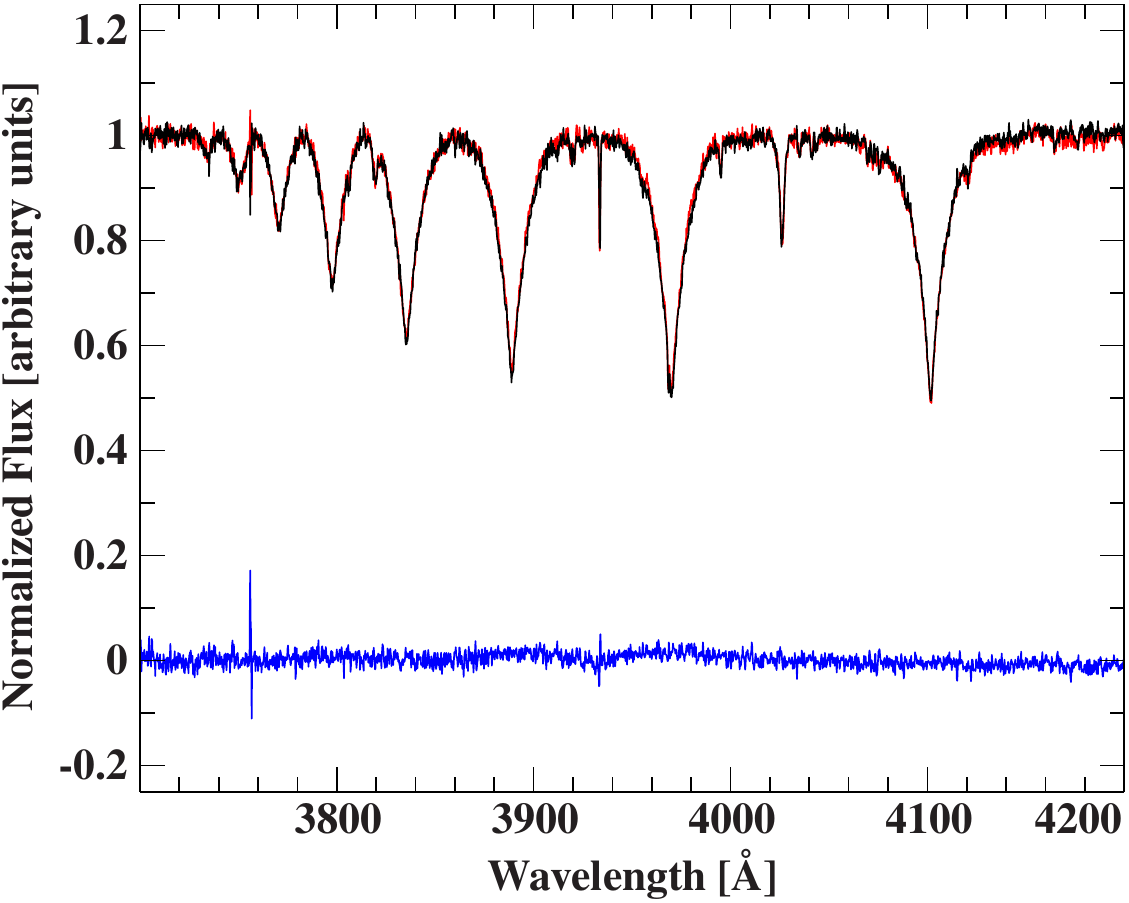}
   \caption{
   Subtraction of the XSHOOTER UVB spectrum in the secondary eclipse (black, orbital phase: 0.018) from the spectra before and after the secondary eclipse (red, orbital phases: 0.978, 0.058). The residuals are given in blue.}
    \label{balmer_companion_uvb}
\end{figure}

\begin{figure}
    \centering
    \includegraphics[width=\linewidth]{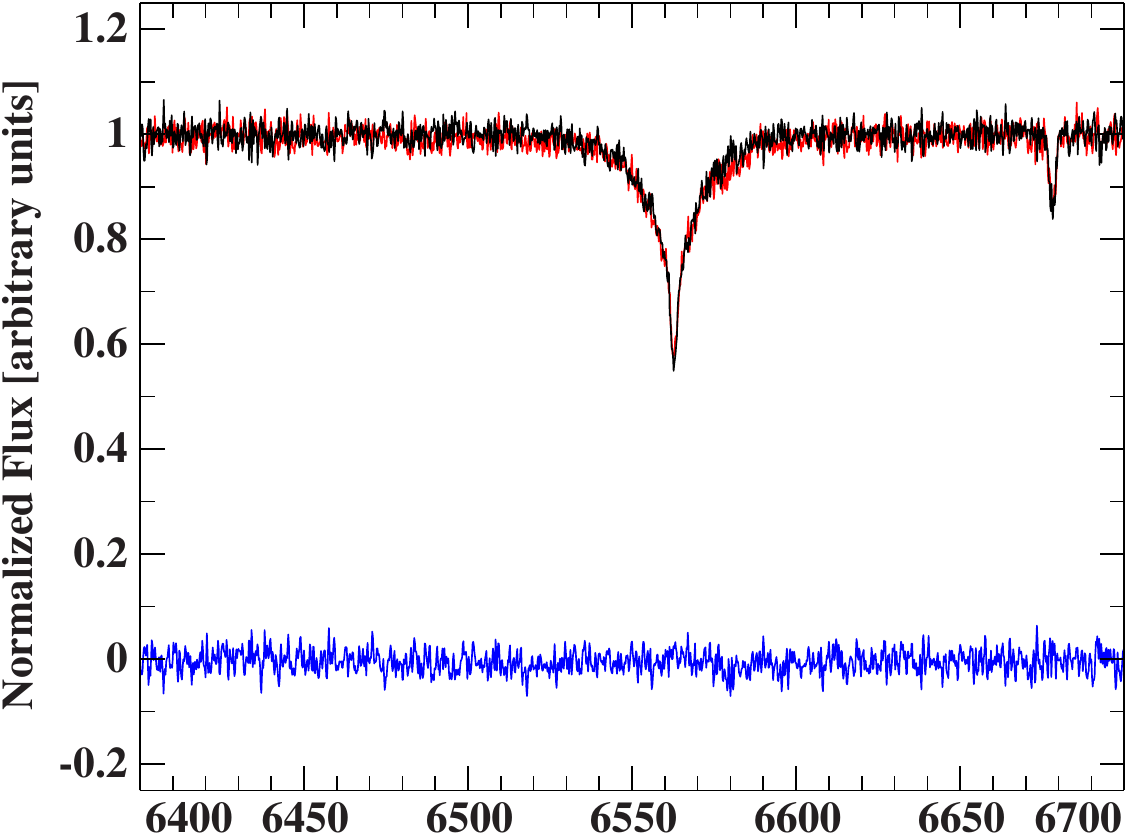}\vspace{2.25mm}
    \includegraphics[width=\linewidth]{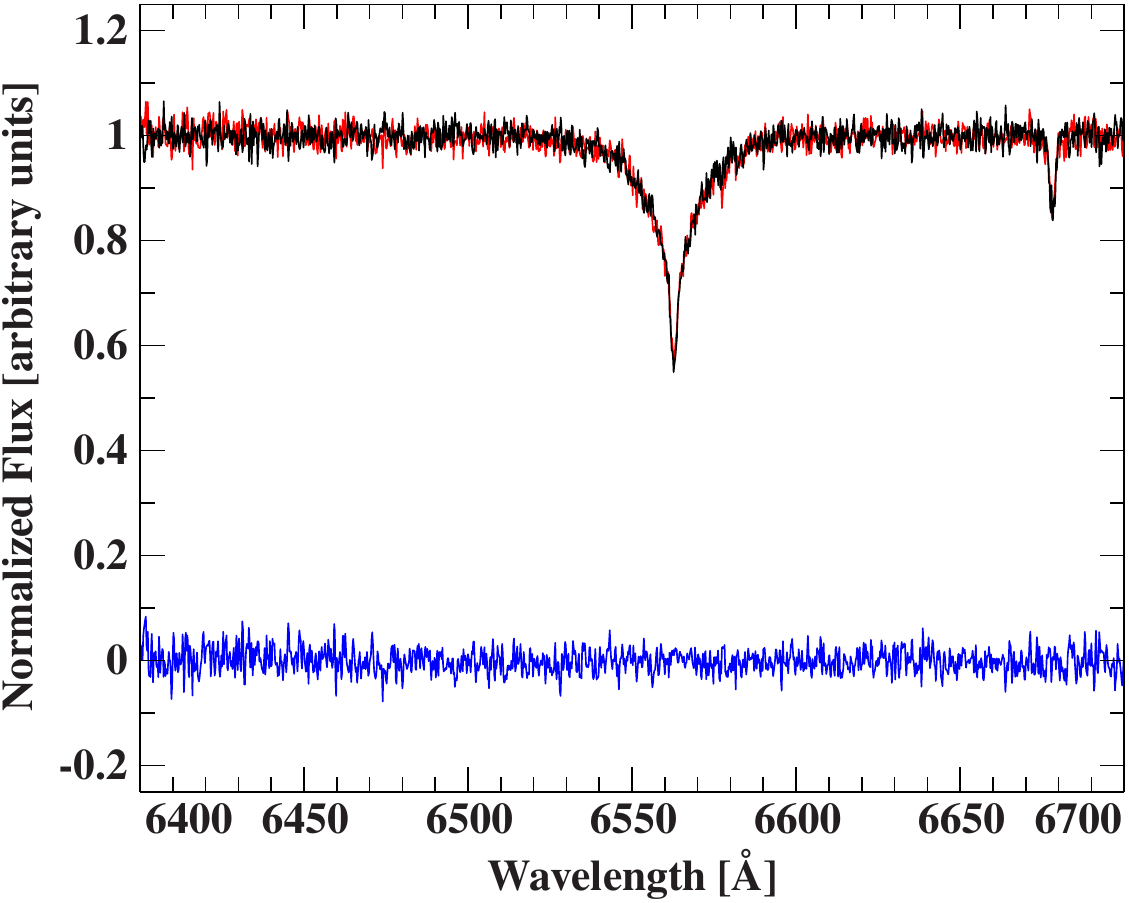}
    \caption{Same figure as Fig. \ref{balmer_companion_uvb}, but for the VIS arm around H$\alpha$.}
    \label{balmer_companion_vis}
\end{figure}

\section{Radial velocities}\label{app:RV}

\begin{table}
\caption{Radial velocities}\label{RVs}
\begin{tabular}{lrl}
\hline\hline
\noalign{\smallskip}
mid$-$BJD$_{TBD}$ & RV [${\rm km\,s^{-1}}$] & Instrument\\
$-2\,450\,000$ & & \\
\noalign{\smallskip}
\hline
\noalign{\smallskip}
3816.608090  &   -12.5 $\pm$ 6.8  & SDSS$^\dagger$ \\
3816.622170  &   -28.4 $\pm$ 6.2  & \\
3816.637894  &   -26.0 $\pm$ 7.1  & \\
3816.653623  &    21.4 $\pm$ 6.7  & \\
3816.669271  &    49.0 $\pm$ 10.5 & \\
3816.684919  &    38.4 $\pm$ 8.8  & \\
\noalign{\smallskip}
\hline
\noalign{\smallskip}
4755.79740  &    58.9 $\pm$  15.8 & EFOSC2$^\dagger$ \\
4755.80127  &    45.1 $\pm$  14.4 & \\
4757.84839  &   -40.7 $\pm$  15.0 & \\
4757.85225  &   -35.4 $\pm$  12.4 & \\
5146.80965  &    47.5 $\pm$  8.0  & \\
5146.82778  &    59.9 $\pm$  8.0  & \\
5146.83743  &    47.5 $\pm$  8.0  & \\
5147.80109  &    19.5 $\pm$  8.9  & \\
5147.81597  &   -14.9 $\pm$  8.1  & \\
5147.82562  &   -34.0 $\pm$  7.9  & \\
5147.84031  &   -29.3 $\pm$  8.9  & \\
5147.84997  &    -6.2 $\pm$  8.6  & \\
5147.86465  &    34.9 $\pm$  9.5  & \\
5147.87430  &    49.2 $\pm$  8.5  & \\
5148.77113  &     4.5 $\pm$  8.5  & \\
5148.77964  &   -20.1 $\pm$  7.9  & \\
5148.79388  &   -38.7 $\pm$  7.8  & \\
5148.80354  &   -37.5 $\pm$  9.0  & \\
\noalign{\smallskip}
\hline      
\noalign{\smallskip}
5657.49252374 &  42.5 $\pm$  5.3  & UVES \\
5657.49654962 &   4.9 $\pm$  3.7  & \\
5657.50057560 &  -0.9 $\pm$  5.2  & \\
5657.50461775 &  -4.8 $\pm$  5.6  & \\
5657.50866197 &  -9.1 $\pm$  3.1  & \\
5657.51270167 & -19.0 $\pm$  4.2  & \\
5657.51674196 & -21.5 $\pm$  2.3  & \\
5657.52107057 & -20.1 $\pm$  3.0  & \\
5657.52509622 & -17.0 $\pm$  7.9  & \\
5657.52912257 & -11.1 $\pm$  6.9  & \\
5657.53316028 &  -2.5 $\pm$  2.5  & \\
5657.53720219 &  12.2 $\pm$  6.8  & \\
5657.54124410 &  22.8 $\pm$  6.1  & \\
5657.54528808 &  33.1 $\pm$  6.7  & \\
5657.54932975 &  43.2 $\pm$  6.5  & \\
5657.55337328 &  60.0 $\pm$  4.5  & \\
5657.55741668 &  60.6 $\pm$  2.2  & \\
5657.56145535 &  69.1 $\pm$  7.6  & \\
5657.56549517 &  74.9 $\pm$  6.4  & \\
5657.56953256 &  73.7 $\pm$  5.2  & \\
5657.57357041 &  78.2 $\pm$  8.8  & \\
5657.57760966 &  65.4 $\pm$  6.9  & \\
5657.58569149 &  46.0 $\pm$  8.0  & \\
5657.58973698 &  29.8 $\pm$  7.4  & \\
5657.59377830 &  13.7 $\pm$  6.7  & \\
5657.59782484 &  -3.2 $\pm$  5.8  & \\
5657.60590841 & -15.8 $\pm$  5.1  & \\
5657.61398712 & -27.9 $\pm$  6.5  & \\
\noalign{\smallskip}               
\hline                             
\noalign{\smallskip}
7801.53891733 &  61.5 $\pm$  1.6  & XSHOOTER \\
7801.54280358 &  52.3 $\pm$  1.0  & \\ 
7801.54661453 &  42.4 $\pm$  0.9  & \\ 
7801.55049162 &  35.6 $\pm$  1.3  & \\ 
7801.55428555 &  10.5 $\pm$  1.2  & \\ 
7801.55816729 &   6.2 $\pm$  1.0  & \\ 
7801.56197835 &  -2.7 $\pm$  1.1  & \\
7801.56585983 & -11.2 $\pm$  1.0  & \\
7801.56967460 & -19.3 $\pm$  1.1  & \\
7801.57356027 & -22.1 $\pm$  1.2  & \\
7801.57736624 & -19.4 $\pm$  0.9  & \\
7801.58125282 & -17.1 $\pm$  0.9  & \\
7801.58774688 &  -5.4 $\pm$  0.9  & \\
7801.59163370 &   5.4 $\pm$  1.0  & \\
7801.59543481 &  14.9 $\pm$  1.0  & \\
7801.59932441 &  26.3 $\pm$  1.0  & \\
7801.60313072 &  38.8 $\pm$  0.8  & \\
7801.60701615 &  47.9 $\pm$  1.0  & \\
7801.61081599 &  58.7 $\pm$  1.0  & \\
7801.61851422 &  72.3 $\pm$  1.3  & \\
7801.62238622 &  74.5 $\pm$  0.9  & \\
7801.62620191 &  74.8 $\pm$  1.1  & \\
7801.63009857 &  69.0 $\pm$  1.2  & \\
\noalign{\smallskip}               
\hline       
    \multicolumn{3}{l}{$\dagger$ \citet{geier11c}}
\end{tabular} 
\end{table}



\section{Times of primary eclipses}

 \begin{table}\caption{Times of the primary eclipse of J08205+0008}\label{ecl_time}
     \centering
     \begin{tabular}{lll}\hline\hline
        eclipse number  & time of primary eclipse & source \\
        & [BJD$_{TDB}$] &\\\hline
 0 & $2455165.709266 \pm 0.000050$ & Merope$^\dagger$\\
31 & $2455168.692622 \pm 0.000050$ & Merope$^\dagger$\\
465 & $2455210.461047 \pm 0.000050$ & Merope$^\dagger$\\
466 & $2455210.557368 \pm 0.000050$ & Merope$^\dagger$\\
467 & $2455210.653586 \pm 0.000050$ & Merope$^\dagger$\\
3980 & $2455548.747324 \pm 0.000020$ & ULTRACAM\\
4704 & $2455618.425621 \pm 0.000050$ & BUSCA\\
4745 & $2455622.371553 \pm 0.000050$ & BUSCA\\
8071 & $2455942.468130 \pm 0.000050$ & SAAO\\
8072 & $2455942.564480 \pm 0.000010$ & SAAO\\
11179 & $2456241.584370 \pm 0.000020$ & SAAO\\
12103 & $2456330.510900 \pm 0.000030$ & SAAO\\
12113 & $2456331.473260 \pm 0.000020$ & SAAO\\
12164 & $2456336.381490 \pm 0.000030$ & SAAO\\
12165 & $2456336.477810 \pm 0.000040$ & SAAO\\
12537 & $2456372.279310 \pm 0.000010$ & SAAO\\
12568 & $2456375.262750 \pm 0.000080$ & SAAO\\
12973 & $2456414.240300 \pm 0.000050$ & SAAO\\
13035 & $2456420.207310 \pm 0.000040$ & SAAO\\
15822 & $2456688.430140 \pm 0.000020$ & SAAO\\
15832 & $2456689.392530 \pm 0.000050$ & SAAO\\
15863 & $2456692.376020 \pm 0.000020$ & SAAO\\
16101 & $2456715.281300 \pm 0.000050$ & SAAO\\
16132 & $2456718.264780 \pm 0.000100$ & SAAO\\
16703 & $2456773.218230 \pm 0.000050$ & SAAO\\
16724 & $2456775.239280 \pm 0.000030$ & SAAO\\
18567 & $2456952.610940 \pm 0.000020$ & SAAO\\
19459 & $2457038.457650 \pm 0.000020$ & SAAO\\
19470 & $2457039.516330 \pm 0.000030$ & SAAO\\
19490 & $2457041.441110 \pm 0.000030$ & SAAO\\
19739 & $2457065.405120 \pm 0.000030$ & SAAO\\
19780 & $2457069.350940 \pm 0.000040$ & SAAO\\
20413 & $2457130.271420 \pm 0.000030$ & SAAO\\
20444 & $2457133.254870 \pm 0.000020$ & SAAO\\
20714 & $2457159.239800 \pm 0.000030$ & SAAO\\
20724 & $2457160.202340 \pm 0.000030$ & SAAO\\
23179 & $2457396.473170 \pm 0.000080$ & SAAO\\
23210 & $2457399.456640 \pm 0.000030$ & SAAO\\
23459 & $2457423.420580 \pm 0.000010$ & SAAO\\
23490 & $2457426.404090 \pm 0.000020$ & SAAO\\
27710 & $2457832.53995 \pm 0.000020$ & ULTRACAM\\
28330 & $2457892.209190 \pm 0.000050$ & SAAO\\
31179 & $2458166.399050 \pm 0.000080$ & SAAO\\
31480 & $2458195.367510 \pm 0.000030$ & SAAO\\
34868 & $2458521.431060 \pm 0.000030$ & SAAO\\
35469 & $2458579.271780 \pm 0.000030$ & SAAO\\
37872 & $2458810.538190 \pm 0.000030$ & SAAO\\
37883 & $2458811.596860 \pm 0.000030$ & SAAO\\
37893 & $2458812.559230 \pm 0.000030$ & SAAO\\
39034 & $2458922.369960 \pm 0.000010$ & SAAO\\
39117 & $2458930.357940 \pm 0.000030$ & SAAO\\
               
        \hline
    \multicolumn{3}{l}{$\dagger$ \citet{geier11c}}
 
     \end{tabular}

 \end{table}

\bsp	
\label{lastpage}
\end{document}